\documentclass[aps,prd,preprint,superscriptaddress,nofootinbib]{revtex4-2}
\usepackage{amsmath, bm}
\usepackage{graphicx}

\begin{document}

\title{Resummation of threshold double logarithms in \\quarkonium fragmentation functions}

\author{Hee Sok Chung}
\email{heesokchung@gwnu.ac.kr}
\affiliation{Department of Mathematics and Physics, Gangneung-Wonju National
University, Gangneung 25457, Korea} 
\affiliation{Department of Physics, Korea University, Seoul 02841, Korea}

\author{U-Rae Kim}
\email{kim87@kma.ac.kr}
\affiliation{Department of Physics, Korea Military Academy, Seoul 01805, Korea}

\author{Jungil Lee}
\email{jungil@korea.ac.kr}
\affiliation{Department of Physics, Korea University, Seoul 02841, Korea}

\date{\today}

\begin{abstract}
We develop a formalism for resumming threshold double logarithms that appear in
fragmentation functions for production of heavy quarkonia. 
Threshold singularities appear in fixed-order calculations of quarkonium 
fragmentation functions in the nonrelativistic QCD factorization formalism due
to radiation of soft gluons. 
Because of this, fixed-order quarkonium fragmentation functions are not 
positive definite, and can lead to unphysically negative cross sections. 
This problem can be resolved by resumming threshold logarithms to all orders in
perturbation theory, which renders the fragmentation functions finite and
ensures the positivity of cross sections. We present a detailed derivation of
the resummation formalism and derive the formula for resummed quarkonium
fragmentation functions, which can be computed entirely within perturbation
theory without the need for nonperturbative model functions. 
\end{abstract}

\maketitle

%==============================================================================
\section{Introduction}
%==============================================================================

It has been known for a long time that short-distance hadronic cross sections 
involve large corrections from radiation of soft gluons, which can be resummed
by exponentiation~\cite{
Kodaira:1981nh, Kodaira:1982az, Davies:1984sp, Collins:1984kg, Sterman:1986aj, 
Catani:1989ne, Catani:1990rp, Catani:1990rr, Laenen:1991af, 
Catani:1991kz, Catani:1991bd, Catani:1992jc, Catani:1992ua, 
Berger:1996ad, Catani:1996yz}. 
In fixed-order perturbation theory, the soft-gluon corrections produce
logarithmic singularities that spoil the convergence of the perturbative
expansion near the boundaries of phase space. 
As the strongest singularities of this type appear as double logarithms, they
are commonly referred to as threshold double logarithms. 
These logarithms can be resummed to all orders in perturbation theory with the
help of perturbative factorization, and only then the cross sections become 
finite and smooth near the threshold. 

It has also been known that fragmentation functions for production of
quarkonia, when computed in the nonrelativistic QCD (NRQCD) factorization 
formalism~\cite{Bodwin:1994jh},
involve singularities of similar nature. The appearance of threshold double
logarithms in quarkonium fragmentation functions was first discovered by
Braaten and Lee in the next-to-leading order (NLO) calculation of the
color-octet $^3S_1$ gluon fragmentation function for heavy quarkonium in
Ref.~\cite{Braaten:2000pc}. Later, threshold double logarithms were also found
in the NLO calculation of $P$-wave fragmentation functions in
Ref.~\cite{Zhang:2020atv}. Because these singularities are
associated with radiation
of soft gluons, quarkonium fragmentation functions computed in fixed-order
perturbation theory involve distributions that become increasingly singular at
the kinematical threshold with increasing order in the strong coupling. 
For some time these singularities in fragmentation functions have been 
considered not much more than a technical nuisance that requires some 
care when computing cross sections~\cite{Bodwin:2014gia, Bodwin:2015iua}, 
as they are integrable and do not produce actual divergences in
sufficiently inclusive observables such as transverse-momentum-dependent cross
sections. This conventional viewpoint faced a serious challenge when it was
found that quarkonium production rates computed at NLO accuracy can turn
unphysically negative at very large transverse momentum. 
This happens because
the threshold double logarithmic singularities in fragmentation functions can
indeed produce dynamical effects in production rates, because they produce 
radiative corrections of the form $\alpha_s \log^2 p_T$, where $\alpha_s$ is
the strong coupling and $p_T$ is the transverse momentum of the heavy
quarkonium. This lead to an effort to theoretically understand the threshold
logarithms in quarkonium fragmentation functions to all orders in perturbation
theory~\cite{Chung:2023ext, Chen:2023gsu}, so that measurements of large-$p_T$
production rates of $J/\psi$ in
Ref.~\cite{ATLAS:2023qnh} could be well described in QCD. 

The first complete calculation of resummation of threshold double logarithms in
inclusive hadroproduction rates of heavy quarkonium was presented in
Ref.~\cite{Chung:2024jfk}. Similarly to the case of hadronic cross sections in
perturbative QCD, after resummation the quarkonium fragmentation functions no
longer involve singularities at the kinematical threshold. Because of this, 
at least for some choice of NRQCD long-distance matrix elements, it is possible
to construct quarkonium fragmentation functions that are positive definite over
the whole kinematically allowed region, which ensures positivity of production
cross sections. In contrast, this is impossible in
fixed-order perturbation theory, because the singularities produce
distributions that change sign rapidly at the kinematical end point. 
Hence, only after threshold resummation can the large-$p_T$ production rates of
heavy quarkonia be under theoretical control. 

In this paper, we present the detailed derivation of the threshold resummation
formalism used in Ref.~\cite{Chung:2024jfk}. The formalism is based on standard
methods of perturbative factorization in QCD, such as the Grammer-Yennie
approximation~\cite{Grammer:1973db} that systematically isolates the source of
infrared singularities of Feynman diagrams in gauge theories. This leads to a
soft factorization of the quarkonium fragmentation function in terms of soft
functions, which completely contain the threshold singularities in the
fragmentation function. The threshold singularities in the soft functions take
an especially simpler form at the double logarithmic level, which allows an
all-orders resummation by exponentiation. This results in fragmentation
functions that are free of threshold singularities, and are smooth and well
behaved at the kinematical end point, similar to nonperturbative model
functions used to describe fragmentation functions for single inclusive hadron
production. 

This paper is organized as follows. In Sec.~\ref{sec:FFs} we collect the known
results for fixed-order calculations of quarkonium fragmentation functions in
the NRQCD factorization formalism. In Sec.~\ref{sec:softfac} we derive the soft
factorization formulas for quarkonium fragmentation functions. The validity of
soft factorization is then verified at leading nonvanishing order in the strong
coupling in Sec.~\ref{sec:softLO}. In Sec.~\ref{sec:softNLO} we compute the
soft functions at NLO in the strong coupling, and identify
the source of the double logarithmic singularities at threshold. 
Based on the NLO results of the soft functions, the resummation formula for 
threshold double logarithms in quarkonium fragmentation functions are derived
in Sec.~\ref{sec:resum}. We conclude in Sec.~\ref{sec:conclusions}.

%==============================================================================
\section{Threshold logarithms in quarkonium fragmentation functions} 
\label{sec:FFs}
%==============================================================================

%==============================================================================
\subsection{NRQCD factorization}
%==============================================================================
In the NRQCD factorization formalism~\cite{Bodwin:1994jh}, 
the transverse-momentum-dependent 
inclusive production rates of a heavy quarkonium $\cal Q$ at large transverse
momentum $p_T$ is described by the factorization formula 
\begin{align}
\label{eq:NRQCDfac}
\frac{d \sigma_{\cal Q}}{d p_T} 
= \sum_{\cal N} \frac{d \sigma_{Q \bar Q ({\cal N})}}{dp_T}
\langle {\cal O}^{\cal Q} ({\cal N}) \rangle, 
\end{align}
where the short-distance coefficient (SDC) 
$d \sigma_{Q \bar Q ({\cal N})}/ dp_T$
corresponds to the $p_T$-differential cross section of a heavy quark $Q$ and 
an antiquark $\bar Q$ pair in a specific color and angular momentum state
${\cal N}$, and $\langle {\cal O}^{\cal Q} ({\cal N}) \rangle$ is the NRQCD
long-distance matrix element (LDME) that describes the nonperturbative
evolution of the $Q \bar Q$ in the state $\cal N$ into a quarkonium ${\cal Q}$
plus light particles. The sum over ${\cal N}$ is usually truncated at a chosen
accuracy in the nonrelativistic expansion; for the $P$-wave quarkonia such as
${\cal Q}=\chi_{cJ}$ with $J=0$, 1, and 2 the $^3P_J^{[1]}$ and $^3S_1^{[8]}$
channels appear at leading order in the nonrelativistic expansion, while for
$S$-wave quarkonia such as ${\cal Q} = J/\psi$ or $\psi(2S)$ the 
$^3S_1^{[1]}$, $^3S_1^{[8]}$, $^1S_0^{[8]}$, and $^3P_J^{[8]}$ channels give
dominant contributions to the cross section. 

The SDCs can be computed perturbatively. At large $p_T$, we can compute the
SDCs in the leading-power (LP) fragmentation formalism~\cite{Collins:1981uw,
Nayak:2005rt}
\begin{align}
\label{eq:LPfac}
\frac{d \sigma^{\rm LP}_{Q \bar Q ({\cal N})}}{dp_T}
= \sum_{i=g,q,\bar{q}} 
\int_0^1 dz \frac{d \hat{\sigma}_{i(K)}}{dp_T} 
D_{i \to Q \bar Q({\cal N})} (z), 
\end{align}
where $\hat{\sigma}_{i(K)}$ is the cross section for production of a massless
parton $i=g$, $q$, or $\bar q$ with momentum $K$, 
$D_{i \to Q \bar Q({\cal N})} (z)$ is the fragmentation function (FF) for
fragmentation of a parton $i$ into $Q \bar Q({\cal N})$, 
$z = P^+/K^+$ is the fraction of the $Q \bar Q$ momentum $P$ along the $+$
direction, which is taken to be the direction of $P$ in the lab frame. 
Both the $\hat{\sigma}_{i(K)}$ and the FFs $D_{i \to Q \bar Q({\cal N})} (z)$
are available to NLO accuracy,\footnote{Very recently, next-to-next-to-leading
order calculation of the parton production cross sections was reported in
Ref.~\cite{Czakon:2025yti}.} 
which allow computation of the SDCs to NLO at large $p_T$. 
The validity of Eq.~(\ref{eq:LPfac}) is based on the collinear factorization
theorem of QCD, which is the foundation of all first-principles calculation 
of single inclusive hadron production at large transverse momentum; for heavy
quarkonium production in NRQCD, the formula has been tested numerically up to
$p_T =400$~GeV at NLO accuracy~\cite{Bodwin:2015iua, Chung:2025gjk}. 
Corrections to Eq.~(\ref{eq:LPfac}), which are
suppressed by $m^2/p_T^2$ with $m$ the heavy quark mass, are given in terms of
next-to-leading-power (NLP) factorization. 

The $D_{i \to Q \bar Q({\cal N})} (z)$ in Eq.~(\ref{eq:LPfac}) can be computed 
perturbatively from the Collins-Soper definition for the FFs~\cite{Collins:1981uw}. 
When computed in fixed-order perturbation theory, the FFs involve distributions
that are singular at threshold $z=1$ such as plus distributions or the delta
function $\delta(1-z)$, because there are diagrams where a $Q \bar Q$ are
produced from a gluon after radiation of a fixed number of soft gluons, or even 
without emitting any partons at all. 
Because the singularities in the FFs are integrable, and the parton production
cross sections $d \hat{\sigma}_{i(K)}$ are regular functions in $z$, 
the cross sections $d \sigma^{\rm LP}_{Q \bar Q({\cal N})}$ are finite despite
these singularities. 
However, the singularities do affect the cross sections because the 
$d \hat{\sigma}_{i(K)}$ steeply rise as $z \to 1$ as a power in $z$, 
which makes the $d \sigma^{\rm LP}_{Q \bar Q({\cal N})}$ very sensitive to the
behavior of the FFs near threshold\footnote{
Note that this problem is not so severe in calculation of single inclusive
production rates of light hadrons, because in such calculations the
nonperturbative model functions for the FFs are taken to be regular functions
in $z$ that usually vanish as $z \to 1$.}. This problem is especially important in
hadroproduction, because the steep rise of the $d \hat{\sigma}_{i(K)}$ in $z$
becomes stronger with increasing $p_T$, due to the decrease of parton densities
in the initial state. Hence, the singular behavior of the FFs affect the $p_T$
shapes of the cross sections $d \sigma^{\rm LP}_{Q \bar Q({\cal N})}$. 

Explicit calculations of the FFs $D_{i \to Q \bar Q({\cal N})} (z)$ at NLO
accuracy show that the strongest of the singularities at NLO appear as 
{\it threshold double logarithms}, which are proportional to 
$\alpha_s \left[ \frac{\log (1-z)}{1-z} \right]_+ \otimes 
D^{\rm LO} (z)$~\cite{Braaten:2000pc, Zhang:2020atv}. 
Here the convolution $\otimes$ is defined by 
\begin{align}
\label{eq:convolution}
(f \otimes g) (z) = \int_0^1 \frac{dz'}{z'} f(z') g(z/z'). 
\end{align}
Appearance of threshold double logarithms in perturbative calculations of FFs
has been known for a long time, such as in $D_{q \to q} (z)$ (see, for example, 
Ref.~\cite{Sterman:1986aj}). These singularities become more severe with
increasing order in the strong coupling $\alpha_s$, as they appear in the form 
$\alpha_s^n \left[ \frac{\log^{2n-1} (1-z)}{1-z} \right]_+ \otimes
D^{\rm LO} (z)$. 
As threshold logarithms at higher orders in $\alpha_s$ will have increasingly 
stronger effects on the $p_T$ shapes of the cross sections 
$d \sigma^{\rm LP}_{Q \bar Q({\cal N})}$, 
the cross sections computed in fixed-order perturbation theory will have very
different $p_T$ shapes depending on at which order the perturbation series has
been truncated. 
Therefore, to obtain a proper theoretical description of
large-$p_T$ production rates of heavy quarkonia, it is imperative to understand
the all-orders structure of threshold double logarithms in quarkonium FFs.

%==============================================================================
\subsection{Threshold singularities in gluon fragmentation functions}
%==============================================================================

We now summarize the threshold double logarithms that appear in 
NLO calculations of gluon FFs. 
The gluon FFs for the $^3S_1^{[8]}$, $^3P_J^{[8]}$,
$^1S_0^{[8]}$, and $^3P_J^{[1]}$ channels to order-$\alpha_s^2$ accuracy 
are given by~\cite{Braaten:1996rp, Bodwin:2003wh, 
Braaten:2000pc, Lee:2005jw, Bodwin:2012xc, Ma:2013yla}
%------------
\begin{subequations}
\label{eq:FOFFs}
\begin{align}
%------------
\label{eq:3S18FF}
D_{g \to Q \bar Q(^3S_1^{[8]})} (z) &= 
\frac{\pi \alpha_s}{(d-1) (N_c^2-1) m^3} \bigg\{ \delta (1-z) +  
\frac{\alpha_s}{\pi} \bigg[
A(\mu_R) \delta (1-z) 
\nonumber \\ & \hspace{4ex}  
+ \left( \log\frac{\mu_F}{2 m} - \frac{1}{2} \right) P_{gg} (z)
+ \frac{3 (1-z)}{z} 
\nonumber \\ & \hspace{4ex}  
+ 6 (2-z+z^2) \log(1-z) 
- \frac{2 N_c}{z} \left( \frac{\log(1-z)}{1-z} \right)_+ 
\bigg] \bigg\} + O(\alpha_s^3),
\\
\label{eq:3PJ8FF}
D_{g \to Q \bar Q(^3P_J^{[8]})} (z) &= 
\frac{8 \alpha_s^2}{3 (d-1) (N_c^2-1) m^5} \frac{N_c^2-4}{4 N_c} 
\bigg[ 
\left( \frac{1}{6}- \log \frac{\mu_\Lambda}{2 m} \right) \delta(1-z) 
\nonumber \\ 
& \quad \quad 
+ \frac{1}{(1-z)_+} 
+ \frac{13-7 z}{4} \log(1-z) 
- \frac{(1-2 z) (8-5 z)}{8} \bigg]
+O(\alpha_s^3), 
\\
D_{g \to Q \bar Q(^1S_0^{[8]})} (z) &= 
\frac{\alpha_s^2}{8 m^3} \frac{N_c^2-4}{4 N_c} 
[3 z-2 z^2+2 (1-z) \log(1-z)]
+O(\alpha_s^3),
\\
\label{eq:3PJ1FF}
D_{g \to Q \bar Q(^3P_J^{[1]})} (z) &= 
\frac{2 \alpha_s^2}{9 N_c^2 m^5} \bigg[ \left( \frac{Q_J}{2 J+1} - 
\log \frac{\mu_\Lambda}{2 m} \right) \delta(1-z) 
\nonumber \\ & \hspace{15ex} 
+ \frac{z}{(1-z)_+} + \frac{P_J(z)}{2 J+1} \bigg]
+O(\alpha_s^3),
%------------
\end{align}
\end{subequations}
%------------
where $d$ is the number of spacetime dimensions, $N_c=3$ is the number of
colors, $A(\mu_R) = \frac{\beta_0}{2} \left( \log \frac{\mu_R}{2 m} +
\frac{13}{6} \right) + \frac{2}{3} - \frac{\pi^2}{2} + 8 \log 2$,
$\beta_0 = \frac{11}{3} N_c+\frac{2}{3} n_f$ is the QCD beta function at one
loop,  
$P_{gg} (z) = 2 C_A \left[ \frac{z}{(1-z)_+} + \frac{1-z}{z} + z (1-z) +
\frac{\beta_0}{12} \delta(1-z) \right]$ the gluon splitting function,
$C_A = N_c$, 
and the $J$-dependent constants and functions are given by 
%------------
\begin{align}
%------------
Q_0 &= \frac{1}{4}, \quad Q_1 = \frac{3}{8}, \quad Q_2 = \frac{7}{8}, 
\\
P_0 (z) &= \frac{z (85-26 z)}{8} + \frac{9 (5-3 z)}{4} \log (1-z),
\\
P_1 (z) &= - \frac{3 z (1+4 z)}{4},
\\
P_2 (z) &= \frac{5 z (11-4 z)}{4} + 9 (2-z) \log (1-z).
%------------
\end{align}
%------------
The scales $\mu_R$ and $\mu_F$ come from the renormalization of the QCD
coupling and the FF, respectively. 
The scale $\mu_\Lambda$ that appears in the $P$-wave FFs 
is the renormalization scale for the $^3S_1^{[8]}$ matrix element in the 
$\overline{\rm MS}$ scheme. The explicit logarithmic dependence on
$\mu_\Lambda$ appears in the $P$-wave FFs because the IR pole that arises from
the phase space integral of the $P$-wave production process is canceled by the
one-loop correction to the $^3S_1^{[8]}$ matrix element. 
The gluon FF for the $^3S_1^{[1]}$ state, which begins at order $\alpha_s^3$, 
has been obtained in Ref.~\cite{Braaten:1993rw} as an integral representation; later, an analytical
result has been obtained in Ref.~\cite{Zhang:2017xoj}. 

Note that the gluon FF for the $^3S_1^{[8]}$ state, as well 
as those for the $^3P_J^{[8]}$ and $^3P_J^{[1]}$ states, contain distributions
that are singular at $z=1$ at leading nonvanishing orders in $\alpha_s$. 
Due to these singularities at the threshold, the cross sections $\sigma_{Q \bar
Q({\cal N})}$ are very sensitive to the behavior of the gluon cross sections
$\hat{\sigma}_{g(K)}$ near $z=1$. The severity of the singularities can be
quantified through the Mellin transform. The definition of the Mellin transform
and some useful formulas can be found in Appendix~\ref{sec:mellin}. 
We see that while the $\delta(1-z)$ term yields a nonzero constant in Mellin
space, the Mellin transform of $1/(1-z)_+$ diverges like $-\log N$, 
which makes the $z=1$ singularity in the plus distribution a more severe one
compared to the Dirac delta function. It is known that QCD radiative
corrections exacerbate these singularities; for example, the most singular term
in the N$^n$LO correction to the FF is proportional to 
%------------
\begin{align}
%------------
\left(\frac{\alpha_s}{\pi} \right)^n 
\left[\frac{\log^{2n-1}(1-z)}{1-z} \right]_+ \otimes D^{\rm LO} (z), 
%------------
\end{align}
%------------
where $D^{\rm LO} (z)$ is the FF at leading nonvanishing order in $\alpha_s$.
As shown in Appendix~\ref{sec:mellin}, the large-$N$ asymptotic behavior of this term in
Mellin space is 
%------------
\begin{align}
%------------
\int_0^1 \frac{dz}{z} z^N \left(\frac{\alpha_s}{\pi} \right)^n 
\left[\frac{\log^{2n-1}(1-z)}{1-z} \right]_+ \otimes D^{\rm LO} (z)
\sim  
\frac{1}{2n}
\left(\frac{\alpha_s}{\pi} 
\log^2N \right)^n \, \tilde{D}^{\rm LO} (N),
%------------
\end{align}
%------------
which becomes more singular as $n$ increases. 
As the leading logarithmic behavior is given by powers of $\alpha_s \log^2 N$,
these singularities are called {\it threshold double logarithms}. 
Due to the appearance of threshold logarithms, the sensitivity to the behavior
of the gluon cross section near $z=1$ becomes stronger with increasing order in
$\alpha_s$, which calls for the need to resummation. 

In the case of the $^1S_0^{[8]}$ and $^3S_1^{[1]}$ channels, the gluon FFs are
regular functions in $z$, meaning that the Mellin-space FFs vanish at least
like $1/N$ as $N \to \infty$. While radiative corrections may involve
logarithms in $N$, these logarithms will always be suppressed by at least a
power of $1/N$. Hence, even at higher orders in $\alpha_s$, the gluon FFs for
these channels will not involve distributions that are singular at $z=1$. 

The appearance of threshold logarithms in quarkonium FFs has been confirmed
explicitly from NLO calculations.  NLO corrections to the gluon FFs have been
computed for all $^3S_1^{[8]}$, $^3P_J^{[8]}$, $^1S_0^{[8]}$, and $^3P_J^{[1]}$
channels. The analytical result for the one-loop correction to $D_{g \to Q \bar
Q(^3S_1^{[8]})} (z)$ is shown in Eq.~(\ref{eq:3S18FF}). 
The radiative correction to $D_{g \to Q
\bar Q(^1S_0^{[8]})} (z)$, which only contains regular functions in $z$, has
been computed numerically~\cite{Artoisenet:2014lpa, Artoisenet:2018dbs}. 
The order-$\alpha_s^3$ corrections to $D_{g \to Q
\bar Q(^3P_J^{[8]})} (z)$ and $D_{g \to Q \bar Q(^3P_J^{[1]})} (z)$ have been
computed semianalytically, where coefficients of plus distributions and the
Dirac delta function are known exactly, while contributions from regular
functions are known numerically~\cite{Zhang:2020atv}. 
The order-$\alpha_s^3$ contribution to 
$D_{g \to Q \bar Q(^3P_J^{[1]})} (z)$ has the form
%------------
\begin{align}
%------------
D_{g \to Q \bar Q(^3P_J^{[1]})} (z) \big|_{\mu_R = \mu_F = \mu_\Lambda = 2 m} 
& = D_{g \to Q \bar Q(^3P_J^{[1]})}^{\rm LO} (z) 
+ \frac{2 \alpha_s^2}{9 N_c^2 m^5} 
\nonumber \\ & \quad \times
\frac{\alpha_s}{\pi} 
\left[ \sum_{n=0}^2 p_n^{J, [1]} \left( \frac{\log^{n} (1-z)}{1-z} \right)_+ 
+ p_\delta^{J, [1]} \delta(1-z) + p^{J, [1]}(z) \right],  
%------------
\end{align}
%------------
where $D_{g \to Q \bar Q(^3P_J^{[1]})}^{\rm LO} (z)$ is the order-$\alpha_s^2$
contribution shown in Eq.~(\ref{eq:3PJ1FF}), 
while the order-$\alpha_s^3$ contribution to
$D_{g \to Q \bar Q(^3P_J^{[8]})} (z)$ has the form
%------------
\begin{align}
%------------
D_{g \to Q \bar Q(^3P_J^{[8]})} (z) \big|_{\mu_R = \mu_F = \mu_\Lambda = 2 m} 
& = D_{g \to Q \bar Q(^3P_J^{[8]})}^{\rm LO} (z)
+ \frac{8 \alpha_s^2}{3 (d-1) (N_c^2-1) m^5} \frac{N_c^2-4}{4 N_c}
\nonumber \\ & \quad \times 
\frac{\alpha_s}{\pi}
\left[ \sum_{n=0}^2 p_n^{[8]} \left( \frac{\log^{n} (1-z)}{1-z} \right)_+
+ p_\delta^{[8]} \delta(1-z) + p^{[8]}(z) \right],
%------------
\end{align}
%------------
where $D_{g \to Q \bar Q(^3P_J^{[8]})}^{\rm LO} (z)$is the order-$\alpha_s^2$
contribution shown in Eq.~(\ref{eq:3PJ8FF}).
The coefficients of the $\left( \frac{\log^2 (1-z)}{1-z} \right)_+$ term 
are independent of the color and $J$ of the final state, and depend only on
$N_c$: 
%------------
\begin{align}
%------------
p_2^{J, [1]} = p_2^{[8]} = -4 N_c,
%------------
\end{align}
%------------
while the $p_1$, $p_0$, $p_\delta$, and the regular functions $p(z)$ in general
depend on $J$ and the color. 
By taking the Mellin transform, we can find the large-$N$ asymptotic behaviors
of the $^3S_1^{[8]}$, $^3P_J^{[8]}$, and $^3P_J^{[1]}$ FFs 
given by 
%------------
\begin{subequations}
\begin{align}
%------------
\tilde D_{g \to Q \bar Q(^3S_1^{[8]})} (N) &\sim 
\frac{\pi \alpha_s}{(d-1) (N_c^2-1) m^3} \bigg[ 1 - 
\frac{\alpha_s N_c}{\pi} \left( \log^2 N + \cdots \right) \bigg] 
+ O(\alpha_s^3),
\\
\tilde D_{g \to Q \bar Q(^3P_J^{[8]})} (N) &\sim 
\frac{8 \alpha_s^2}{3 (d-1) (N_c^2-1) m^5} \frac{N_c^2-4}{4 N_c}
\bigg[ 
- \log N 
+ O(1/N^0) 
\nonumber \\
& \hspace{32ex} 
+ \frac{4}{3} \frac{\alpha_s N_c}{\pi} \left( \log^3 N + \cdots \right) 
\bigg] 
+O(\alpha_s^4),
\\
\tilde D_{g \to Q \bar Q(^3P_J^{[1]})} (N) &\sim
\frac{2 \alpha_s^2}{9 N_c^2 m^5} \bigg[ 
- \log N 
+ O(1/N^0) 
+ \frac{4}{3} \frac{\alpha_s N_c}{\pi} \left( \log^3 N + \cdots \right)
\bigg] 
+O(\alpha_s^4),
%------------
\end{align}
\end{subequations}
%------------
where the $\cdots$ represent terms that are less divergent than the preceding
term. While the threshold double logarithm in the $^3S_1^{[8]}$
gluon FF is $- \frac{\alpha_s}{\pi} N_c \log^2 N$ times the
leading-order piece, the threshold double logarithm in the $P$-wave FFs 
is $- \frac{4}{3} \frac{\alpha_s}{\pi} N_c \log^2 N$ times the leading-order 
piece for both $^3P_J^{[8]}$ and $^3P_J^{[1]}$ channels. 
That is, if we only collect the leading threshold double logarithms in the NLO
correction, we can write 
%------------
\begin{subequations}
\label{eq:doublethresholdFF}
\begin{align}
%------------
\label{eq:doublethreshold3S1}
\tilde D_{g \to Q \bar Q(^3S_1^{[8]})} (N) &\sim
\left( 1 -
\frac{\alpha_s N_c}{\pi} \log^2 N + O(\alpha_s^2) \right) 
\tilde D^{\rm LO}_{g \to Q \bar Q(^3S_1^{[8]})} (N), 
\\
\label{eq:doublethreshold3PJ8}
\tilde D_{g \to Q \bar Q(^3P_J^{[8]})} (N) &\sim
\left( 1 -
\frac{4}{3} 
\frac{\alpha_s N_c}{\pi} \log^2 N + O(\alpha_s^2) \right)
\tilde D^{\rm LO}_{g \to Q \bar Q(^3P_J^{[8]})} (N), 
\\
\label{eq:doublethreshold3PJ1}
\tilde D_{g \to Q \bar Q(^3P_J^{[1]})} (N) &\sim
\left( 1 -
\frac{4}{3}
\frac{\alpha_s N_c}{\pi} \log^2 N + O(\alpha_s^2) \right)
\tilde D^{\rm LO}_{g \to Q \bar Q(^3P_J^{[1]})} (N). 
%------------
\end{align}
\end{subequations}
%------------
We see that not only the threshold double logarithms do appear in
quarkonium FFs, they begin to appear at different orders in
$\alpha_s$, because while $D_{g \to Q \bar Q(^3S_1^{[8]})}$ begins at order
$\alpha_s$, $D_{g \to Q \bar Q(^3P_J^{[8]})}$ and $D_{g \to Q \bar
Q(^3P_J^{[1]})}$ begin at order $\alpha_s^2$. 
Hence, in a fixed-order calculation, threshold logarithms will be always 
truncated inconsistently; for example, a fixed-order NLO 
calculation of quarkonium production rates will include contributions from FFs 
to order-$\alpha_s^2$ accuracy, which will include the first
threshold double logarithm in the $^3S_1^{[8]}$ channel, while the threshold
logarithms in the $P$-wave channels will be neglected. This inconsistency can
cause the quarkonium production rates computed at NLO to turn negative at large
transverse momentum as the negative correction from the threshold double
logarithm in the $^3S_1^{[8]}$ gluon FF becomes amplified.
Because of this, an all-orders resummation is the only proper way to treat the
threshold double logarithms in production of heavy quarkonia.

%==============================================================================
\subsection{Quark fragmentation and polarized fragmentation functions}
%==============================================================================

Before we move on to the next section, for completeness let us inspect the
quark FFs and polarized FFs for production of polarized quarkonia. 
The results for quark FFs and polarized FFs are summarized in
Refs.~\cite{Ma:2013yla, Ma:2015yka}. 

Quark FFs appear from order $\alpha_s^2$ in the $^3S_1^{[8]}$ and $^3S_1^{[1]}$ 
channels; they are continuous functions in $z$, and their Mellin transforms 
vanish like $1/N$ as $N \to \infty$. Hence, similarly to the case of the 
$^1S_0^{[8]}$ and $^3S_1^{[1]}$ gluon FFs, logarithmic 
corrections to the quark FF will be power suppressed and will not lead to
singular distributions. Due to the quark-gluon mixing in
the splitting functions, the threshold logarithms in the $^3S_1^{[8]}$ gluon
FF may mix into the quark FF at higher orders in $\alpha_s$; however, because
the quark-gluon mixing contribution in the splitting function vanishes like
$1/N$, these logarithms are always power suppressed when they appear in the
quark FF, and do not produce singular distributions in $z$ space. 

Finally, let us discuss the singularities in the polarized FFs. 
It is known from explicit calculations that to order $\alpha_s^2$ accuracy that
the polarized $^3S_1^{[8]}$, $^1S_0^{[8]}$, and $^3P_J^{[8]}$ gluon FFs that
produce longitudinally polarized $S$-wave vector quarkonium are regular
functions in $z$. The same is true for the polarized gluon FFs for the
$^3S_1^{[1]}$ channel, which begin at order $\alpha_s^3$, and also for the
quark FFs, which are known through order $\alpha_s^2$. 
Hence, the polarized FFs for production of longitudinally polarized $S$-wave
vector quarkonium do not involve singular distributions in $z$. 
In particular, the explicit NLO calculation of polarized $^3S_1^{[8]}$ gluon FF
shows that the relation in Eq.~(\ref{eq:doublethreshold3S1}) for leading 
threshold double logarithms in the $^3S_1^{[8]}$ gluon FF still holds for
production of transversely polarized $J/\psi$ or $\psi(2S)$~\cite{Braaten:2000pc}. 
In the case of $\chi_{cJ}$ or $\chi_{bJ}$ production, the helicity
decomposition for the $^3S_1^{[8]}$ and $^3P_J^{[1]}$ FFs shown in
Appendix~\ref{sec:helicitydecomp}, 
indicates that the gluon FF for production of polarized $\chi_{cJ}$ or
$\chi_{bJ}$ contains singularities for every helicity state, 
and the relation in Eq.~(\ref{eq:doublethreshold3S1}) for the $^3S_1^{[8]}$
gluon FF holds for every helicity state in the production of polarized
$\chi_{cJ}$ or $\chi_{bJ}$. 

While loop corrections to the $^3P_J^{[8]}$ and $^3P_J^{[1]}$ gluon FFs
have not been computed for the polarized case, we expect the relation in
Eq.~(\ref{eq:doublethreshold3PJ8}) to hold for production of transversely
polarized $S$-wave vector quarkonium, and
Eq.~(\ref{eq:doublethreshold3PJ1}) to hold for production of $\chi_{cJ}$ or
$\chi_{bJ}$ for every helicity state, because the attachments of soft gluons
that are responsible for generating threshold logarithms do not change the
heavy quark spin. We will show this in the following sections 
using the general analysis of threshold logarithms in quarkonium FFs.

%==============================================================================
\section{Soft factorization}
\label{sec:softfac}
%==============================================================================

In this section, we investigate the $z \to 1$ behavior of the gluon FFs 
in the presence of an arbitrary number of soft gluons, 
which will allow us to compute threshold logarithms to
arbitrarily high orders in $\alpha_s$.

\subsection{Grammer-Yennie approximation}

We begin with the operator definition of the gluon FF for production of a
hadron $H$, which is given by~\cite{Collins:1981uw}
%---------------
\begin{align}
\label{eq:gFFdef}
%---------------
D_{g \to H} (z) & = \frac{-g^{\mu \nu} z^{d-3}}{2 \pi K^+ (N_c^2-1) (d-2)}
\int_{-\infty}^{+\infty} dx^- e^{-i K^+ x^-} 
\nonumber \\ &  \quad \times 
\langle 0 | G_{+\mu}^c (0) \Phi_n^{\dag bc} (\infty,0) a^\dag_{H(P)} a_{H(P)} 
\Phi_n^{ba} (\infty,x^-) G_{+\nu}^a (n x^-) | 0 \rangle , 
%---------------
\end{align}
%---------------
where $K$ is the momentum of the initial-state gluon, 
$G_{\mu \nu} = G_{\mu \nu}^a T^a$ is the QCD field-strength tensor, 
$\Phi_k^{ab} (y,x)$ is a
path-ordered exponential of the gluon field in the $k$ direction 
in the adjoint representation defined by 
%---------------
\begin{align}
%---------------
\Phi_k (y,x) = {\cal P} \exp \left[ -i g \int_x^y d \lambda \, 
k \cdot A^{\rm adj} (k \lambda) \right], 
%---------------
\end{align}
%---------------
with ${\cal P}$ the path ordering, 
$n$ is a lightlike vector defined through $n \cdot K = K^+$, 
and $a_{H(P)}$ is an operator that annihilates an $H$ with momentum $P$, 
so that $a^\dag_{H(P)} a_{H(P)}$ projects onto states that contain an $H$ with
momentum $P$. 
Our goal is to find the $z \to 1$ behavior for production of a $Q \bar Q$ in a
specific color and angular momentum state. 
The production of a $Q \bar Q$ from a fragmenting gluon occurs through the
process $g^* \to Q \bar Q$, followed by radiation of an arbitrary number of
gluons from $Q$ and $\bar Q$ lines. Because we are interested in the region $z
\approx 1$, we only consider the case where the gluon attachments to $Q$ or
$\bar Q$ are soft. In this case, we can apply the Grammer-Yennie approximation
(soft approximation) to simplify the Feynman rule for soft-gluon vertices~\cite{Grammer:1973db, Collins:1981uk, Nayak:2005rt}. 
The outermost gluon attachment to the $Q$ line can be simplified as 
%---------------
\begin{align}
%---------------
\bar{u} (p_1) \gamma^\mu T^a \frac{i(p\!\!\!/_1+k\!\!\!/+m) }{(p_1+k)^2-m^2+i \varepsilon}
&=
\bar{u}(p_1) T^a \frac{i[2 (p_1^\mu + k^\mu) - (p\!\!\!/_1+k\!\!\!/) \gamma^\mu + m
\gamma^\mu] }{2 p_1 \cdot k + k^2 +i \varepsilon}
\nonumber \\
&=
\bar{u}(p_1) T^a \frac{i[2 (p_1^\mu + k^\mu) - k\!\!\!/ \gamma^\mu ]}{2 p_1 \cdot k + k^2 +i \varepsilon} 
\nonumber \\
&\approx
\bar{u}(p_1) T^a \frac{i p_1^\mu }{p_1 \cdot k + i \varepsilon}, 
%---------------
\end{align}
%---------------
where $p_1$ is the momentum of the $Q$, and $k$ is the outgoing momentum of the
gluon. In the last line, we neglected $k$ compared to $p_1$ in the numerator,
and neglected $k^2$ compared to $p_1 \cdot k$ in the denominator. 
Similarly, the outermost gluon attachment to the $\bar Q$ line can be
simplified as 
%---------------
\begin{align}
%---------------
\frac{i(-p\!\!\!/_2-k\!\!\!/+m) }{(p_2+k)^2-m^2+i \varepsilon} \gamma^\mu T^a v(p_2)
&=
\frac{i [ -2 (p_2^\mu + k^\mu) + \gamma^\mu (p\!\!\!/_2+k\!\!\!/)
+ m \gamma^\mu]}{2 p_2 \cdot k + k^2 +i \varepsilon} T^a v(p_2)
\nonumber \\
&=
- \frac{i [ 2 (p_2^\mu + k^\mu) - \gamma^\mu k\!\!\!/
]}{2 p_2 \cdot k + k^2 +i \varepsilon} T^a v(p_2)
\nonumber\\
&\approx
- \frac{i p_2^\mu }{p_2 \cdot k +i \varepsilon} T^a v(p_2), 
%---------------
\end{align}
%---------------
with $p_2$ the momentum of the $\bar Q$. 
For multiple gluon attachments, the same approximation can be applied
sequentially, beginning with the outermost gluon attachment. 
An $n$-gluon attachment to a $Q$ line in the soft approximation leads to
%---------------
\begin{align}
%---------------
\bar{u}(p_1) 
\frac{i p_1^{\mu_1} T^{a_1} }{p_1 \cdot k_1 + i \varepsilon}
\frac{i p_1^{\mu_2} T^{a_2} }{p_1 \cdot k_2 + i \varepsilon}
\cdots 
\frac{i p_1^{\mu_n} T^{a_n} }{p_1 \cdot k_n + i \varepsilon}, 
%---------------
\end{align}
%---------------
where $\mu_i$ and $a_i$ are the Lorentz and color indices of the $i$th gluon,
respectively, 
and $k_i$ is the momentum carried by the $i$th gluon. This is equivalent to the
Feynman rule for $n$-gluon attachment to the path ordered, time ordered Wilson
line in the $p_1$ direction, defined by 
%---------------
\begin{align}
%---------------
W_{p_1} (t',t) = {\cal P} \exp \left[ -i g \int_t^{t'} d \lambda \, 
p_1 \cdot A(p_1 \lambda) \right]. 
%---------------
\end{align}
%---------------
Similarly, soft-gluon attachments to the $\bar Q$ line are equivalent to the 
antipath ordered, time ordered Wilson line in the $p_2$ direction:
%---------------
\begin{align}
%---------------
W_{p_2}^\dag (t',t) = \bar{\cal P} \exp \left[ +i g \int_t^{t'} d \lambda \, 
p_2 \cdot A(p_2 \lambda) \right],
%---------------
\end{align}
%---------------
with $\bar{\cal P}$ the antipath ordering. This lets us write the amplitude
for $g^* \to Q \bar Q$ followed by radiation of an arbitrary number of soft
gluons as the operator 
%---------------
\begin{align}
%---------------
{\cal A}_{\rm soft}^{\mu,a} = 
T \left[ \bar{u} (p_1) W_{p_1} (\infty,0) (-i g \gamma^\mu T^a) 
W_{p_2}^\dag (\infty,0) v (p_2)\right], 
%---------------
\end{align}
%---------------
where $T$ is the time ordering, and $\mu$ and $a$ are the Lorentz and color
indices of the fragmenting gluon, respectively. 
Then the gluon fragmentation function in the soft approximation can be written
as 
%---------------
\begin{align}
\label{eq:gFFsoft}
%---------------
D^{\rm soft}_{g \to Q \bar Q} (z)
&= 2 M (-g^{\mu \nu}) C_{\rm frag} 
\left| \frac{-i}{K^2 +i \varepsilon} \right|^2
\left( g_{\mu \alpha} - \frac{K_\mu n_\alpha}{K^+} \right)
\left( g_{\nu \beta} - \frac{K_\nu n_{\beta}}{K^+} \right)
\nonumber \\ & \quad \times
\langle 0 | \bar{T} \left[ {\cal A}_{\rm soft}^{\beta,c} 
\Phi_n^{bc} (\infty,0) \right]^\dag
2 \pi \delta(n \cdot \hat{p} - P^+ (1-z))
T \left[{\cal A}_{\rm soft}^{\alpha,a} \Phi_n^{ba} (\infty,0)  \right]
| 0 \rangle,
%---------------
\end{align}
%---------------
where $M = \sqrt{P^2}$, $C_{\rm frag} = z^{d-3} K^+/[2 \pi (N_c^2-1)(d-2)]$, 
$\bar T$ is antitime ordering, 
and $\hat p$ is the operator that reads off the momentum of the operator to the
right. 
The $-i/(K^2+i \varepsilon)$ is the propagator of the fragmenting gluon, 
and tensors $g_{\mu \alpha} - K_\mu n_\alpha/K^+$ and 
$g_{\nu \beta} - K_\nu n_{\beta}/K^+$ arise from the Feynman rule of the
operators $G^c_{+\mu}$ and $G^a_{+\nu}$ that create and annihilate the gluon in
the definition of the gluon FF. 
The delta function arises from the Fourier transform in Eq.~(\ref{eq:gFFdef}). 

Equation~(\ref{eq:gFFsoft}) can be considered a soft factorization formula, where
the $z \approx 1$ behavior of the FF is contained in the soft function, which
is given by the vacuum expectation value. 
Hence, we can investigate the 
singularities in the gluon FF at $z=1$ by computing the soft function, 
rather than the gluon FF itself. Note that while the matching coefficient of
this soft factorization formula can also receive radiative corrections, our
determination at leading nonvanishing order in $\alpha_s$ is sufficient for
computing the leading threshold logarithms, which is the goal of this work. 
Note that, because now the singularities in $z$ are completely included in the
soft function, any residual $z$ dependence in the coefficient can be neglected
by setting $K = P$ in Eq.~(\ref{eq:gFFsoft}). 
We can already calculate the tensors in Eq.~(\ref{eq:gFFsoft}) as
%---------------
\begin{align}
\label{eq:gFFsoftten}
%---------------
(-g_{\mu \nu}) 
\left( g^{\mu \alpha} - \frac{P^\mu n^\alpha}{P^+} \right)
\left( g^{\nu \beta} - \frac{P^\nu n^{\beta}}{P^+} \right)
= - g^{\alpha \beta} + \frac{P^\alpha n^\beta + P^\beta n^\alpha}{P^+}
- \frac{P^2 n^\alpha n^\beta}{(P^+)^2} \equiv I_T^{\alpha \beta}, 
%---------------
\end{align}
%---------------
which satisfies $I_T^{\alpha \beta} = I_T^{\beta \alpha}$ and 
$I_T^{\alpha \beta} P_\alpha = I_T^{\alpha \beta} n_\alpha = 0$.
The tensor $I_T^{\alpha \beta}$ can be identified as the transverse
polarization sum for a spin-1 boson with momentum $P$~\cite{Cho:1994gb}. 
Hence, 
%---------------
\begin{align}
\label{eq:gFFsoft2}
%---------------
D^{\rm soft}_{g \to Q \bar Q} (z)
&= 2 M C_{\rm frag}
\left| \frac{-i}{K^2 +i \varepsilon} \right|^2 I_T^{\alpha \beta} 
S_{\alpha \beta} (z), 
%---------------
\end{align}
%---------------
where $S_{\alpha \beta} (z)$ is the soft function defined by 
%---------------
\begin{align}
\label{eq:softfuncuniv}
%---------------
S^{\alpha \beta} (z) &\equiv
\langle 0 | \bar{T} \left[ {\cal A}_{\rm soft}^{\beta,c}
\Phi_n^{bc} (\infty,0) \right]^\dag
2 \pi \delta(n \cdot \hat{p} - P^+ (1-z))
T \left[{\cal A}_{\rm soft}^{\alpha,a} \Phi_n^{ba} (\infty,0)  \right]
| 0 \rangle,
%---------------
\end{align}
%---------------
While it is in principle possible to tackle the soft function $S^{\alpha \beta}
(z)$ directly, the calculation can be further simplified by
projecting the final-state $Q \bar Q$ onto specific color and angular momentum
states, which will also allow a more direct comparison with fixed-order results
for the FFs. 

%==============================================================================
\subsection{Spin and color projectors}
%==============================================================================

We use the projector method to 
project the Dirac spinors in the soft amplitude
${\cal A}^{\mu, a}_{\rm soft}$ onto
specific color and angular momentum.  
We take the nonrelativistic normalization for the
spinors, and take the conventions used in Ref.~\cite{Bodwin:2010fi}. 
In the spin-triplet case, this is done by the replacement
%---------------
\begin{align}
%---------------
\bar{u}(p_1) \Gamma v(p_2) 
&\to \frac{N_1 N_2}{2 \sqrt{2} M} {\rm tr} \left[ (p\!\!\!/_2-m) 
\epsilon\!\!\!/_S^* (P\!\!\!/ + M) (p\!\!\!/_1+m) \Lambda_{\rm c} \Gamma 
\right] , 
%---------------
\end{align}
%---------------
where the trace is over Dirac and color matrices, 
$\Gamma= W_{p_1} (\infty,0) (-i g \gamma^\mu T^a)
W_{p_2}^\dag (\infty,0)$, 
$p_1^2=p_2^2=m^2$, $P^2 = M^2$, 
$N_i = 1/\sqrt{2 E_i (E_i+m)}$ with $E_i = p_i \cdot P/M$, 
$\epsilon_S$ is the polarization vector
for the $Q \bar Q$ spin in the spin-triplet state, 
and $\Lambda_{\rm c}$ is the color projector for color-singlet (${\rm c}=1$) 
and octet (${\rm c}=8$) states given by 
%---------------
\begin{align}
%---------------
\Lambda_1   &= \frac{{\bf 1}}{\sqrt{N_c}}, 
\\
\Lambda_8^a &= \frac{T^a}{\sqrt{T_F}}, 
%---------------
\end{align}
%---------------
with ${\bf 1}$ the $SU(N_c)$ unit matrix. 
The projection onto specific orbital angular momentum states is done by
expressing the soft amplitude as a function of $p = (p_1+p_2)/2=P/2$ and 
the relative momentum $q = (p_1-p_2)/2$, and then expanding in powers of $q$. 
The $S$-wave
contribution is the $q=0$ term, and the $P$-wave contribution comes from the
term linear in $q$. 
As the only Dirac matrix in $\Gamma$ is $\gamma^\mu$, 
the only Dirac trace that we need is 
%---------------
\begin{align}
\label{eq:Diractrace}
%---------------
%& \hspace{-5ex} 
\frac{N_1 N_2}{2 \sqrt{2} M} 
{\rm tr}_{\rm Dirac} \left[ (p\!\!\!/_2-m)
\epsilon\!\!\!/_S^* (P\!\!\!/ + M) (p\!\!\!/_1+m) \gamma^\mu 
\right] 
%\nonumber \\
&= 
- \sqrt{2} 
\epsilon_S^{*\mu} - \frac{4 \sqrt{2} q^\mu \epsilon_S^* \cdot q}{M (M+2 m)}, 
%---------------
\end{align}
%---------------
where we used $p \cdot \epsilon_S^* = 0$. 
The right-hand side is equivalent to Eq.~(A9b) of Ref.~\cite{Braaten:1996rp} up
to a phase and normalization convention. 
Note that since $M=2 \sqrt{m^2-q^2}$, the right-hand side does not involve
terms linear in $q$ when expanded in powers of $q$, as the second term on the
right-hand side is already quadratic in $q$. 
Since we are only interested in the $S$- and $P$-wave states, we only need to
retain the first term on the right-hand side in our calculation of the soft
function. The color trace in the soft amplitude is given by 
%---------------
\begin{align}
\label{eq:colortrace}
%---------------
{\rm tr}_{\rm color} \left[ \Lambda_{\rm c} W_{p_1} (\infty,0) T^a
W_{p_2}^\dag (\infty,0) \right], 
%---------------
\end{align}
%---------------
which is in general a function of $p$ and $q$. When expanded in powers of $q$,
this can be expressed in terms of the adjoint Wilson line and the
field-strength tensor by using Polyakov's identity~\cite{Polyakov:1980ca}. 
We carry out this calculation for each color and angular momentum state.

%==============================================================================
\subsection{\boldmath $^3S_1^{[8]}$}
%==============================================================================

In order to project onto the $S$-wave state, we expand the soft amplitude in
powers of the relative momentum $q$ and retain only the $q=0$ contribution. 
As the Dirac trace is already computed in the previous section, we only need to
consider the color trace, which we compute at $q=0$. 
We make use of the following identities for products of Wilson lines 
%---------------
\begin{subequations}
\label{eq:WLidentities}
\begin{eqnarray}
%---------------
W_p (\lambda_f, \lambda_i) &=& 
W_p (\lambda_f, \lambda) W_p (\lambda, \lambda_i), 
\\
W_p (\lambda_f,\lambda_i) T^a W^\dag_p (\lambda_f,\lambda_i)
&=& T^b \Phi_p^{ba} (\lambda_f,\lambda_i), 
\\
W^\dag_p (\lambda_f,\lambda_i)
T^a 
W_p (\lambda_f,\lambda_i) 
&=& \Phi_p^{ab} (\lambda_f,\lambda_i)T^b , 
\\
W_p (\lambda_f,\lambda_i) W^\dag_p (\lambda_f,\lambda_i)
&=& 1, 
%---------------
\end{eqnarray}
\end{subequations}
%---------------
which hold for $\lambda_f>\lambda>\lambda_i$. From these we obtain
%---------------
\begin{align}
%---------------
\frac{1}{\sqrt{T_F}} 
{\rm tr}_{\rm color} \left( W_{p} (\infty,0) T^a W_{p}^\dag 
(\infty,0) T^d \right)
&= 
\frac{1}{\sqrt{T_F}} 
\Phi_p^{ba} (\infty,0) {\rm tr}_{\rm color} (T^b T^d)
\nonumber \\
&= 
\sqrt{T_F}
\Phi_p^{da} (\infty,0), 
%---------------
\end{align}
%---------------
where $d$ is the adjoint color index for the $^3S_1^{[8]}$ state. 
By combining this with the result for the Dirac trace we obtain 
%---------------
\begin{equation}
%---------------
{\cal A}_{\rm soft}^{\mu,a;d} (^3S_1^{[8]}) =
\sqrt{2 T_F} i g \epsilon_S^*{}^\mu \Phi_p^{da} (\infty,0).
%---------------
\end{equation}
%---------------
By plugging this into Eq.~(\ref{eq:gFFsoft}) we obtain
%---------------
\begin{eqnarray}
%---------------
D^{\rm soft}_{g \to Q \bar Q (^3S_1^{[8]})} (z)
&=& 
2 M \left|\frac{-i}{P^2}\right|^2
C_{\rm frag} 2 T_F g^2
I_T^{\alpha \beta} 
\sum_\lambda
\epsilon^*_{S \alpha} (\lambda)
\epsilon_{S \beta} (\lambda)
S_{^3S_1^{[8]}} (z), 
%---------------
\end{eqnarray}
%---------------
where we define the $^3S_1^{[8]}$ soft function as 
%---------------
\begin{eqnarray}
\label{eq:softfunc3S18}
%---------------
S_{^3S_1^{[8]}} (z) &\equiv&
\langle 0| [{\cal W}(^3S_1^{[8]})^{cb} ]^\dag 2 \pi
\delta(n \cdot \hat{p} - P^+ (1-z))
{\cal W}(^3S_1^{[8]})^{cb} | 0 \rangle,
%---------------
\end{eqnarray}
%---------------
with 
%---------------
\begin{eqnarray}
%---------------
{\cal W}(^3S_1^{[8]})^{cb} \equiv 
T \left[ \Phi_p^{ca} (\infty,0) \Phi_n^{ba} (\infty,0) \right].
%---------------
\end{eqnarray}
%---------------
The polarization sum is evaluated by using 
%---------------
\begin{eqnarray}
%---------------
\sum_\lambda
\epsilon^*_{S \alpha} (\lambda)
\epsilon_{S \beta} (\lambda) = I_{\alpha \beta}
\equiv -g_{\alpha \beta} + \frac{P_\alpha P_\beta}{P^2}, 
%---------------
\end{eqnarray}
%---------------
which yields 
%---------------
\begin{eqnarray}
\label{eq:softfac3S18}
%---------------
D^{\rm soft}_{g \to Q \bar Q (^3S_1^{[8]})} (z) 
&=&
\frac{
C_{\rm frag} (d-2) g^2 
}{4m^3} 
S_{^3S_1^{[8]}} (z).
%---------------
\end{eqnarray}
%---------------
We note that, because the tensor $I_T^{\alpha \beta}$ vanishes when contracted
with a longitudinal polarization vector, there is no contribution to 
$D^{\rm soft}_{g \to Q \bar Q (^3S_1^{[8]})}$ coming from the longitudinally
polarized $Q \bar Q$ in the $^3S_1^{[8]}$ state.

%==============================================================================
\subsection{\boldmath $^3P^{[8]}$}
%==============================================================================

We next consider the $^3P^{[8]}$ case. 
As we have discussed previously, the Dirac trace does not involve terms linear
in $q$, so that we can set $q=0$ in Eq.~(\ref{eq:Diractrace}). 
Hence, the only term linear in $q$ can come from the expansion of the color
trace in powers of $q$. 
The expansion can be carried out by using a
straightforward generalization of Polyakov's identity~\cite{Polyakov:1980ca, 
Nayak:2005rt} : 
%---------------
\begin{eqnarray}
%---------------
&& \hspace{-5ex} 
\delta {\cal P} \exp\left[- ig \oint_C d\lambda A_\alpha (x(\lambda))
\frac{dx^\alpha(\lambda)}{d\lambda}\right] 
= {\cal P} \oint_C ds \, 
\exp \left[ -ig \int_s^{\lambda_f} d\lambda A_\alpha (x(\lambda)) 
\frac{dx^\alpha(\lambda)}{d\lambda} \right]
\nonumber \\ && \times (-ig) G_{\mu \nu} (x(s)) \frac{d x^\nu(s)}{ds} 
\exp \left[ -ig \int^s_{\lambda_i} d\lambda' A_\alpha (x(\lambda')) 
\frac{dx^\alpha(\lambda')}{d\lambda'} \right] \delta x^\mu (s),
%---------------
\end{eqnarray}
%---------------
where $C$ is a closed path in spacetime parametrized by $x(\lambda)$, 
$\delta x$ is the variation of the path, 
and $x(\lambda_i) = x(\lambda_f)$ is an arbitrary point on $C$. 
We use this identity to expand the $W_{p_1} (\infty,0)$ and 
$W_{p_2}^\dag (\infty,0)$ in the color trace to linear power in $q$. 
We obtain 
%---------------
\begin{eqnarray}
%---------------
&& \hspace{-5ex} W_{p_1} (\infty,0) T^a W_{p_2}^\dag (\infty,0)
\nonumber \\ &=& 
\bigg[ W_{p}(\infty,0)
-i g \int_0^\infty d\lambda \, \lambda
W_{p} (\infty,\lambda) 
G_{\mu \nu} (p \lambda) p^\mu q^\nu 
W_{p} (\lambda,0)
+ O(q^2/p^2)
\bigg] T^a W_{p_2}^\dag (\infty,0)
\nonumber \\ &=& 
\bigg[ W_{p}(\infty,0) T^a W^\dag_p(\infty,0)
-i g \int_0^\infty d\lambda \, \lambda
W_{p} (\infty,\lambda) 
G_{\mu \nu} (p \lambda) p^\mu q^\nu 
W_{p} (\lambda,0)T^a W^\dag_p(\infty,0)
\nonumber \\ && 
\quad 
+i g \int_0^\infty d\lambda \, \lambda 
W_{p}(\infty,0) T^a
W_{p}^\dag (\lambda,0)
G_{\mu \nu} (p \lambda) p^\mu (-q^\nu )
W_{p}^\dag (\infty,\lambda) 
+ O(q^2/p^2)
\bigg].
%---------------
\end{eqnarray}
%---------------
The first term in the square brackets is the $q=0$ term, and the remaining
terms are linear in $q$. The $P$-wave contribution corresponds to the
coefficient of $q$ in the linear term. We take the normalization used in 
Ref.~\cite{Bodwin:2012xc}, where the $P$-wave contribution in the linear term 
$A_\alpha q^\alpha$ in the amplitude is given by $A_\alpha
\epsilon_L^\alpha/\sqrt{d-1}$, with $\epsilon_L$ the polarization vector for
the orbital angular momentum of the $Q \bar Q$. 
In this normalization, the color-octet $P$-wave contribution of the color trace
is given by 
%---------------
\begin{eqnarray}
%---------------
&& \hspace{-5ex} 
{\rm tr}_{\rm color} \left[W_{p_1} (\infty,0) T^a W_{p_2}^\dag (\infty,0) 
\Lambda_8^{a'}
\right]_{P{\rm-wave}}
\nonumber \\
&=&
\frac{i g \epsilon_L^{* \nu}}{\sqrt{T_F (d-1)}} \int_0^\infty d\lambda \, \lambda
\, {\rm tr}_{\rm color} \bigg[
W_{p}^\dag (\infty,0) T^{a'}
W_{p} (\infty,\lambda)
p^\mu
G_{\mu \nu}^b (p \lambda) T^b
W_{p} (\lambda,0) T^a
\nonumber \\ && \hspace{12ex} +
T^a W_{p}^\dag (\lambda,0)
p^\mu
G_{\mu \nu}^b (p \lambda) T^b
W_{p}^\dag (\infty,\lambda)
T^{a'}
W_p(\infty,0)
\bigg]
.
%---------------
\end{eqnarray}
%---------------
We can now use the identities in Eqs.~(\ref{eq:WLidentities}) and the cyclic
property of the trace to reduce the fundamental Wilson lines into adjoint ones.
We have 
%---------------
\begin{eqnarray}
%---------------
{\rm tr}_{\rm color} [ W_{p}^\dag (\infty,0) T^{a'}
W_{p} (\infty,\lambda)
T^b W_{p} (\lambda,0) T^a]
&=&
{\rm tr}_{\rm color} [
\Phi^{a'c} (\infty,\lambda) T^c
T^b T^d
\Phi_p^{da} (\lambda,0)
], 
\\
{\rm tr}_{\rm color} [
T^a W_{p}^\dag (\lambda,0)
T^b
W_{p}^\dag (\infty,\lambda)
T^{a'}
W_p(\infty,0)]
&=& 
{\rm tr}_{\rm color} [
\Phi_p^{da} (\lambda,0)
T^d
T^b
\Phi_p^{a'c} (\infty,\lambda) T^c
],
%---------------
\end{eqnarray}
%---------------
which lead to 
%---------------
\begin{eqnarray}
%---------------
&& \hspace{-5ex} 
{\rm tr}_{\rm color} \left[ W_{p_1} (\infty,0) T^a W_{p_2}^\dag (\infty,0) 
\Lambda_8^{a'} \right]_{P{\rm -wave}}
\nonumber \\ &=&
\frac{i g \epsilon_L^{*\nu} }{\sqrt{2(d-1)}} 
\int_0^\infty d\lambda \, \lambda
\Phi_p^{a'c} (\infty, \lambda) 
d^{bcd} 
p^\mu 
G_{\mu \nu}^b (p \lambda) 
\Phi_p^{da} (\lambda,0),
%---------------
\end{eqnarray}
%---------------
where we used ${\rm tr}_{\rm color} ( \{ T^b, T^{c} \}T^d ) = T_F d^{bcd}$ and
$T_F = 1/2$. 
Note that $d^{bcd} p^\mu G_{\mu \nu}^b (p \lambda)$ is a rank-2 tensor in the
adjoint representation of $SU(N_c)$, and is a Lorentz vector that is orthogonal
to $p^\nu$. 
By using this result we obtain the fragmentation function in the soft
approximation 
%---------------
\begin{eqnarray}
%---------------
D^{\rm soft}_{g\to Q \bar Q(^3P^{[8]})}(z) &=&
\frac{1}{4 m^3}
\frac{g^2}{2 (d-1)}
2 g^2
C_{\rm frag} 
I_T^{\alpha \alpha'} 
\sum_{\lambda_S}
\epsilon^*_{S \alpha} (\lambda_S) \epsilon_{S \alpha'} (\lambda_S)
\sum_{\lambda_L}
\epsilon_L^{*\beta} (\lambda_L) \epsilon_L^{\beta'}(\lambda_L)
\nonumber \\ &&
\times 
\langle 0 |
[{\cal W}^{yx}_{\beta'} (^3P^{[8]})]^\dag
2 \pi \delta(n \cdot \hat{p} - P^+(1-z))
{\cal W}^{yx}_\beta (^3P^{[8]})
| 0 \rangle , 
%---------------
\end{eqnarray}
%---------------
where
%---------------
\begin{eqnarray}
%---------------
{\cal W}^{yx}_\beta (^3P^{[8]}) &\equiv&
T \left[
\int_0^\infty d\lambda \, \lambda
\Phi_p^{yc} (\infty, \lambda)
p^\mu
G_{\mu \beta}^b (p \lambda)
d^{bcd}
\Phi_p^{da} (\lambda,0)
\Phi_n^{xa} (\infty,0) \right]. 
%---------------
\end{eqnarray}
%---------------
The polarization sums are given by
%---------------
\begin{eqnarray}
%---------------
\sum_{\lambda_S} \epsilon_S^{*\alpha} (\lambda_S) \epsilon_S^{\alpha'}
(\lambda_S)
=
\sum_{\lambda_L} \epsilon_L^{*\alpha} (\lambda_L) \epsilon_L^{\alpha'} 
(\lambda_L)
= I^{\alpha \alpha'} 
=- g^{\alpha \alpha'} + \frac{P^\alpha P^{\alpha'}}{P^2}. 
%---------------
\end{eqnarray}
%---------------
Since ${\cal W}^{yx}_\beta (^3P^{[8]})$ is orthogonal to $p^\beta$, we may
neglect the $P^\beta P^{\beta '}/P^2$ term in the polarization sum of 
$\epsilon_L^{*\beta} (\lambda_L) \epsilon_L^{\beta'}(\lambda_L)$. 
From this we obtain 
%---------------
\begin{eqnarray}
\label{eq:softfac3P8}
%---------------
D^{\rm soft}_{g\to Q \bar Q(^3P^{[8]})}(z) 
&=&
- \frac{1}{4m^3}
\frac{g^4 (d-2)}{(d-1)}
C_{\rm frag}
S_{^3P^{[8]}} (z),
%---------------
\end{eqnarray}
%---------------
where $S_{^3P^{[8]}} (z)$ is the $^3P^{[8]}$ soft function defined by 
%---------------
\begin{eqnarray}
\label{eq:softfunc3P8}
%---------------
S_{^3P^{[8]}} (z) \equiv 
\langle 0 |
[{\cal W}^{yx}_{\beta'} (^3P^{[8]})]^\dag
2 \pi \delta(n \cdot \hat{p} - P^+ (1-z))
{\cal W}^{yx}_\beta (^3P^{[8]})
| 0 \rangle g^{\beta \beta'}. 
%---------------
\end{eqnarray}
%---------------
Similarly to the $^3S_1^{[8]}$ case, the contribution from longitudinal 
$Q \bar Q$ spin vanishes because $\epsilon_{S\alpha}^*(\lambda_S)$ is
contracted with the tensor $I_T^{\alpha \alpha'}$. 

%==============================================================================
\subsection{\boldmath $^3P_J^{[1]}$}
%==============================================================================

Finally, we consider the $^3P_J^{[1]}$ state. 
The calculation is similar to the $^3P^{[8]}$ state, except that we 
project onto the color-singlet state. We have 
%---------------
\begin{eqnarray}
%---------------
&& \hspace{-5ex} 
{\rm tr}_{\rm color} \left[W_{p_1} (\infty,0) T^a W_{p_2}^\dag
(\infty,0) \Lambda_1 \right]_{P{\rm -wave}} 
\nonumber \\
&=& 
\frac{i g \epsilon_L^{*\nu}}{\sqrt{N_c (d-1)}} 
\int_0^\infty d\lambda' \, \lambda'
{\rm tr}_{\rm color} \bigg[
W_{p}^\dag (\infty,0)
W_{p} (\infty,\lambda')
p^\mu G_{\mu \nu}^b (p \lambda') T^b 
W_{p} (\lambda',0) T^a 
\nonumber \\ && \hspace{12ex} +
T^a W_{p}^\dag (\lambda',0)
p^\mu G_{\mu \nu}^b (p \lambda') T^b 
W_{p}^\dag (\infty,\lambda')
W_p(\infty,0) 
\bigg], 
%---------------
\end{eqnarray}
%---------------
where again $\epsilon_L$ is the polarization vector for the orbital angular
momentum of the $Q \bar Q$. 
By using the identities in Eqs.~(\ref{eq:WLidentities}) we obtain 
%---------------
\begin{align}
%---------------
& \hspace{-5ex} 
{\rm tr}_{\rm color} [W_{p_1} (\infty,0) T^a W_{p_2}^\dag (\infty,0) 
\Lambda_1]_{P{\rm -wave}}
\nonumber \\
&= 
\frac{2 T_F \epsilon_L^{*\nu}}{\sqrt{N_c (d-1)}} 
i g \int_0^\infty d\lambda' \, \lambda'
p^\mu 
G_{\mu \nu}^b (p \lambda') 
\Phi_p^{ba} (\lambda',0). 
%---------------
\end{align}
%---------------
From this we obtain
%---------------
\begin{eqnarray}
\label{3PJsofttensorform}
%---------------
D^{\rm soft} _{g \to Q \bar Q(^3P^{[1]})} (z) &=&
\frac{1}{4 m^3}
\frac{2 g^4 }{N_c(d-1)}
C_{\rm frag} 
I_T^{\alpha \alpha'} 
\epsilon_{S \alpha}^* \epsilon_{S \alpha'}
\epsilon^*_{L\beta}   \epsilon_{L \beta'} 
S_{^3P^{[1]}}^{\beta \beta'} (z)
,
%---------------
\end{eqnarray}
%---------------
where 
%---------------
\begin{align}
%---------------
S_{^3P^{[1]}}^{\beta \beta'} &\equiv
\langle 0 |
[ {\cal W}^b{}^{\beta'} (^3P^{[1]}) ]^\dag
2 \pi \delta(n\cdot \hat{p}-P^+ (1-z))
{\cal W}^b{}^{\beta} (^3P^{[1]})
| 0 \rangle, 
\\
{\cal W}^b_{\beta} (^3P^{[1]}) &\equiv 
\int_0^\infty d\lambda \, \lambda \,
p^\mu
G_{\mu \beta}^{d} (p \lambda) \Phi_p^{da} (\lambda,0) \Phi_n^{ba} (\infty, 0).
%---------------
\end{align}
%---------------
For the $^3P_J^{[1]}$ channel, the polarization sums need to be done for
specific $J = 0$, 1, and 2. 
This can be done by making the replacements 
$\epsilon_{S \alpha}^* \epsilon_{S \alpha'} \epsilon_{L \beta}^{*}
\epsilon_{L\beta}  \to \Pi^J_{\alpha \beta \alpha' \beta'}$, where 
%---------------
\begin{eqnarray}
%---------------
\Pi^{J=0}_{\alpha \beta \alpha' \beta'}
&=& \frac{1}{d-1} I_{\alpha \beta} I_{\alpha' \beta'}, 
\nonumber\\
\Pi^{J=1}_{\alpha \beta \alpha' \beta'}
&=& \frac{1}{2} \left( 
I_{\alpha \alpha'} I_{\beta \beta'}
-I_{\alpha \beta'} I_{\alpha' \beta}
\right)
\nonumber\\
\Pi^{J=2}_{\alpha \beta \alpha' \beta'}
&=& \frac{1}{2} \left( 
I_{\alpha \alpha'} I_{\beta \beta'}
+I_{\alpha \beta'} I_{\alpha' \beta}
\right)
- \frac{1}{d-1} I_{\alpha \beta} I_{\alpha' \beta'}. 
%---------------
\end{eqnarray}
%---------------
Note that $\sum_J \Pi^J_{\alpha \beta \alpha' \beta'}
= I _{\alpha \alpha'} I_{\beta \beta'}$. 
This gives 
%---------------
\begin{align}
%---------------
I_T^{\alpha \alpha'}
\Pi^{J}_{\alpha \beta \alpha' \beta'}
S_{^3P^{[1]}}^{\beta \beta'}
&= 
S_{^3P^{[1]}}^{\beta \beta'} \times 
\begin{cases}
\displaystyle
-\frac{1}{d-1} \left( g_{\beta \beta'} + \frac{P^2 n_\beta n_{\beta'}}{P_+^2}
\right) & (J=0)\\
\displaystyle
-\frac{d-3}{2} g_{\beta \beta'} + \frac{1}{2} \frac{P^2 n_\beta n_{\beta'}}{P_+^2}
& (J=1) \\
\displaystyle
-\frac{d^2-2 d-1}{2 (d-1)} g_{\beta \beta'} 
- \frac{d-3}{2(d-1)} \frac{P^2 n_\beta n_{\beta'}}{P_+^2}
& (J=2)
\end{cases}.
%---------------
\end{align}
%---------------
Here, we eliminated the terms involving $P^\beta$ and $P^{\beta'}$ in 
$\Pi^{J}_{\alpha \beta \alpha' \beta'}$
by using $S_{^3P^{[1]}}^{\beta \beta'} P_\beta = 
S_{^3P^{[1]}}^{\beta' \beta} P_{\beta} = 0$. 
The $n_\beta n_{\beta'}$ terms can be rewritten as 
%---------------
\begin{align}
%---------------
n_{\beta} n_{\beta'} &= 
- 
\frac{(n \cdot p)^2}{p^2 (d-1)}
g_{\beta \beta'}
+  \left( n_\beta n_{\beta'}  + \frac{(n \cdot p)^2}{p^2 (d-1)}
g_{\beta \beta'} \right),
%---------------
\end{align}
%---------------
where the terms in the parentheses vanish when contracted with the isotropic
tensor $I^{\beta \beta'}$. By using this we can write 
the isotropic and anisotropic contributions as
%---------------
\begin{eqnarray}
%---------------
I_T^{\alpha \alpha'} 
\Pi^{J}_{\alpha \beta \alpha' \beta'}
S_{^3P^{[1]}}^{\beta \beta'} 
\Big|_{\rm isotropic}
=
-(d-2)
S_{^3P^{[1]}}^{\beta \beta'} g_{\beta \beta'} \times 
\begin{cases}
\displaystyle
\frac{1}{(d-1)^2} 
 & (J=0)\\
\displaystyle
\frac{(d-2)}{2 (d-1)}
& (J=1) \\
\displaystyle
\frac{(d-2) (d+1)}{2 (d-1)^2}
& (J=2)
\end{cases},
%---------------
\end{eqnarray}
%---------------
and
%---------------
\begin{eqnarray}
%---------------
I_T^{\alpha \alpha'} 
\Pi^{J}_{\alpha \beta \alpha' \beta'}
S_{^3P^{[1]}}^{\beta \beta'} 
\Big|_{\rm anisotropic}
=
S_{^3P^{[1]}}^{\beta \beta'} 
\left( \frac{p^2 n_\beta n_{\beta'}}{p_+^2} + \frac{ g_{\beta \beta'} }{d-1} 
\right) 
\times 
\begin{cases}
\displaystyle
-\frac{1}{d-1} 
& (J=0)\\
\displaystyle
\frac{1}{2} 
& (J=1) \\
\displaystyle
- \frac{d-3}{2(d-1)} 
& (J=2)
\end{cases}. 
%---------------
\end{eqnarray}
%---------------
From these we can write
%---------------
\begin{eqnarray}
\label{eq:softfac3Psinglet}
%---------------
D^{\rm soft}_{g \to Q \bar Q (^3P_J^{[1]})} (z) &=&
-
\frac{1}{4 m^3}
\frac{2 (d-2) g^4 }{N_c(d-1)}
C_{\rm frag}
\left[ c^J S_{^3P^{[1]}} (z) + c_{TT}^J S^{TT}_{^3P^{[1]}} (z)  \right]
,
%---------------
\end{eqnarray}
%---------------
where 
%---------------
\begin{subequations}
\label{eq:softfunc3Psinglet}
\begin{align}
%---------------
S_{^3P^{[1]}} (z)
&=
\langle 0 |
[ {\cal W}^b_{\beta'} (^3P^{[1]}) ]^\dag
2 \pi \delta(n\cdot \hat{p}-P^+ (1-z))
{\cal W}^b_{\beta} (^3P^{[1]})
| 0 \rangle
g^{\beta \beta'}
,
\\
S_{^3P^{[1]}}^{TT} (z)
&=
\langle 0 |
[ {\cal W}^b_{\beta'} (^3P^{[1]}) ]^\dag
2 \pi \delta(n\cdot \hat{p}-P^+ (1-z))
{\cal W}^b_{\beta} (^3P^{[1]})
| 0 \rangle
\left( \frac{p^2 n^\beta n^{\beta'}}{p_+^2} + \frac{ g^{\beta \beta'} }{d-1} 
\right) ,
%---------------
\end{align}
\end{subequations}
%---------------
with 
%---------------
\begin{subequations}
\begin{align}
%---------------
c^J &=
\begin{cases}
\displaystyle
\frac{1}{(d-1)^2} 
 & (J=0)\\
\displaystyle
\frac{(d-2)}{2 (d-1)}
& (J=1) \\
\displaystyle
\frac{(d-2) (d+1)}{2 (d-1)^2}
& (J=2)
\end{cases},
\\
-(d-2) \times c_{TT}^J &=
\begin{cases}
\displaystyle
-\frac{1}{d-1}
& (J=0)\\
\displaystyle
\frac{1}{2}
& (J=1) \\
\displaystyle
- \frac{d-3}{2(d-1)}
& (J=2)
\end{cases}.
%---------------
\end{align}
\end{subequations}
%---------------
Note that $\sum_{J=0}^2 c^J = 1$, $c^J = (2 J+1)/9 + O(\epsilon)$, 
and $\sum_{J=0}^2 c^J_{TT} = 0$. 

We also consider the polarized case. 
The polarized spin sums are computed by using~\cite{Ma:2015yka}
%---------------
\begin{align}
%---------------
\Pi^{J=1,|h|=1}_{\alpha \beta \alpha' \beta'}
&= \frac{1}{2} \left(
  I_T{}_{\alpha \alpha'} I_L{}_{\beta \beta'}
+ I_L{}_{\alpha \alpha'} I_T{}_{\beta \beta'}
- I_T{}_{\alpha \beta'} I_L{}_{\alpha' \beta}
- I_L{}_{\alpha \beta'} I_T{}_{\alpha' \beta}
\right),
\\
\Pi^{J=1,h=0}_{\alpha \beta \alpha' \beta'}
&= \frac{1}{2} \left(
  I_T{}_{\alpha \alpha'} I_T{}_{\beta \beta'}
- I_T{}_{\alpha \beta'} I_T{}_{\alpha' \beta}
\right), 
\\
\Pi^{J=2,|h|=2}_{\alpha \beta \alpha' \beta'}
&= \frac{1}{2} \left( 
  I_T{}_{\alpha \alpha'} I_T{}_{\beta \beta'} 
+ I_T{}_{\alpha \beta'} I_T{}_{\alpha' \beta} 
\right)
- \frac{1}{2} I_T{}_{\alpha \beta} I_T{}_{\alpha' \beta'}, 
\\
\Pi^{J=2,|h|=1}_{\alpha \beta \alpha' \beta'}
&= \frac{1}{2} \left( 
  I_T{}_{\alpha \alpha'} I_L{}_{\beta \beta'} 
+ I_L{}_{\alpha \alpha'} I_T{}_{\beta \beta'} 
+ I_T{}_{\alpha \beta'} I_L{}_{\alpha' \beta}
+ I_L{}_{\alpha \beta'} I_T{}_{\alpha' \beta}
\right), 
\\
\Pi^{J=2,h=0}_{\alpha \beta \alpha' \beta'}
&= \frac{d-2}{d-1} 
\left( I_L{}_{\alpha \beta} - \frac{1}{d-2} I_T{}_{\alpha \beta} \right) 
\left( I_L{}_{\alpha' \beta'} - \frac{1}{d-2} I_T{}_{\alpha' \beta'} \right) ,
%---------------
\end{align}
%---------------
where 
$ \Pi^{J,|h|=1}_{\alpha \beta \alpha' \beta'}
= \sum_{h=\pm 1} \Pi^{J,h}_{\alpha \beta \alpha' \beta'}$, 
$ \Pi^{J,|h|=2}_{\alpha \beta \alpha' \beta'}
= \sum_{h=\pm 2} \Pi^{J,h}_{\alpha \beta \alpha' \beta'}$,
and
$I_L^{\alpha \beta} = I^{\alpha \beta} - I_T^{\alpha \beta}$. 
We have 
%---------------
\begin{eqnarray}
\label{eq:softfac3Psingletpol}
%---------------
D^{\rm soft}_{g \to Q \bar Q (^3P_{J,h}^{[1]})} &=&
-
\frac{1}{4 m^3}
\frac{2 (d-2) g^4 }{N_c(d-1)}
C_{\rm frag}
\left[ c^{J,h} S_{^3P^{[1]}} (z) + c_{TT}^{J,h} S^{TT}_{^3P^{[1]}} (z) \right]
,
%---------------
\end{eqnarray}
%---------------
where 
%---------------
\begin{subequations}
\begin{align}
%---------------
c^{J=1,|h|=1} &= 
\frac{1}{2 (d-1)},
\\
c^{J=1,h=0} &= 
\frac{(d-3)}{2 (d-1)},
\\
c^{J=2,|h|=2} &= 
\frac{d (d-3)}{2 (d-1)(d-2)} ,
\\
c^{J=2,|h|=1} &= 
\frac{1}{2 (d-1)},
\\
c^{J=2,h=0} &= 
\frac{1}{(d-1)^2 (d-2)},
\end{align}
\begin{align}
-(d-2) \times 
c_{TT}^{J=1,|h|=1} &=
\frac{d-2}{2},
\\
-(d-2) \times 
c_{TT}^{J=1,h=0} &=
- \frac{d-3}{2},
\\
-(d-2) \times c_{TT}^{J=2,|h|=2} &=
- \frac{d (d-3)}{2 (d-2)} ,
\\
-(d-2) \times c_{TT}^{J=2,|h|=1} &=
\frac{d-2}{2},
\\
-(d-2) \times c_{TT}^{J=2,h=0} &=
-\frac{1}{(d-1) (d-2)}. 
%---------------
\end{align}
\end{subequations}
%---------------

%==============================================================================
\subsection{Summary of soft factorization}
%==============================================================================

We now summarize the results of the soft factorization in this section. 
The general soft factorization formula in Eq.~(\ref{eq:gFFsoft2})
is obtained by applying the Grammer-Yennie approximation (soft approximation)
to gluon attachments to the $Q$ and $\bar Q$ lines in the process 
$g^* \to Q \bar Q$. The soft function we obtain is given in
Eq.~(\ref{eq:softfuncuniv}), which is valid for $Q \bar Q$ in an arbitrary
color and angular momentum state. 
While it is in principle possible to tackle the general soft function in
Eq.~(\ref{eq:softfuncuniv}) directly, as was done in
Ref.~\cite{Bodwin:2019bpf}, the calculation can be further simplified by
projecting the final-state $Q \bar Q$ onto specific color and angular momentum
states. 
The resulting soft factorization formulas are given in
Eqs.~(\ref{eq:softfac3S18}), (\ref{eq:softfac3P8}), and
(\ref{eq:softfac3Psinglet}) for the $^3S_1^{[8]}$, $^3P^{[8]}$, and 
$^3P^{[1]}_J$ states, respectively, 
with the corresponding soft functions given in 
Eqs.~(\ref{eq:softfunc3S18}), (\ref{eq:softfunc3P8}), and 
(\ref{eq:softfunc3Psinglet}). These soft functions for specific color and
angular momentum states are given by vacuum expectation values of adjoint
Wilson lines and field-strength tensors. 
In the following sections, we will compute these soft functions to reproduce
the $z \to 1$ singularities in the fragmentation functions, from which we can
identify the threshold logarithms. 

We note that the expression for the isotropic $^3P^{[1]}$ soft function
$S_{^3P^{[1]}}(z)$ has first been found in Ref.~\cite{Chung:2023ext} in the
context of shape functions. Its first moment $\int_0^1 dz \, S_{^3P^{[1]}}(z)$ has
been obtained in Ref.~\cite{Nayak:2005rt} for the study of infrared divergences
in quarkonium LDMEs; this was also used in the analysis of universality
relations for $^3P^{[1]}_J$ LDMEs in Refs.~\cite{Brambilla:2020ojz,
Brambilla:2021abf}. 

We can see that in the projection of the soft function onto specific color and
angular momentum states that the soft function vanishes for the color-singlet
$S$-wave state; this is because the color trace in Eq.~(\ref{eq:colortrace})
vanishes for color singlet when we set $q=0$. This is consistent with the fact
that the gluon fragmentation function for $^3S_1^{[1]}$ state is free of 
singularities at $z=1$. We also note that the Dirac trace in
Eq.~(\ref{eq:Diractrace}) vanishes if we use the spin-singlet projector instead
of the spin-triplet one (see, for example, Eq.~(A9b) of
Ref.~\cite{Braaten:1996rp}). 
Hence, the soft function also vanishes when the final-state $Q \bar Q$ is in a
spin-singlet state. This is consistent with the fact that the gluon
fragmentation functions for spin-singlet $S$-wave and $P$-wave states 
do not involve singular distributions at $z=1$. 
Hence, the nonvanishing soft functions that are relevant for production of a
heavy quarkonium come from the $^3S_1^{[8]}$, $^3P^{[8]}$, and $^3P_J^{[1]}$
states. 

As we have mentioned previously, the matching coefficient of the soft
factorization formula that we obtain in Eq.~(\ref{eq:gFFsoft2}) is valid at
leading order in the strong coupling. Loop corrections to this coefficient can
arise from gluon attachments to the fragmenting gluon; because the fragmenting
gluon has a virtuality of at least $M^2$, gluon attachments to the fragmenting
gluon do not produce additional singularities at $z \to 1$. Hence, while
radiative corrections to the matching coefficient can in principle affect the
subleading singularities, the leading singularities are completely contained in
the soft function in Eq.~(\ref{eq:softfuncuniv}) to all orders in the strong
coupling. Because we are only interested in the leading threshold singularities
in the fragmentation functions, we only need to consider loop corrections to 
the soft functions.

%==============================================================================
\section{Leading-order fragmentation functions in the soft approximation}
\label{sec:softLO}
%==============================================================================

\begin{figure}[t]
\includegraphics[width=.9\textwidth]{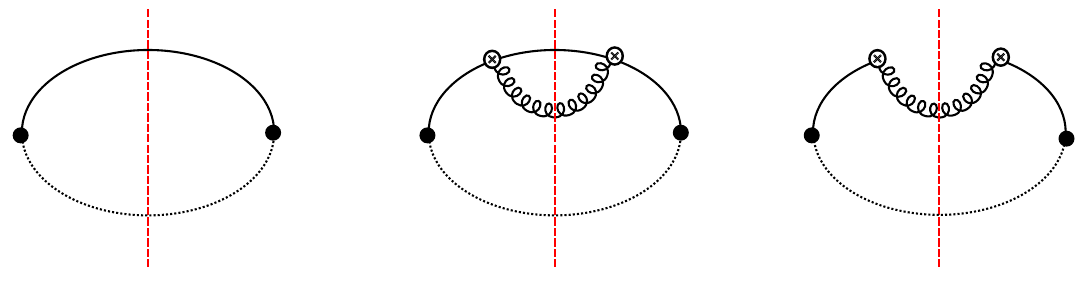}%
\caption{\label{fig:LOdiags} 
Feynman diagrams for soft functions $S_{^3S_1^{[8]}}(z)$ (left), 
$S_{^3P^{[8]}}(z)$ (middle), and $S_{^3P^{[1]}}(z)$ (right) at leading
nonvanishing order in the strong coupling. Black solid lines are timelike
Wilson lines in the $p$ direction, dotted lines are lightlike Wilson lines in
the $n$ direction, vertical dashed lines are final-state cuts, 
filled circles are the spacetime origins, 
curly lines are gluons, and the symbol $\otimes$ stands for insertion of the
field-strength tensor. 
}
\end{figure}

In this section, we compute the soft functions at leading nonvanishing orders
in the strong coupling, and verify that the soft factorization formulas we
obtained in the previous section reproduce the $z\to1$ singularities in the
leading-order fragmentation functions. 
The leading-order Feynman diagrams are shown in Fig.~\ref{fig:LOdiags}.

%%%%%%%%%%%%%%%%%%%%%%%%%%%%%%%%%%%%%%%%%%%%%%%%%%%%%%%%%%%%%%%%%%%%%%%%%%%%%
%%%%%%%%%%%%%%%%%%%%%%%%%%%%%%%%%%%%%%%%%%%%%%%%%%%%%%%%%%%%%%%%%%%%%%%%%%%%%

%==============================================================================
\subsection{\boldmath $^3S_1^{[8]}$}
%==============================================================================

We first consider the $^3S_1^{[8]}$ channel. The leading nonvanishing
contribution to the soft function comes from the $g^0$ terms of the adjoint
Wilson lines. That is, 
%---------------
\begin{align}
%---------------
{\cal W}(^3S_1^{[8]})^{cb} 
= \delta^{ca} \delta^{ba} +O(g) = \delta^{cb}+O(g), 
%---------------
\end{align}
%---------------
so that 
%---------------
\begin{align}
%---------------
S_{^3S_1^{[8]}}(z) = 2 \pi \delta(P^+ (1-z)) \delta^{cb} \delta^{cb}
+O(\alpha_s)
= \frac{2 \pi (N_c^2-1)}{P^+} \delta(1-z) 
+O(\alpha_s), 
%---------------
\end{align}
%---------------
which leads to 
%---------------
\begin{align}
%---------------
D^{\rm soft}_{g \to Q \bar Q (^3S_1^{[8]})}(z) &=
\frac{
C_{\rm frag} (d-2) g^2
}{4m^3}
\frac{2 \pi 2 \pi (N_c^2-1)}{P^+} \delta(1-z)
+ O(\alpha_s^2)
\nonumber \\
&= 
\frac{\pi \alpha_s}{m^3} \delta(1-z) + O(\alpha_s^2). 
%---------------
\end{align}
%---------------
By using the tree-level NRQCD matrix element 
$\langle 0 | {\cal O}^{Q \bar Q(^3S_1^{[8]})} (^3S_1^{[8]}) |0 \rangle =
(d-1) (N_c^2-1)$ [see Eq.~(6.1c) of Ref.~\cite{Bodwin:2012xc}], 
we obtain the $^3S_1^{[8]}$ FF in the soft approximation
%---------------
\begin{align}
\label{eq:d3s8soft}
%---------------
D^{\rm soft}_{g \to Q \bar Q (^3S_1^{[8]})}(z)
&= 
\frac{\pi \alpha_s}{m^3 (d-1) (N_c^2-1)} 
\delta(1-z) + O(\alpha_s^2),
%---------------
\end{align}
%---------------
which agrees with the full result. 
As is already clear from the soft factorization formula, 
the $^3S_1^{[8]}$ FF for longitudinally polarized $Q \bar Q$ vanishes in the
soft approximation, in agreement with the full result.

\subsection{\boldmath $^3P^{[8]}$}

Now we consider the $^3P^{[8]}$ channel. Since the soft function 
$S_{^3P^{[8]}}(z)$ explicitly contains the field-strength tensors, the
contribution at leading nonvanishing order in $\alpha_s$ comes from the matrix
element of ${\cal W}^{yx}_\beta(^3P^{[8]})$ on the vacuum and a single-gluon
state. We have 
%---------------
\begin{align}
%---------------
\langle g(k) | {\cal W}^{yx}_\beta (^3P^{[8]}) | 0 \rangle 
&= \langle g(k) | \int_0^\infty d\lambda \, \lambda \, \delta^{yc} p^\mu
G_{\mu \beta}^b (p \lambda) d^{bcd} \delta^{ad} \delta^{ax}| 0 \rangle
+ O(g^2) 
\nonumber \\ &= 
i \left(\frac{i}{ k \cdot p +i \varepsilon} \right)^2
p^\mu ( k_\mu \delta^\alpha_\beta - k_\beta \delta^\alpha_\mu )
d^{byx} + O(g^2) . 
%---------------
\end{align}
%---------------
By squaring this we obtain 
%---------------
\begin{align}
%---------------
S_{^3P^{[8]}}(z) 
&= 
- (d-2) 
\frac{(N_c^2-4) (N_c^2-1)}{N_c} 
\int d {\rm PS}_k 
\frac{
2 \pi (k^+ - P^+ (1-z))
(k \cdot p)^2 
}{(k\cdot p+i \varepsilon)^2 
(k\cdot p-i \varepsilon)^2 } + O(\alpha_s), 
%---------------
\end{align}
%---------------
where we used $d^{abc} d^{abc} = (N_c^2-4) (N_c^2-1)/N_c$ and 
$d {\rm PS}_k$ is the massless phase space integral given by 
%---------------
\begin{align}
%---------------
\int d{\rm PS}_k = 
\mu'^{2 \epsilon}
\int \frac{d^dk}{(2 \pi)^d} 2 \pi \delta(k^2) \theta(k_0), 
%---------------
\end{align}
%---------------
where $\mu'^2 = \mu^2 e^{\gamma_{\rm E}}/(4 \pi)$ and $\mu$ is the
$\overline{\rm MS}$ scale. 
The phase space integral can be evaluated in light-cone coordinates as 
%---------------
\begin{align}
%---------------
& \hspace{-5ex}
\int d {\rm PS}_k \frac{ 2 \pi (k^+ - P^+ (1-z)) (k \cdot p)^2
}{(k\cdot p+i \varepsilon)^2 (k\cdot p-i \varepsilon)^2 } 
\nonumber \\
&= 
\mu'^{2 \epsilon}
\int_0^\infty dk^+ 
\int_0^\infty dk^- 
\int \frac{d^{d-2} k_\perp}{(2 \pi)^{d-2}}
\frac{ \delta(k^2) \delta(k^+ - P^+ (1-z)) }{(k\cdot p)^2}
\nonumber \\
&= 
\frac{
\mu'^{2 \epsilon}}{2 \pi^{1-\epsilon} P_+ (p^2)^{1+\epsilon}}
\Gamma(1+\epsilon) 
\frac{1}{(1-z)^{1+2 \epsilon}}. 
%---------------
\end{align}
%---------------
Here, we integrated over $k^-$ and $k^+$ by using the delta functions, and then
computed the $k_\perp$ integral by using Eq.~(\ref{eq:kperpint}). 
From this we obtain 
%---------------
\begin{align}
%---------------
S_{^3P^{[8]}} (z)
&= 
-(d-2)
\frac{(N_c^2-4) (N_c^2-1)}{N_c}
\frac{\Gamma(1+\epsilon) \mu'^{2 \epsilon}}{2 \pi^{1-\epsilon} 
m^{2+2\epsilon} P^+}
\frac{1}{(1-z)^{1+2 \epsilon}}
+O(\alpha_s). 
%---------------
\end{align}
%---------------
This gives the fragmentation function in the soft approximation
%---------------
\begin{align}
\label{eq:Dg3P8bare}
%---------------
D^{\rm soft, \, bare}_{g \to Q \bar Q (^3P^{[8]})}(z) &=
\frac{2 \alpha_s^2 (N_c^2-4)}{3 m^5 N_c} 
\left[1+2 \epsilon \left( \log \frac{\mu}{2 m} -\frac{1}{6} \right)
+O(\epsilon^2) \right] \frac{1}{(1-z)^{1+2 \epsilon}} +O(\alpha_s^3) 
\\
&= 
\frac{2 \alpha_s^2 (N_c^2-4)}{3 m^5 N_c}
\bigg[
- \frac{\delta(1-z)}{2 \epsilon_{\rm IR}} 
+ 
\left(\frac{1}{6} -\log \frac{\mu}{2 m} \right)
\delta(1-z)
\nonumber \\ & \hspace{20ex} 
+ \frac{1}{(1-z)_+} 
+O(\epsilon) \bigg] + O(\alpha_s^3).
%---------------
\end{align}
%---------------
In expanding the function $1/(1-z)^{1+2 \epsilon}$ in powers of $\epsilon$ we
used the identity 
%---------------
\begin{align}
\label{eq:plusidentity}
%---------------
\frac{1}{(1-z)^{1+n \epsilon}}
= - \frac{1}{n \epsilon_{\rm IR}} \delta(1-z) 
+ \left[ \frac{1}{(1-z)^{1+n \epsilon}}\right]_+ . 
%---------------
\end{align}
%---------------
In NRQCD factorization, the IR pole in Eq.~(\ref{eq:Dg3P8bare}) is canceled by
the one-loop correction to the $^3S_1^{[8]}$ matrix element, 
which involves both UV and IR poles. 
When the NRQCD matrix element is renormalized in the $\overline{\rm MS}$
scheme, this effectively amounts to subtracting the IR pole in 
Eq.~(\ref{eq:Dg3P8bare}) and replacing $\mu$ by the $\overline{\rm MS}$ scale 
$\mu_\Lambda$. 
Then the renormalized result for $D^{\rm soft}_{g \to Q \bar Q
(^3P^{[8]})}(z)$ is obtained by dividing by the normalization for the 
$^3P^{[8]}$ matrix element, which is given by $(d-1) (N_c^2-1)$ in the
normalization used in Ref.~\cite{Bodwin:2010fi}. 
We thus obtain 
%---------------
\begin{align}
\label{eq:dg3P8soft}
%---------------
D^{\rm soft}_{g \to Q \bar Q (^3P^{[8]})}(z) &=
\frac{8 \alpha_s^2 }{3 (d-1) (N_c^2-1) m^5}
\frac{N_c^2-4}{4N_c}
\nonumber \\ & \quad
\times 
\bigg[
\left(\frac{1}{6} -\log \frac{\mu_\Lambda}{2 m} \right)
\delta(1-z)
+ \frac{1}{(1-z)_+}
\bigg] + O(\alpha_s^3),
%---------------
\end{align}
%---------------
which reproduces exactly the singular distributions in the 
$D_{g \to Q \bar Q (^3P^{[8]})}(z)$. 
Similarly to the $^3S_1^{[8]}$ case, the contribution from longitudinally
polarized $Q \bar Q$ spin vanishes in the soft approximation, which is
consistent with the fact that the $^3P_J^{[8]}$ FF for longitudinal
$Q \bar Q$ spin polarization does not involve singular
distributions~\cite{Bodwin:2012xc}.

%==============================================================================
\subsection{\boldmath $^3P_J^{[1]}$}
%==============================================================================

We next consider the $^3P_J^{[1]}$ channel. Similarly to the $^3P^{[8]}$ case, 
the contribution at leading nonvanishing order in $\alpha_s$ comes from the 
matrix element of ${\cal W}_\beta^b (^3P_J^{[1]})$ on the vacuum and a
single-gluon state. We have 
%---------------
\begin{align}
%---------------
\langle g(k) | {\cal W}^b_{\beta} (^3P_J^{[1]}) | 0 \rangle
&=
\langle g(k) | \int_0^\infty d\lambda \, \lambda\, 
p^\sigma G_{\sigma \beta}^{d} (p \lambda) \delta^{da} \delta^{ba} 
| 0 \rangle + O(g^2)
\nonumber \\ 
&=
i \left(\frac{i}{ k \cdot p +i \varepsilon}\right)^2
p^\mu ( k_\mu \delta^\alpha_\beta - k_\beta \delta^\alpha_\mu )
\delta^{zb} + O(g^2) ,
%---------------
\end{align}
%---------------
which leads to 
%---------------
\begin{align}
%---------------
S_{^3P^{[1]}}(z) &= 
-(d-2) (N_c^2-1) \int d{\rm PS}_k \frac{2 \pi \delta(k^+ - P^+ (1-z))}
{(k \cdot p)^2}+ O(\alpha_s),
\\
S_{^3P^{[1]}}^{TT}(z) &=
-(N_c^2-1) \int d{\rm PS}_k \frac{2 \pi \delta(k^+ - P^+ (1-z))}{(k \cdot p)^2}
\nonumber \\ & \hspace{18ex} \times 
\left( 
\frac{p^2}{p_+^2}
\frac{k_+^2 p^2 -2 p^+ k^+ k \cdot p}{( k \cdot p)^2}
+ \frac{d-2}{d-1} \right) + O(\alpha_s). 
%---------------
\end{align}
%---------------
It is clear from this expression that 
at leading nonvanishing order in $\alpha_s$,
$S_{^3P^{[1]}}(z)$ is obtained from
$S_{^3P^{[8]}}(z)$ by multiplying $N_c/(N_c^2-4)$. 
The anisotropic part $S_{^3P^{[1]}}^{TT}(z)$ can again be computed in
light-cone coordinates. 
The result is 
%---------------
\begin{align}
%---------------
S_{^3P^{[1]}}^{TT}(z) &=
\frac{(N_c^2-1) \epsilon (1-\epsilon) (1-2 \epsilon) \Gamma(1+\epsilon) \mu'^{2
\epsilon}}
{3 (3-2 \epsilon) \pi^{1-\epsilon} m^2 P^+ (1-z)^{1+2 \epsilon}} 
+ O(\alpha_s)
\nonumber \\
&= - \frac{N_c^2-1}{18 \pi m^2 P^+} \delta(1-z) +O(\epsilon, \alpha_s). 
%---------------
\end{align}
%---------------
Because of the explicit factor of $\epsilon$, 
the $S_{^3P^{[1]}}^{TT}(z)$ is free of poles at leading nonvanishing order in
$\alpha_s$. From these we obtain 
%---------------
\begin{align}
\label{eq:Dg3P1bare}
%---------------
D^{\rm soft, \, bare}_{g \to Q \bar Q (^3P_J^{[1]})}(z) &=
\frac{4 \alpha_s^2}{3 m^5 N_c} c^J 
\left[1+2 \epsilon \left( \log \frac{\mu}{2 m} - \frac{Q_J}{2 J+1}
\right)
+O(\epsilon^2) \right] \frac{1}{(1-z)^{1+2 \epsilon}} +O(\alpha_s^3)
\\
&= 
\frac{4 \alpha_s^2}{3 m^5 N_c} c^J 
\bigg[
- \frac{\delta(1-z)}{2 \epsilon_{\rm IR}} 
+ \left( \frac{Q_J}{2 J+1} - \log \frac{\mu}{2 m} \right) \delta(1-z) 
\nonumber \\
& \hspace{18ex} + \frac{1}{(1-z)_+} 
+O(\epsilon) \bigg] 
+O(\alpha_s^3), 
%---------------
\end{align}
%---------------
where we used $-2 \epsilon (1-z)^{-1-2 \epsilon} = 1 + O(\epsilon)$ 
to combine the order-$\epsilon$ coefficient of $(1-z)^{-1-2 \epsilon}$
in $S_{^3P^{[1]}}(z)$ and the finite piece in $S_{^3P^{[1]}}^{TT}(z)$ 
into $Q_J/(2 J+1)$. 
Here the $\epsilon$ dependence in the overall factor $c_J$ is kept, because 
this corresponds to the $d$-dimensional definition of the $^3P_J^{[1]}$ matrix
element. 
In the polarized case, we obtain
%---------------
\begin{align}
\label{eq:Dg3P1polbare}
%---------------
D^{\rm soft, \, bare}_{g \to Q \bar Q (^3P_{J,h}^{[1]})}(z) &=
\frac{4 \alpha_s^2}{3 m^5 N_c} c^{J,h}
\left[
1+2 \epsilon \left( 
\log \frac{\mu}{2 m} - 
\frac{c^J}{c^{J,h}} 
\frac{Q_{J,h}}{2 J+1}
\right)
+O(\epsilon^2) \right] \frac{1}{(1-z)^{1+2 \epsilon}} +O(\alpha_s^3)
\\
&=
\frac{4 \alpha_s^2}{3 m^5 N_c} c^{J,h}
\bigg[
- \frac{\delta(1-z)}{2 \epsilon_{\rm IR}}
+ \left( 
\frac{c^J}{c^{J,h}} 
\frac{Q_{J,h}}{2 J+1} - 
\log \frac{\mu}{2 m} \right) \delta(1-z)
\nonumber \\
& \hspace{18ex} + 
\frac{1}{(1-z)_+}
+O(\epsilon) \bigg]
+O(\alpha_s^3),
%---------------
\end{align}
%---------------
where $Q_{J=1,|h|=1} = 0$, $Q_{J=1,h=0} = 3/8$, 
$Q_{J=2,h=|2|} = 3/4$, $Q_{J=2,|h|=1} = 0$, and $Q_{J=2,h=0} = 1/8$. 
Again, the $\epsilon$ dependences in the $c^{J,h}$ are kept, because they
correspond to the $d$-dimensional definition of the polarized $^3P_J^{[1]}$
matrix element. 
Similarly to the $^3P^{[8]}$ case, the IR pole is canceled by the one-loop
correction to the $^3S_1^{[8]}$ matrix element, which involves an IR pole
proportional to the tree-level $^3P_J^{[1]}$ matrix element. 
The $^3P_J^{[1]}$ short-distance coefficient in the soft approximation is then
obtained by subtracting the IR pole multiplied by the normalization of the
$^3P_J^{[1]}$ matrix element that is proportional to $c^J$ ($c^{J,h}$ in
the polarized case), replacing $\mu$ by the $\overline{\rm MS}$ scale 
$\mu_\Lambda$ for renormalization of the $^3S_1^{[8]}$ matrix element, 
and dividing by the normalization $c_J \times 2 N_c (d-1)$ of the 
polarization-summed $^3P_J^{[1]}$ matrix element. We obtain 
%---------------
\begin{align}
\label{eq:dg3P1soft}
%---------------
D^{\rm soft}_{g \to Q \bar Q (^3P_J^{[1]})}(z) &=
\frac{4 \alpha_s^2}{3 (d-1) m^5 N_c^2} 
\bigg[
\left( \frac{Q_J}{2 J+1} - \log \frac{\mu_\Lambda}{2 m} \right) \delta(1-z)
+
\frac{1}{(1-z)_+}
\bigg]
+O(\alpha_s^3),
%---------------
\end{align}
%---------------
which reproduces exactly the singular distributions in 
$D_{g \to Q \bar Q (^3P_J^{[1]})}(z)$. 
Similarly, in the polarized case, we obtain
%---------------
\begin{align}
\label{eq:dg3P1softpol}
%---------------
D^{\rm soft}_{g \to Q \bar Q (^3P_{J,h}^{[1]})}(z) &=
\frac{4 \alpha_s^2}{3 (d-1) m^5 N_c^2}
\bigg[
\left( \frac{Q_{J,h}}{2 J+1} - a_{J,h}\log \frac{\mu_\Lambda}{2 m} \right) \delta(1-z)
+
\frac{a_{J,h}}{(1-z)_+}
\bigg]
+O(\alpha_s^3).
%---------------
\end{align}
%---------------
where $a_{J,h} = c^{J,h}/c^J|_{d=4}$. 
These again exactly reproduce the singular distributions in the polarized
results for $D_{g \to Q \bar Q (^3P_{J,h}^{[1]})}(z)$ in
Refs.~\cite{Cho:1994gb, Ma:2015yka}, which are shown in
Appendix~\ref{sec:helicitydecomp}.

%==============================================================================
\subsection{Summary of leading-order results} 
%==============================================================================

We computed the fragmentation functions in the soft approximation by using the
soft factorization formulas we obtained in Sec.~\ref{sec:softfac}. 
The results in Eqs.~(\ref{eq:d3s8soft}), (\ref{eq:dg3P8soft}), and 
(\ref{eq:dg3P1soft}) reproduce exactly the singular distributions, namely, the 
Dirac delta functions and plus distributions, in the leading-order
fragmentation functions in Eqs.~(\ref{eq:FOFFs}). 
This implies that in Mellin space, 
%---------------
\begin{align}
%---------------
\lim_{N \to \infty} 
\left[ 
\tilde{D}^{\rm soft}_{g \to Q \bar Q ({\cal N})}(N)
- \tilde{D}_{g \to Q \bar Q ({\cal N})}(N) \right] 
= 0, 
%---------------
\end{align}
%---------------
meaning that the terms in ${D}_{g \to Q \bar Q ({\cal N})}(z)$
that are not reproduced by the soft approximation are regular functions in $z$. 
This explicitly confirms the validity of the soft factorization at leading
nonvanishing order in $\alpha_s$.

%==============================================================================
\section{Soft functions at NLO}
\label{sec:softNLO}
%==============================================================================

In this section, we compute the soft functions that appear in the soft
factorization formulas for the $^3S_1^{[8]}$, $^3P^{[8]}$, and $^3P_J^{[1]}$
fragmentation functions at NLO in $\alpha_s$. We work in the Feynman gauge. 

%==============================================================================
\subsection{\boldmath $^3S_1^{[8]}$}
%==============================================================================

\begin{figure}[t]
\includegraphics[width=.9\textwidth]{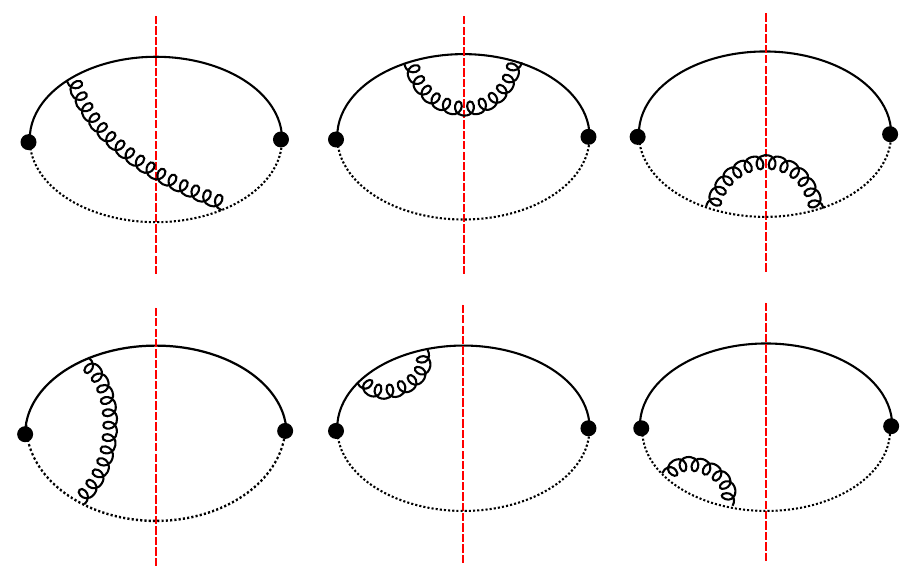}%
\caption{\label{fig:NLO_3s1}
Feynman diagrams that appear in NLO calculation of the 
soft function $S_{^3S_1^{[8]}}(z)$. There are additional diagrams that can be
obtained by complex conjugation. 
}
\end{figure}

The Feynman diagrams for the soft function $S_{^3S_1^{[8]}}(z)$ at NLO are
shown in Fig.~\ref{fig:NLO_3s1}. We compute the real and virtual diagrams
separately.

%%%%%%%%%%%%%%%%%%%%%%%%%%%%%%%%%%%%%%%%%%%%%%%%%%%%%%%%%%%%%%%%%%%%%%%%%%%%%
%%%%%%%%%%%%%%%%%%%%%%%%%%%%%%%%%%%%%%%%%%%%%%%%%%%%%%%%%%%%%%%%%%%%%%%%%%%%%

%==============================================================================
\subsubsection{Real diagrams}
%==============================================================================

We first consider the real diagrams, where a gluon attaches to the
lightlike or timelike Wilson lines. 
The matrix element of ${\cal W}(^3S_1^{[8]})^{cb}$ on the vacuum and
the one-gluon state is given at leading nonvanishing order by 
%---------------
\begin{align}
\label{eq:W3S8_realamp}
%---------------
\langle g(k) | {\cal W}(^3S_1^{[8]})^{cb} | 0 \rangle
&= 
- g \frac{i p \cdot \epsilon^*(k) }{k\cdot p+i \varepsilon} f^{xca} \delta^{ba} 
- g \frac{i n \cdot \epsilon^*(k)}{k\cdot n+i \varepsilon} f^{xba} \delta^{ca} 
+O(g^3), 
%---------------
\end{align}
%---------------
where $\epsilon(k)$ is the polarization vector of the outgoing gluon. 
Then the contribution from the real diagrams to $S_{^3S_1^{[8]}} (z)$ is given
by
%---------------
\begin{align}
%---------------
S_{^3S_1^{[8]}}^{\rm NLO} (z) \Big|_{\rm real} &= 
g^2 C_A (N_c^2-1) 
\int d{\rm PS}_k 2 \pi \delta(k^+-P^+(1-z)) 
\left[ -\frac{p^2}{(k \cdot p)^2} + \frac{2 p^+}{k^+ k \cdot p} \right] .
%---------------
\end{align}
%---------------
Here the first term in the square brackets comes from the square of the first
term on the right-hand side of Eq.~(\ref{eq:W3S8_realamp}), and the second term
comes from the cross term. The square of the second term on the right-hand side
of Eq.~(\ref{eq:W3S8_realamp}) vanishes because it is proportional to $n^2 =
0$. 
The phase-space integration is straightforward:
%---------------
\begin{align}
%---------------
\int d{\rm PS}_k 
\frac{2 \pi \delta(k^+-P^+(1-z)) p^2}{(k \cdot p)^2}
&= 
\frac{\mu'^{2 \epsilon}}{(2 \pi)^{1-\epsilon} [P_+ (1-z)]^{1+2 \epsilon}}
\Gamma(1+\epsilon) ,
\\
\int d{\rm PS}_k 
\frac{2 \pi \delta(k^+-P^+(1-z)) 2 p^+}{k^+ (k \cdot p)}
&= 
\frac{\mu'^{2 \epsilon}}{(2 \pi)^{1-\epsilon} [P_+ (1-z)]^{1+2 \epsilon}}
\Gamma(\epsilon_{\rm UV}) ,
%---------------
\end{align}
%---------------
which give 
%---------------
\begin{align}
\label{eq:S3S8real}
%---------------
S_{^3S_1^{[8]}}^{\rm NLO} (z) \Big|_{\rm real} &=
4 \pi \alpha_s C_A (N_c^2-1)
\frac{\mu'^{2 \epsilon}}{(2 \pi)^{1-\epsilon} [P_+ (1-z)]^{1+2 \epsilon}} 
\left[ \Gamma(\epsilon_{\rm UV}) - \Gamma(1+\epsilon) \right] 
\\
&= \frac{\alpha_s C_A (N_c^2-1)}{P_+} 
\left( \frac{\mu}{2 m} \right)^{2 \epsilon} 
\left[ \frac{2}{\epsilon_{\rm UV}} - 2 + \frac{\pi^2}{6} \epsilon +
O(\epsilon^2) \right] 
\nonumber \\ & \quad \times 
\left\{ - \frac{\delta(1-z)}{2 \epsilon_{\rm IR}} + 
\left[ \frac{1}{(1-z)^{1+2 \epsilon}} \right] \right\}.
%---------------
\end{align}
%---------------

%==============================================================================
\subsubsection{Virtual diagrams}
%==============================================================================

The contribution from the virtual diagrams that come from the exchange of a
gluon between the lightlike and timelike Wilson lines is 
%---------------
\begin{align}
%---------------
S_{^3S_1^{[8]}}^{\rm NLO} (z) \Big|_{\rm virtual} &=
2 {\rm Re} 
\int_k
\left(- g n^\mu \frac{i}{k \cdot n +i \varepsilon} f^{xba} \right) 
\frac{-i g_{\mu \nu} \delta^{bc} }{k^2+i \varepsilon} 
\nonumber \\ & \quad \quad \times 
\left(- g p^\nu \frac{i}{-k \cdot p +i \varepsilon} f^{xca} \right)
2 \pi \delta (P^+ (1-z))
\nonumber\\
&= 
{\rm Re}  \, 
i g^2 (N_c^2-1) C_A 
\int_k
\frac{
2 \pi \delta (1-z)
}{(k^2+i \varepsilon)(k \cdot n +i \varepsilon) (-k \cdot p+i
\varepsilon)}. 
%---------------
\end{align}
%---------------
We use the shorthand 
%---------------
\begin{align}
%---------------
\int_k = \mu'^{2 \epsilon} \int \frac{d^dk}{(2 \pi)^d}. 
%---------------
\end{align}
%---------------
It is easy to compute this scaleless one-loop integral by using Schwinger
parametrization. We have 
%---------------
\begin{align}
%---------------
& \hspace{-5ex} 
\int_k
\frac{1}{(k^2+i \varepsilon)(k \cdot n +i \varepsilon) (-k \cdot p+i
\varepsilon)}
\nonumber \\ &= 
\int_0^\infty dx dy dz \int_k
\exp \left( x k^2 + y k \cdot n -z k \cdot p +i \varepsilon \right)
\nonumber \\ &= 
\int_0^\infty dx dy dz \int_k
\exp \left( x k^2 - \frac{(n y-z p)^2}{4 x}
 +i \varepsilon \right)
\nonumber \\ &= 
\frac{i \mu'^{2 \epsilon}}{(4 \pi)^{d/2}} 
\int_0^\infty dx dy dz \, x^{-d/2}
\exp \left( - \frac{(n y-z p)^2}{4 x} \right)
\nonumber \\ &= 
\frac{i}{(4 \pi)^2} 
\frac{1}{p_+} \frac{1}{\epsilon_{\rm UV}} 
\left( 
\frac{1}{\epsilon_{\rm UV}} 
- \frac{1}{\epsilon_{\rm IR}} 
\right),
%---------------
\end{align}
%---------------
which gives 
%---------------
\begin{align}
\label{eq:S3S8virtual}
%---------------
S_{^3S_1^{[8]}}^{\rm NLO} (z)|_{\textrm{virtual}} =
\frac{ \alpha_s C_A (N_c^2-1) }{P_+} \delta (1-z)
\frac{1}{\epsilon_{\rm UV}}
\left( \frac{1}{\epsilon_{\rm IR}} - \frac{1}{\epsilon_{\rm UV}} \right).
%---------------
\end{align}
%---------------

The remaining diagrams have to do with the self energies of the
lightlike and timelike Wilson lines. The self-energy diagram for the lightlike
Wilson line vanishes in Feynman gauge because it is proportional to $n^2 =0$. 
The self-energy of the eikonal line gives the following correction term
in the eikonal-line propagator 
%---------------
\begin{align}
%---------------
\frac{i}{l \cdot p+i \varepsilon} 
\left[ g^2 \int_k \frac{i}{(k+l) \cdot p+i \varepsilon}
f^{bcx} f^{bxa}
\frac{-i p^2}{k^2+i \varepsilon} 
\right]
\frac{i}{l \cdot p+i \varepsilon}
\equiv 
\frac{i}{l \cdot p+i \varepsilon} \Sigma_p(l \cdot p) \frac{i}{l \cdot p+i
\varepsilon}.
%---------------
\end{align}
%---------------
This gives the field renormalization of the eikonal-line propagator 
%---------------
\begin{align}
%---------------
\frac{i}{l \cdot p -i\Sigma_p(l \cdot p)+i \varepsilon} 
= 
\frac{i}{l \cdot p+i \varepsilon} 
\left[
1-i
\frac{\partial}{\partial l \cdot p} \Sigma_p(l \cdot p)\right]_{l \cdot
p=0}^{-1} + \cdots,
%---------------
\end{align}
%---------------
where the $\cdots$ are terms regular at $l \cdot p=0$. 
The calculation of $\Sigma_p(l \cdot p)$ is straightforward. By using Feynman
parametrization we obtain 
%---------------
\begin{align}
%---------------
\Sigma_p(l \cdot p)
&= 
-C_A g^2 p^2 \int_k \frac{1}{(k+l) \cdot p+i \varepsilon}
\frac{1}{k^2+i \varepsilon}
\nonumber\\ &= 
-C_A g^2 p^2 
\frac{i}{(4 \pi)^{2-\epsilon}} \Gamma(\epsilon_{\rm UV})
\int_0^\infty dx 
\frac{1}{[ x^2 p^2/4- x l \cdot p-i \varepsilon]^{\epsilon} }, 
\\
-i \frac{\partial}{\partial l \cdot p} \Sigma_p(l \cdot p)\Big|_{l \cdot p=0}
&= 
-C_A g^2 p^2 \frac{1}{(4 \pi)^{2-\epsilon}} \Gamma(\epsilon_{\rm UV})
\int_0^\infty dx \frac{4^{1+\epsilon} \epsilon x}{( x^2 p^2)^{1+\epsilon} }
\nonumber \\
&= 
-C_A g^2 (p^2)^{-\epsilon} \frac{4^{1+\epsilon}}{(4 \pi)^{2-\epsilon}} \Gamma(1+\epsilon)
\int_0^\infty dx \frac{1}{x^{1+2 \epsilon} } 
\nonumber \\
&= 
- \frac{\alpha_s C_A}{\pi} 
\left( \frac{1}{2\epsilon_{\rm UV}} - \frac{1}{2 \epsilon_{\rm IR}} \right).
%---------------
\end{align}
%---------------
Hence, the self-energy diagrams give the following contribution to the 
$^3S_1^{[8]}$ soft function
%---------------
\begin{align}
\label{eq:S3S8self}
%---------------
S_{^3S_1^{[8]}}^{\rm NLO} (z) \Big|_{\textrm{self-energy} }
&=
\frac{\alpha_s C_A}{\pi}
\left( \frac{1}{2\epsilon_{\rm UV}} - \frac{1}{2 \epsilon_{\rm IR}} \right)
S_{^3S_1^{[8]}}^{\rm LO} (z) 
\nonumber \\
&=
\frac{\alpha_s C_A (N_c^2-1)}{P^+} \delta(1-z)
\left( \frac{1}{\epsilon_{\rm UV}} - \frac{1}{\epsilon_{\rm IR}} \right).
%---------------
\end{align}
%---------------

%==============================================================================
\subsubsection{Results and discussion}
%==============================================================================

The $^3S_1^{[8]}$ soft function at NLO is given by the sum of
Eqs.~(\ref{eq:S3S8real}), (\ref{eq:S3S8virtual}), and (\ref{eq:S3S8self}). 
The result is 
%---------------
\begin{align}
\label{eq:S3S8NLO}
%---------------
S_{^3S_1^{[8]}}^{\rm NLO} (z) 
&=
\frac{\alpha_s C_A (N_c^2-1)}{P^+} 
\left( \frac{\mu}{2 m} \right)^{2 \epsilon}
\bigg[ 
\frac{2 \epsilon_{\rm UV}^{-1} - 2 + \epsilon \pi^2/6 + O(\epsilon^2) 
}{(1-z)^{1+2 \epsilon}} 
\nonumber \\ & \quad 
- \delta(1-z) \frac{1}{\epsilon_{\rm UV}} 
\left( \frac{1}{\epsilon_{\rm UV}} - \frac{1}{\epsilon_{\rm IR}} \right)
+ \delta(1-z) 
\left( \frac{1}{\epsilon_{\rm UV}} - \frac{1}{\epsilon_{\rm IR}} \right)
\bigg].
%---------------
\end{align}
%---------------
The term involving $(1-z)^{-1-2 \epsilon}$ comes from the real diagram, while
the terms proportional to $\epsilon_{\rm UV}^{-1} - \epsilon_{\rm IR}^{-1}$
come from the virtual and self-energy diagrams. 
By using the expansion of $(1-z)^{-1-2 \epsilon}$ in powers of $\epsilon$ 
given by 
%---------------
\begin{align}
%---------------
\frac{1}{(1-z)^{1+2 \epsilon}} 
= - \frac{\delta(1-z)}{2 \epsilon_{\rm IR}} + 
\frac{1}{(1-z)_+} + \epsilon \left[ \frac{-2 \log(1-z)}{1-z} \right]_+ 
+ O(\epsilon^2),
%---------------
\end{align}
%---------------
it is straightforward to see that all IR poles cancel in 
$S_{^3S_1^{[8]}}^{\rm NLO} (z)$. 
In particular, the mixed double pole $\epsilon_{\rm UV}^{-1} \epsilon_{\rm
IR}^{-1}$ cancels between the real and virtual diagram contributions. 
The single IR pole $\epsilon_{\rm IR}^{-1}$ cancels between the real and
self-energy diagram contributions. 
By using $S_{^3S_1^{[8]}}^{\rm LO} (z) = 2 \pi (N_c^2-1) \delta(1-z)/P^+$,
we can write
%---------------
\begin{align}
\label{eq:S3S8NLO3}
%---------------
S_{^3S_1^{[8]}} (z)
&=
\frac{2 \pi (N_c^2-1)}{P^+}
\left( \frac{\mu}{2 m} \right)^{2 \epsilon}
\bigg\{ \delta(1-z) 
+ \frac{\alpha_s C_A }{\pi}
\bigg[
- \frac{\delta(1-z) }{2 \epsilon_{\rm UV}^2}
\nonumber \\ & \hspace{10ex}
+ \frac{1}{\epsilon_{\rm UV}} \left[ \frac{1}{(1-z)_+}
+ \frac{1}{2} \delta(1-z) \right]
\nonumber \\ & \hspace{10ex}
+ \left[ \frac{-2 \log(1-z)}{1-z} \right]_+ - \frac{1}{(1-z)_+}
- \frac{\pi^2}{24} \delta(1-z) 
\bigg] \bigg\}+ O(\epsilon,\alpha_s^2).
%---------------
\end{align}
%---------------
Note that now all poles are of UV origin. 
The order-$\epsilon^0$ term $\alpha_s C_A/\pi [ -2 \log(1-z)/(1-z)]_+$
is exactly the threshold double logarithmic term 
that appears in the NLO correction to the gluon fragmentation function in the
$^3S_1^{[8]}$ channel. Its origin can be traced back to the order-$\epsilon$
term in the expansion of $(1-z)^{-1-2 \epsilon}$ in powers of $\epsilon$.
Because of this, the coefficient of the threshold double logarithm is tied to
the double UV pole, and also to the term proportional to $1/(1-z)_+$ in the
single UV pole. 
We also note that the order-$\epsilon^0$ term $-\alpha_s C_A/\pi \times 
1/(1-z)_+$ is exactly the
threshold single logarithmic term in $D_{g \to Q \bar Q(^3S_1^{[8]})}(z)$ at
the factorization scale $\mu_F = 2 m$, which comes from the 
$2 C_A z/(1-z)_+$ term in $P_{gg}(z)$. Similarly to the threshold double
logarithm, its coefficient is tied to the $\delta(1-z)$ term in the single UV
pole. Hence, the threshold logarithm in the gluon fragmentation function in the
$^3S_1^{[8]}$ channel is determined by the coefficient of the single UV pole of
the soft function, which corresponds to its anomalous dimension. 
To see this more clearly for the threshold double logarithm, let us take the
Mellin transform and consider its large $N$ behavior. We have 
%---------------
\begin{align}
\label{eq:S3S8NLOMellin}
%---------------
\tilde{S}_{^3S_1^{[8]}} (N)|_{\rm threshold} 
&=
\frac{2 \pi (N_c^2-1)}{P^+}
\bigg[ 1 
+ \frac{\alpha_s C_A }{\pi}
\bigg(- \log^2 N 
+ \frac{\log N}{\epsilon_{\rm UV}} 
- \frac{1}{2 \epsilon_{\rm UV}^2} 
\bigg) \bigg]+ O(\alpha_s^2),
%---------------
\end{align}
%---------------
where at each order in $\alpha_s$ we only kept the leading terms at large $N$. 
We may identify the coefficient of the $1/\epsilon_{\rm UV}$ pole as the cusp
anomalous dimension in the adjoint representation at leading order in
$\alpha_s$.

%==============================================================================
\subsection{\boldmath $^3P^{[8]}$}
%==============================================================================

We now turn to the case of the $^3P^{[8]}$ channel. 
The calculation of the $^3P^{[8]}$ soft function is a lot more involved
compared to the $^3S_1^{[8]}$ case, as the
operator definition of $S_{^3P^{[8]}}(z)$ explicitly contains gluon field
operators, and it already involves an IR pole at leading nonvanishing order in
$\alpha_s$. Fortunately, the amount of calculation needed is reduced
significantly if we only consider the threshold double logarithmic
contributions. As we will see later, because nonvanishing contributions to 
the $^3P^{[8]}$ soft function always involve at least one gluon crossing
the cut, the loop integrals involve only the scale $P^+(1-z)$, so that from
dimensional considerations 
the next-to-leading-order results are proportional to $(1-z)^{-1-4\epsilon}$. 
Hence, the threshold double logarithmic contributions can only arise from 
terms proportional to $\epsilon^{-2} (1-z)^{-1-4\epsilon}$, and so we can only
retain contributions that produce double poles in $\epsilon$. 
Similarly to the $^3S_1^{[8]}$ case, the $^3P^{[8]}$ soft function can be 
regarded as the NRQCD matrix element of the $^3S_1^{[8]}$ operator on the
$^3P^{[8]}$ state at leading power in the nonrelativistic expansion, 
when the delta function $2 \pi \delta( n \cdot \hat{p} - P^+ (1-z))$ 
is replaced with unity. 
That is, the soft function integrated over the unbounded momentum 
$\ell^+ \equiv P^+ (1-z)$ is proportional to the NRQCD matrix element. 
%which is already known to next-to-leading order in $\alpha_s$. 
The next-to-leading-order correction to the 
$^3S_1^{[8]}$ operator on the $^3P^{[8]}$ state 
is known to involve only a single IR pole\footnote{The double poles in Eqs.~(3.9) 
and (3.10) in Ref.~\cite{Zhang:2020atv} are coming from the vacuum polarization
correction to the matrix element.}.
This means that, because integration over $\ell^+$ does not change the infrared
behavior, the double pole in the soft function must be of ultraviolet nature. 
This significantly simplifies the calculation, 
as for the purpose of identifying threshold double logarithms, 
it suffices to look only for UV poles.

%==============================================================================
\subsubsection{Planar diagrams}
%==============================================================================

%%%%%%%%%%%%%%%%%%%%%%%%%%%%%%%%%%%%%%%%%%%%%%%%%%%%%%%%%%%%%%%%%%%%%%%%%%%%%
\begin{figure}[t]
\includegraphics[width=.6\textwidth]{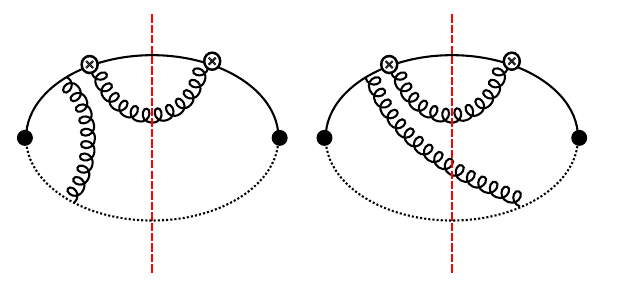}%
\caption{\label{fig:NLO_3pj}
Planar Feynman diagrams that appear in NLO calculation of the
soft function $S_{^3P^{[8]}}(z)$. There are additional diagrams that can be
obtained by complex conjugation.
}
\end{figure}
%%%%%%%%%%%%%%%%%%%%%%%%%%%%%%%%%%%%%%%%%%%%%%%%%%%%%%%%%%%%%%%%%%%%%%%%%%%%%

From the per-diagram results available in Ref.~\cite{Bodwin:2019bpf}, 
we know that the double UV pole arises from the diagram where an additional
virtual gluon is exchanged between the timelike and lightlike Wilson lines. 
As we will see later, these double UV poles will only arise from planar
diagrams shown in Fig.~\ref{fig:NLO_3pj}. Here we first consider the
contributions from the virtual and real planar diagrams.

The virtual correction gives the following contribution to the matrix element 
of the operator ${\cal W}^{yx}_\beta (^3P^{[8]})$ between the vacuum and the
one-gluon state:
%---------------
\begin{align}
\label{eq:3p8virtual}
%---------------
& \hspace{-5ex} 
\langle g^{b,\alpha} (l) | {\cal W}^{yx}_\beta (^3P^{[8]}) | 0\rangle
|_{\textrm{virtual}}
\nonumber \\
&= 
i g^2 C_A d^{bxy}
p^\mu
( l_\mu \delta_\beta^{\alpha} - l_\beta \delta^{\alpha}_\mu ) 
\int_k
\frac{-i}{k^2+i \varepsilon}
p \cdot n
\frac{i}{-k \cdot n+i \varepsilon} 
\nonumber \\ & \quad \times
\left[ 
\left( \frac{i}{l \cdot p+i \varepsilon}\right)^2 
\frac{i}{(l+k) \cdot p +i \varepsilon} 
+
 \frac{i}{l \cdot p+i \varepsilon}
\left(\frac{i}{(l+k) \cdot p +i \varepsilon} \right)^2 
\right], 
%---------------
\end{align}
%---------------
where $b$ and $\alpha$ are the color and Lorentz indices of the outgoing gluon, 
respectively. 
Note the squaring of the timelike Wilson-line denominators, which follows from
expanding the Wilson lines in the $p+q$ and $p-q$ directions to linear power in
$q$. 
As we will see later, only the first term in the square brackets involves UV
divergences. 
We thus need to compute the one-loop integral 
%---------------
\begin{align}
%---------------
I (l) &= p \cdot n \int_k \frac{-i}{k^2+i \varepsilon} 
\frac{i}{-k \cdot n +i \varepsilon} 
\frac{i}{(l +k) \cdot p +i \varepsilon} 
%---------------
\end{align}
%---------------
This integral is computed straightforwardly by using the parametrization 
%---------------
\begin{align}
%---------------
\frac{1}{ABC} = \int_0^\infty dx dy \frac{2}{(A+x B+y C)^3},
%---------------
\end{align}
%---------------
which leads to 
%---------------
\begin{align}
%---------------
I (l) &=
2 i p\cdot n 
\int_{k} 
\int_0^\infty dx dy \frac{1}{[k^2 - x k \cdot n +y (l + k) \cdot p 
+i \varepsilon]^3}
\nonumber \\ &=
\frac{p\cdot n \mu'^{2 \epsilon} }{(4 \pi)^{2-\epsilon}} 
\Gamma(1+\epsilon)
\int_0^\infty dx dy \frac{1}{[
-x y n \cdot p/2 + y p^2 (y - 4 l \cdot p/p^2)/4 -i \varepsilon]^{1+\epsilon}}. 
%---------------
\end{align}
%---------------
We first integrate over $x$ to obtain 
%---------------
\begin{align}
%---------------
I (l) &=
- 
\frac{ \mu'^{2 \epsilon} 2^{1+2 \epsilon}}{(4 \pi)^{2-\epsilon} (p^2)^\epsilon} 
\Gamma(\epsilon_{\rm UV})
\int_0^\infty dy 
\frac{(y-4 l \cdot p /p^2-i \varepsilon)^{-\epsilon}}{y^{1+\epsilon}}, 
%---------------
\end{align}
%---------------
where we used $\int_0^\infty dx (x+a)^{-1-\epsilon} =
a^{-\epsilon} \epsilon_{\rm UV}^{-1}$, which is valid for complex $a$. 
The integral over $y$ yields
%---------------
\begin{align}
\label{eq:iintresult}
%---------------
I (l) &=
- 
\frac{ \mu'^{2 \epsilon} }{(4 \pi)^{2-\epsilon}} 
2^{1-2 \epsilon} (p^2)^{\epsilon}
( l \cdot p)^{-2 \epsilon} 
\nonumber \\ & \quad \times
\Gamma(\epsilon_{\rm UV})
\left[ \frac{\Gamma(1-\epsilon) \Gamma(2 \epsilon_{\rm UV})}{\Gamma(1+\epsilon)}
+ \frac{e^{i \pi \epsilon} 2^{2 \epsilon} \Gamma(1/2) \Gamma(-\epsilon_{\rm
IR})}{\Gamma(1/2-\epsilon)} 
\right], 
%---------------
\end{align}
%---------------
where the first term in the square brackets comes from the region 
$y> 4 l \cdot p/p^2$, while the second term comes from 
$0 < y < 4 l \cdot p/p^2$. The factor $e^{i \pi \epsilon}$ produces imaginary
parts of this integral, but since this cancels when the complex conjugated
diagram is added, it may be discarded without affecting the poles of $I(l)$. 
By expanding the last line of Eq.~(\ref{eq:iintresult}) in powers of
$\epsilon$, we find 
%---------------
\begin{align}
\label{eq:iintresultresult}
%---------------
I (l) &=
-
\frac{ \mu'^{2 \epsilon} }{(4 \pi)^{2-\epsilon}}
2^{1-2 \epsilon} (p^2)^{\epsilon}
( l \cdot p)^{-2 \epsilon}
\left( 
\frac{1}{2 \epsilon_{\rm UV}^2} - \frac{1}{\epsilon_{\rm UV} \epsilon_{\rm IR}}
+ O(\epsilon^{-1}) 
\right).
%---------------
\end{align}
%---------------
The order-$\epsilon^{-1}$ terms do not affect the threshold double logarithms,
and can be discarded in this work. Hence, the UV-divergent contribution of the
virtual planar diagram to the soft function $S_{^3P^{[8]}}(z)$ is given by 
%---------------
\begin{align}
\label{eq:S3p8_virtual}
%---------------
S_{^3P^{[8]}}(z) |_{\textrm{virtual, UV}} 
&= - g^2 (d-2) \frac{(N_c-4) (N_c^2-1)}{N_c} 
\nonumber \\ & \quad \times
C_A \int d{\rm PS}_l 
\frac{2 \pi \delta(l^+ - P^+(1-z))I(l)}{(l \cdot p)^2} +{\rm c.c.} , 
%---------------
\end{align}
%---------------
where c.c. denote the complex conjugation of preceding terms. 
The phase-space integration is almost identical to the tree-level calculation. 
We have 
%---------------
\begin{align}
%---------------
\int d {\rm PS}_l \frac{ 2 \pi \delta (l^+ - P^+ (1-z)) (p^2)^\epsilon
}{(l\cdot p)^{2+2 \epsilon}}
&=
\frac{\mu'^{2 \epsilon}}{2 \pi^{1-\epsilon}P^+ (p^2)^{1+2 \epsilon}} 
\frac{\Gamma(1+3 \epsilon)}{\Gamma(2+2 \epsilon)}
\frac{1}{(1-z)^{1+4 \epsilon}},  
%---------------
\end{align}
%---------------
which yields
%---------------
\begin{align}
%---------------
S_{^3P^{[8]}}(z) |_{\textrm{virtual, UV}}
&= 2 g^2 (d-2) \frac{(N_c-4) (N_c^2-1) }{N_c} 
\nonumber \\ & \quad \times C_A
\frac{\mu'^{4 \epsilon}}{16 \pi^{3-2\epsilon}P^+ (p^2)^{1+2 \epsilon}}
\frac{1}{(1-z)^{1+4 \epsilon}}
\left(
\frac{1}{2 \epsilon_{\rm UV}^2} - \frac{1}{\epsilon_{\rm UV} \epsilon_{\rm IR}}
+ O(\epsilon^{-1})
\right). 
%---------------
\end{align}
%---------------

The mixed double pole $1/(\epsilon_{\rm UV} \epsilon_{\rm IR})$ in the virtual
planar diagram contribution cancels against the real diagram. To see this, 
it is instructive to recompute the integral $I(l)$ in light-cone coordinates,
which also allows us to show that the last term in the last line of
Eq.~(\ref{eq:3p8virtual}) does not produce double logarithms. We have 
%---------------
\begin{align}
%---------------
I (l) &= i \mu'^{2 \epsilon}
\int_{-\infty}^{+\infty} \frac{dk^+}{2 \pi} 
\int_{-\infty}^{+\infty} \frac{dk^-}{2 \pi} 
\int \frac{d^{d-2} k_\perp}{(2 \pi)^{d-2}} 
\frac{1}{(2 k^+ k^- - \bm{k}_\perp^2 +i \varepsilon)}
\nonumber \\ & \quad \times 
\frac{1}{ (-k^+ +i \varepsilon) (l^++l^-+k^++k^- +i \varepsilon)}. 
%---------------
\end{align}
%---------------
We first integrate over $k^-$ by using contour integration. 
For $k^+>0$, both $k^2+i \varepsilon$ and $(l+k) \cdot p+i \varepsilon$ 
denominator poles are on the same side of the real line, so that the contour 
integral vanishes. Hence we have 
%---------------
\begin{align}
\label{eq:virtual_lightconecoordinates}
%---------------
I (l) &= \frac{\mu'^{2 \epsilon}}{(2 \pi)^{d-1}} 
\int_{-\infty}^{0} dk^+ \int d^{d-2} k_\perp
\frac{1}{k^+ [ \bm{k}_\perp^2+2 k^+{}^2+2 k^+ (l^+ + l^-)-i \varepsilon] }. 
%---------------
\end{align}
%---------------
Note that the $k_\perp$ integral has a UV-divergent power count. If we first
integrate over $k_\perp$, then the $k^+$ integral involves IR and UV poles
arising from small and large $|k^+|$. 
The last term in the last line of Eq.~(\ref{eq:3p8virtual}) results in a 
similar integral, except that the $[ \bm{k}_\perp^2+2 k^+{}^2+2 k^+ (l^+ +
l^-)-i \varepsilon]$ denominator factor is now squared. In this case, the 
$k_\perp$
integral does not have a UV-divergent power count, and the integral can at most
have a single pole. Since such contributions cannot produce threshold double
logarithms, the last term in the last line of Eq.~(\ref{eq:3p8virtual}) can be
discarded in this work. 

By plugging the expression in Eq.~(\ref{eq:virtual_lightconecoordinates})
into Eq.~(\ref{eq:S3p8_virtual}) we obtain
%---------------
\begin{align}
\label{eq:S3p8_virtual_lightcone}
%---------------
S_{^3P^{[8]}} |_{\textrm{virtual, UV}} &=
-g^2 (d-2) \frac{(N_c^2-4) (N_c^2-1)}{N_c}
\nonumber \\ & \quad 
\times 
\frac{C_A \mu'^{2 \epsilon}}{(2 \pi)^{2d-3} (p_+)^2}
\int_0^\infty dl^+ \int_0^\infty dl^- \int d^{d-2} l_\perp
\int_{-\infty}^0 dk^+ \int d^{d-2} k_\perp
\nonumber\\ & \quad \times
\frac{\delta(l_+-P^+(1-z)) \delta(l^2)}
{k_+ (l_+ + l_-)^2 
[\bm{k}_\perp^2 + 2 k_+^2 + 2 k_+ (l_++l_-) -i \varepsilon]}
+{\rm c.c.}.
%---------------
\end{align}
%---------------
Now we consider the corresponding real diagram. 
The matrix element of the operator ${\cal W}^{yx}_\beta (^3P^{[8]})$
between the vacuum and a two-gluon state from the gluon attachment on the
timelike Wilson line is given by 
%---------------
\begin{align}
\label{eq:3p8real_a}
%---------------
& \hspace{-5ex} \langle g^{a_1,\mu_1}(k_1) g^{a_2,\mu_2}(k_2) 
| {\cal W}^{yx}_\beta (^3P^{[8]}) | 0
\rangle|_{\textrm{real (a)}}
\nonumber \\ 
&=
-ig
\left[
\frac{i}{p \cdot k_1+i \varepsilon} 
\left(\frac{i}{p \cdot (k_1+k_2)+i \varepsilon} \right)^2
+
\left(
\frac{i}{p \cdot k_1+i \varepsilon} \right)^2
\frac{i}{p \cdot (k_1+k_2)+i \varepsilon} 
\right]
\nonumber \\ & \quad \times
p^\mu  
(k_1{}_\mu \delta^{\mu_1}_\beta - k_1{}_\beta \delta^{\mu_1}_\mu ) 
p^{\mu_2} 
d^{a_1 yd} f^{a_2 dx}. 
%---------------
\end{align}
%---------------
The one from the gluon attachment on the lightlike Wilson line is 
%---------------
\begin{align}
\label{eq:3p8real_b}
%---------------
& \hspace{-5ex} \langle g^{a_1,\mu_1}(k_1) g^{a_2,\mu_2}(k_2)
| {\cal W}^{yx}_\beta (^3P^{[8]}) | 0
\rangle|_{\textrm{real (b)}}
\nonumber \\ &=
-i g 
\frac{i}{n \cdot k_2 + i\varepsilon}
\left( \frac{i}{p \cdot k_1 +i \varepsilon} \right)^2
p^\mu 
(k_1{}_\mu \delta^{\mu_1}_\beta - k_1{}_\beta \delta^{\mu_1}_\mu )
n^{\mu_2} 
d^{a_1 dy}
f^{a_2 xd}. 
%---------------
\end{align}
%---------------
Similarly to the virtual diagram, only the first term in the square brackets in
Eq.~(\ref{eq:3p8real_a}) will give rise to a UV-divergent power count, so that
the remaining term can be discarded. 
The UV-divergent contribution to $S_{^3P^{[8]}} (z)$ from the planar real
diagram is then given by 
%---------------
\begin{align}
\label{eq:S3p8_real_lightcone}
%---------------
S_{^3P^{[8]}} (z) |_{\textrm{real, UV}} &=
- \int d {\rm PS}_{k_1} \int d {\rm PS}_{k_2} 
2 \pi \delta (k_1^++k_2^+-P^+ (1-z))
\nonumber \\ & \quad \times 
g^2 (d-2) \frac{(N_c^2-1) (N_c^2-4)}{N_c} 
\frac{C_A p^+}
{ k_2^+ (k_1 \cdot p)^2 [ (k_1+k_2) \cdot p]}
+{\rm c.c.}
\nonumber \\
&= - 
g^2 (d-2) \frac{(N_c^2-1) (N_c^2-4)}{N_c} 
\nonumber \\ & \quad \times \frac{C_A\mu'^{2 \epsilon}}{(2 \pi)^{2 d-3} (p^+)^2}
\int_0^\infty dl^+ \int_0^\infty dl^- \int d^{d-2} l_\perp
\int_0^\infty dk^+ \int_0^\infty dk^- \int d^{d-2} k_\perp
\nonumber \\ & \quad \times 
\frac{\delta(l^++k^+-P^+(1-z)) \delta(l^2) \delta(k^2)}
{k^+ (l^++l^-)^2 (l^+ + l^- + k^+ + k^-)}
+{\rm c.c.}
\nonumber \\ 
&= - g^2 (d-2) \frac{(N_c^2-1) (N_c^2-4)}{N_c}
\nonumber \\ & \quad \times \frac{C_A\mu'^{2 \epsilon}}{(2 \pi)^{2 d-3} (p^+)^2}
\int_0^\infty dl^+ \int_0^\infty dl^- \int d^{d-2} l_\perp
\int_0^\infty dk^+ \int d^{d-2} k_\perp
\nonumber \\ & \quad \times
\frac{\delta(l^++k^+-P^+(1-z)) \delta(l^2)}
{k^+ (l^++l^-)^2 [ \bm{k}_\perp^2 + 2 k^+{}^2+2 k^+ (l^++l^-)]}
+{\rm c.c.}, 
%---------------
\end{align}
%---------------
where in the second equality we relabeled momenta $k_1$ and $k_2$ as $l$ and
$k$ to allow direct comparison with the expression in 
Eq.~(\ref{eq:S3p8_virtual_lightcone}), and in the last equality we integrated
over $k^-$ by using $\delta(k^2)$. 
It is now easy to see that the integrands of 
Eqs.~(\ref{eq:S3p8_virtual_lightcone}) and (\ref{eq:S3p8_real_lightcone})
coalesce at $k^+ = 0$, so that the IR divergence at $k^+ \to 0$ cancels in the
sum of virtual and real planar diagrams.
While the $k_\perp$ integral in Eq.~(\ref{eq:S3p8_real_lightcone}) is UV
divergent, there is no source of additional UV poles 
because the $k^+$ and $l^+$ are bounded, unlike the virtual diagram. 
Therefore, in the sum of real and virtual planar diagrams, only the double UV
pole in the virtual diagram produces the threshold double logarithm. 
We obtain 
%---------------
\begin{align}
%---------------
S_{^3P^{[8]}}(z) |_{\textrm{NLO planar} }
&= g^2 (d-2) \frac{(N_c^2-4) (N_c^2-1) }{N_c}
\nonumber \\ & \quad \times C_A
\frac{\mu'^{4 \epsilon}}
{16 \pi^{3-2\epsilon}P^+ (p^2)^{1+2 \epsilon}}
\frac{1}{(1-z)^{1+4 \epsilon}}
\left( \frac{1}{\epsilon_{\rm UV}^2} + O(\epsilon^{-1}) \right).
%---------------
\end{align}
%---------------

%==============================================================================
\subsubsection{UV power counting}
%==============================================================================

Now we consider the remaining NLO diagrams that contribute to
$S_{^3P^{[8]}}(z)$. As we have discussed previously, threshold double
logarithms can only arise in $S_{^3P^{[8]}}(z)$ at NLO from double UV poles
multiplying $(1-z)^{-1-n \epsilon}$. We will show in this section that none
of the remaining diagrams can produce double UV poles, so that they can be
discarded if we only look for threshold double logarithms. 

Note that all doubly virtual diagrams vanish because they are proportional to a
vanishing color factor $d^{bcd} f^{bda} = 0$, where $d^{bcd}$ comes from the
operator definition of the ${\cal W}^{yx}_\beta (^3P^{[8]})$ and $f^{bda}$
comes from the attachment of a gluon onto a Wilson line on the same side of the
cut. Some virtual-real diagrams also vanish if a gluon from the field-strength 
tensor attaches to a Wilson line on the same side of the cut without
additional vertices on the Wilson line between the gluon attachment and the
field-strength tensor, because they are also proportional to the same vanishing
color factor. This leaves us with only a handful of virtual-real diagrams and
doubly real diagrams. 

The virtual-real diagrams involve a one-loop integral, followed by a
phase-space integral over a lightlike momentum $\ell$. The phase-space integral 
cannot produce double UV poles, because the $\ell^+$ and $\ell^-$ integrals
have finite bounds, and so only the $\ell_\perp$ integral can produce UV poles. 
In practice, virtual-real diagrams can produce double UV poles only when 
the virtual one-loop integral contains a double UV pole. This is because, due
to dimensional considerations, the phase-space integral cannot have a
UV-divergent power count unless the gluon with momentum $\ell$ from the
field-strength tensor attaches to the lightlike Wilson line on the other side 
of the cut, so that the phase-space integrand carries factors of $\ell \cdot n$
in the denominator; in such case, the virtual one-loop integral is UV finite
because its integrand carries too many propagator denominator factors. 

We have checked that, with the exception of the planar virtual diagram we
computed in the previous section, all other UV-divergent virtual-real diagrams
contain only single UV poles. This is clear for self-energy and vertex
correction diagrams, as we know from renormalization of the QCD Lagrangian that
they only involve single UV poles at one loop. The only diagram that needs
attention is the one involving the triple-gluon vertex, where the additional
gluon is attached to the timelike Wilson line. This diagram involves the
one-loop integral given by 
%---------------
\begin{align}
%---------------
\int_k \left[ \frac{1}{\ell \cdot p (k \cdot p)^2} + \frac{1}{(\ell \cdot p)^2 k
\cdot p} \right] \frac{n(k)}{k^2 (k-\ell)^2},
%---------------
\end{align}
%---------------
where the numerator $n(k)$ is proportional to 
%---------------
\begin{align}
%---------------
& 
(p \cdot \ell g^{\mu \beta} - \ell^\beta p^\mu) 
(p \cdot k \delta^{\nu}_{\beta} - k_\beta p^\nu) p^\sigma 
\left[ g_{\mu \nu} (-k-\ell)_\sigma + g_{\sigma \mu} (2 \ell-k)_\nu
+ g_{\nu \sigma} (2 k-\ell)_\mu \right] 
\nonumber \\
&= (2-d) [(\ell \cdot p)^2 k \cdot p + \ell \cdot p (k \cdot p)^2] 
- \frac{m^2}{2} (k-\ell)^2 ( \ell \cdot p + k \cdot p )
+ \frac{m^2}{2} k^2 ( k \cdot p -3 \ell \cdot p ). 
%---------------
\end{align}
%---------------
It is easy to evaluate the resulting integrals with standard methods, for
example by using Feynman parametrization, to confirm that
none contain double UV poles, so this diagram can be discarded as well.

%%%%%%%%%%%%%%%%%%%%%%%%%%%%%%%%%%%%%%%%%%%%%%%%%%%%%%%%%%%%%%%%%%%%%%%%%%%%%
\begin{figure}[t]
\includegraphics[width=.99\textwidth]{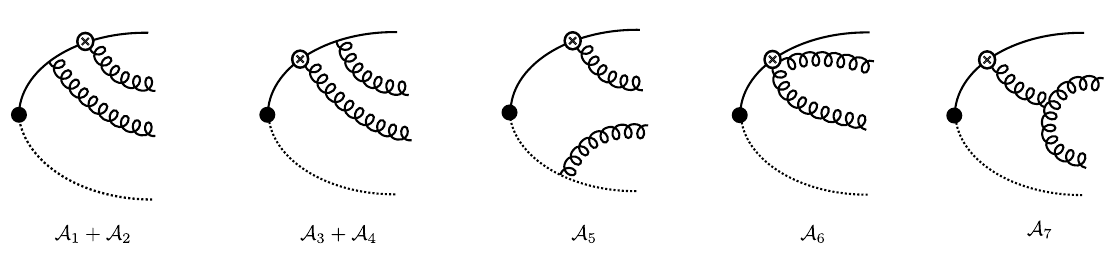}%
\caption{\label{fig:doublereal}
Feynman diagrams for amplitudes that give rise to the doubly real contributions
in the soft function $S_{^3P^{[8]}}(z)$ at NLO. 
See Eq.~(\ref{eq:Amps}) and text for definitions of ${\cal A}_1$--${\cal A}_7$.
}
\end{figure}
%%%%%%%%%%%%%%%%%%%%%%%%%%%%%%%%%%%%%%%%%%%%%%%%%%%%%%%%%%%%%%%%%%%%%%%%%%%%%

Now we are left with doubly real diagrams. 
All diagrams that contribute to the matrix element of the operator 
${\cal W}^{yx}_\beta (^3P^{[8]})$
between the vacuum and a two-gluon state from the gluon attachment on the
timelike Wilson line $\langle g^{a_1,\mu_1}(k_1) g^{a_2,\mu_2}(k_2)
| {\cal W}^{yx}_\beta (^3P^{[8]}) | 0\rangle$ are shown in 
Fig.~\ref{fig:doublereal}. 
Their values are 
%------------
\begin{subequations}
\label{eq:Amps}
\begin{align}
%------------
\label{eq:A12}
{\cal A}_1+{\cal A}_2
&=
-ig
\left[
\frac{i}{p \cdot k_1+i \varepsilon} 
\left(\frac{i}{p \cdot (k_1+k_2)+i \varepsilon} \right)^2
+
\left(
\frac{i}{p \cdot k_1+i \varepsilon} \right)^2
\frac{i}{p \cdot (k_1+k_2)+i \varepsilon} 
\right]
\nonumber \\ & \quad \times
p^\mu  
(k_1{}_\mu \delta^{\mu_1}_\beta - k_1{}_\beta \delta^{\mu_1}_\mu ) 
p^{\mu_2} 
\delta^{yc} 
\delta^{a_1 b}
d^{b cd} f^{a_2 dx}, \\
\label{eq:A34}
{\cal A}_3+{\cal A}_4
&= 
-ig
\left[
\frac{i}{p \cdot k_2+i \varepsilon} 
\left(\frac{i}{p \cdot (k_1+k_2)+i \varepsilon} \right)^2
+
\left(
\frac{i}{p \cdot k_2+i \varepsilon} \right)^2
\frac{i}{p \cdot (k_1+k_2)+i \varepsilon} 
\right]
\nonumber \\ & \quad \times
p^\mu  
(k_1{}_\mu \delta^{\mu_1}_\beta - k_1{}_\beta \delta^{\mu_1}_\mu ) 
p^{\mu_2} 
f^{a_2 yc}
\delta^{b a_1}
d^{b cd} 
\delta^{da} \delta^{xa},  \\
{\cal A}_5 &= 
-i g
\frac{i}{n \cdot k_2 + i\varepsilon}
\left( \frac{i}{p \cdot k_1 +i \varepsilon} \right)^2
p^\mu
(k_1{}_\mu \delta^{\mu_1}_\beta - k_1{}_\beta \delta^{\mu_1}_\mu )
n^{\mu_2}
\delta^{yc}
\delta^{a_1 b} 
d^{b cd}
f^{a_2 xd},\\
{\cal A}_6
&=
\frac{g}{[p \cdot (k_1+k_2)]^2} 
(p^{\mu_1} \delta^{\mu_2}_\beta-p^{\mu_2} \delta^{\mu_1}_\beta)
d^{bcd} f^{b a_1 a_2} \delta^{da} \delta^{xa} \delta^{yc} , \\
{\cal A}_7
&=
\frac{g}{[p \cdot (k_1+k_2)]^2} \frac{1}{(k_1+k_2)^2} p^\mu 
[(k_1+k_2)^{\mu} \delta^\nu_\beta- (k_1+k_2)_\beta g^{\mu \nu}] 
\nonumber \\ & \quad \times 
G_3^{\nu \mu_1 \mu_2} (k_1+k_2,-k_1,-k_2) d^{bcd} f^{ba_1a_2} \delta^{da}
\delta^{xa} \delta^{yc}, 
%---------------
\end{align}
\end{subequations}
%---------------
with $G_3^{\mu_1 \mu_2 \mu_3}(k_1,k_2,k_3)=g^{\mu_1 \mu_2} (k_1-k_2)_{\mu_3} + 
g^{\mu_2 \mu_3} (k_2-k_3)_{\mu_1} + g^{\mu_3 \mu_1} (k_3-k_1)_{\mu_2}$.
In Eq.~(\ref{eq:A12}), ${\cal A}_1$ and ${\cal A}_2$ are given by the
contributions from the first and second terms in the square brackets,
respectively; similarly, in Eq.~(\ref{eq:A34}), ${\cal A}_3$ and ${\cal A}_4$ 
are given by the contributions from the first and second terms in the square
brackets, respectively.
The amplitude ${\cal A}_6$ comes from the $A^\mu A^\nu$ term of the
field-strength tensor, while the ${\cal A}_7$ arises from the triple-gluon
vertex. 
In ${\cal A}_1$ through ${\cal A}_5$, the gluon attached to the field-strength
tensor carries the momentum $k_1$. 
If we define the crossed amplitudes as 
$\bar{\cal A}_i \equiv {\cal A}_i|_{1 \leftrightarrow 2}$,
the contributions to $S_{^3P^{[8]}}(z)$ from the doubly real diagrams are given
by ${\cal A}_i^* {\cal A}_j$ and 
${\cal A}_i^* \bar{\cal A}_j$ for $i$, $j=1$, 2, $\cdots,$ 7, 
multiplied by $2 \pi \delta(k_1^+ + k_2^+ - P^+ (1-z))$, and integrated over
the phase space of $k_1$ and $k_2$. 
We now examine the UV divergences in each contribution.

The calculation of the planar diagrams in the previous section leads to a
simple way to analyze UV divergences in these phase-space integrals. 
If we evaluate a phase-space integral over $k$ in light-cone coordinates, the
only source of UV divergences is the integral over $k_\perp$, because the $k^-$
integral is done by the on-shell delta function $\delta(k^2)$, and the $k^+$ 
integral has a finite bound. Hence, for doubly real diagrams where the final 
state involves two gluons with momenta $k_1$ and $k_2$, it suffices to consider
only the power counting of the $k_{1\perp}$ and $k_{2\perp}$ integrals, after
integrating out the $\delta(k_1^2)$ and $\delta(k_2^2)$ with the $k_1^-$ and
$k_2^-$ integrals. 
In most cases, general form of the $k_{1\perp}$ and $k_{2\perp}$ integrals at
large $|\bm{k}_{1 \perp}|$ and $|\bm{k}_{2 \perp}|$ can be written as 
%------------
\begin{align}
\label{eq:doublereal_UVpowercounting}
%------------
\int 
\frac{ d^{d-2} k_{1\perp} d^{d-2} k_{2\perp} }{
\bm{k}_{1\perp}^{2a} \bm{k}_{2\perp}^{2b} 
(\bm{k}_{1\perp}^2 + \bm{k}_{2\perp}^2)^c }, 
%------------
\end{align}
%------------
after suitable rescaling of $k_{1\perp}$ and $k_{2\perp}$. 
There are exceptional cases when a factor of $k_1 \cdot k_2$ appears in the
numerator or in the denominator; we will later examine these cases explicitly. 
We can easily identify the UV behaviors of the $k_{1\perp}$ and $k_{2\perp}$
integrals in Eq.~(\ref{eq:doublereal_UVpowercounting}) by counting the degrees
of divergence. If an integral over $k_\perp$ behaves like 
$\int \frac{d^{d-2} k_\perp}{\bm{k}_\perp^{2n}}$ at large $|\bm{k}_\perp|$,
then the degree of divergence of this integral is $d-2-2n$. 
If this number vanishes or exceeds 0 at $d=4$, then the $k_\perp$ integral can
be UV divergent. 
Let us now apply this analysis to integrals of the form
Eq.~(\ref{eq:doublereal_UVpowercounting}). 
The case $c=0$ is trivial, in which case the degrees of divergence of the
$k_{1\perp}$ and $k_{2\perp}$ integrals are simply given by $d-2-2 a$ and
$d-2-2 b$, respectively.
In the general case $c\neq 0$, if we integrate over $k_{2 \perp}$ first, the
degree of divergence of the $k_{2 \perp}$ integral is $d-2-2 (b+c)$. Since the
only scale of the $k_{2 \perp}$ integral is $|\bm{k}_{1 \perp}|$, 
the remaining $k_{1\perp}$ integral is proportional to $\int d^{d-2} k_{1
\perp} |\bm{k}_{1\perp}|^{d-2-2 (a+b+c)}$. The degree of divergence of this
integral is $2 (d-2-a-b-c)$. 

Now let us examine the case where a factor of $k_1 \cdot k_2$ appears in the
numerator. From
%---------------
\begin{align}
%---------------
k_1 \cdot k_2 = k_1^+ k_2^- + k_1^- k_2^+ - \bm{k}_{1 \perp} \cdot \bm{k}_{2
\perp}, 
%---------------
\end{align}
%---------------
after integrating out the on-shell delta functions to eliminate $k_1^-$ and
$k_2^-$, the first and second terms become proportional to $\bm{k}_{2\perp}^2$
and $\bm{k}_{1\perp}^2$, respectively. The last term is odd in both 
$k_{1\perp}$ and $k_{2\perp}$, while the remaining integrand factors are even,
so that it can be discarded. Hence, integrals with a factor of $k_1 \cdot k_2$
in the numerator can always be reduced to the form in
Eq.~(\ref{eq:doublereal_UVpowercounting}). 

The remaining case is when the factor $k_1 \cdot k_2$ appears in the
denominator. These integrals appear in the form 
%---------------
\begin{align}
\label{eq:k1k2integrals}
%---------------
\int d{\rm PS}_{k_1} \int d{\rm PS}_{k_2} 
\frac{2 \pi \delta(k_1^+ + k_2^+ - P^+ (1-z))}
{(k_1 \cdot p)^a (k_2 \cdot p)^b [(k_1 + k_2) \cdot p]^c k_1 \cdot k_2}. 
%---------------
\end{align}
%---------------
After integration over the on-shell delta functions by eliminating $k_1^-$ and
$k_2^-$, the factors $k_1 \cdot p$, $k_2 \cdot p$, and $(k_1 + k_2) \cdot p$
involve $\bm{k}_{1\perp}^2$, $\bm{k}_{2\perp}^2$, and
$\bm{k}_{1\perp}^2+\bm{k}_{2\perp}^2$, respectively, while 
$k_1 \cdot k_2$ involves a linear combination of $\bm{k}_{1\perp}^2$,
$\bm{k}_{2\perp}^2$, and $\bm{k}_{1 \perp} \cdot \bm{k}_{2 \perp}$.
Explicit calculation shows that we never encounter cases where both $a$ and $b$
are positive. If $b \leq 0$ and $a \geq 0$, we can combine the 
$[(k_1 + k_2) \cdot p]^c$ and $k_1 \cdot k_2$ denominator factors by using
Feynman parametrization and shift $k_{2\perp}$ to eliminate the 
$\bm{k}_{1 \perp} \cdot \bm{k}_{2 \perp}$ term in the denominator. 
Hence, this integral reduces to the form given in
Eq.~(\ref{eq:doublereal_UVpowercounting}), and the UV power counting can be
done in the same way. The same applies to the case $a \leq 0 $ and $b \geq 0$. 
The case where both $a$ and $b$ are negative can be rewritten as a linear
combination of integrals with either $a=0$ or $b=0$ by using $2 (k_1 \cdot p)
(k_2 \cdot p) = [(k_1 + k_2) \cdot p]^2 - (k_1 \cdot p)^2-(k_2 \cdot p)^2$. 
As a result, the UV power counting of all integrals of the form in
Eq.~(\ref{eq:k1k2integrals}) can be done in a similar way. 

By applying the UV power counting analysis of all doubly real diagrams, we find
that the only diagrams that have a UV-divergent power count are:
${\cal A}_2^* {\cal A}_5$ which is the 
real planar diagram we computed in the previous section, 
${\cal A}_{5}^* {\cal A}_{7}$, 
and 
${\cal A}_{5}^* \bar{\cal A}_{5}$.
Note that the contribution from ${\cal A}_2^* \bar {\cal A}_4$
is proportional to a vanishing color factor, similarly to its
virtual-real counterpart. 
Note that none of these diagrams involve double UV divergences, as only one of
the $k_{1\perp}$ and $k_{2\perp}$ integrals are UV divergent. 
In fact, it can be shown by direct evaluation that the contributions from 
${\cal A}_{5}^* \bar{\cal A}_{5}$ and 
${\cal A}_{5}^* {\cal A}_{7}$ are actually UV finite. 
The diagram ${\cal A}_{5}^* \bar{\cal A}_{5}$ 
is proportional to 
%---------------
\begin{align}
%---------------
& \hspace{-5ex} 
\int d{\rm PS}_{k_1} \int d{\rm PS}_{k_2} 
\frac{k_1^+ k_2^0 + k_2^+ k_1^0 - k_1 \cdot k_2/\sqrt{2}}
{ ( k_1^0)^2 (k_2^0)^2 k_1^+ k_2^+} 
2 \pi \delta(k_1^+ + k_2^+ - P^+ (1-z))
\nonumber \\ &= 
\int d{\rm PS}_{k_1} \int d{\rm PS}_{k_2} 
\frac{k_1^+ k_2^0 + k_2^+ k_1^0 - (k_1^+ k_2^- + k_1^- k_2^+)/\sqrt{2}}
{ ( k_1^0)^2 (k_2^0)^2 k_1^+ k_2^+} 
2 \pi \delta(k_1^+ + k_2^+ - P^+ (1-z))
\nonumber \\ &= 
\sqrt{2} \int d{\rm PS}_{k_1} \int d{\rm PS}_{k_2} 
\frac{ 2 \pi \delta(k_1^+ + k_2^+ - P^+ (1-z)) }
{ ( k_1^0)^2 (k_2^0)^2}, 
%---------------
\end{align}
%---------------
where in the first equality we eliminated the $\bm{k}_{1\perp} \cdot
\bm{k}_{2\perp}$ term because it is odd in both $k_{1\perp}$ and $k_{2\perp}$, 
and in the second equality we used $k_{1,2}^0 = (k_{1,2}^+ +
k_{1,2}^-)/\sqrt{2}$. The resulting integral is now UV finite in both
$\bm{k}_{1\perp}$ and $\bm{k}_{2\perp}$. 

The diagram ${\cal A}_{5}^* {\cal A}_{7}$ needs more work. 
If we only collect the terms that provide a UV-divergent power count, we have
%---------------
\begin{align}
%---------------
\int d{\rm PS}_{k_1} \int d{\rm PS}_{k_2} 
\frac{k_2^0 /\sqrt{2}- k_1^+ (k_2^0)^2/k_1\cdot k_2
}{(k_1^0)^2 k_2^+ (k_1^0+k_2^0)^2} 
2 \pi \delta(k_1^+ + k_2^+ - P^+ (1-z)). 
%---------------
\end{align}
%---------------
If we first integrate over $k_1^-$ and $k_2^-$ by
using the on-shell delta functions, and then integrate over $k_{2\perp}$
followed by integration over $k_{1\perp}$, only the $k_{2\perp}$ integral
involves UV divergences. 
This means that, only the highest powers of $k_{2\perp}$ in the numerator 
$k_2^0$, which comes from the $k_2^-$ component, produces UV divergences. 
Hence, we can replace $k_2^0$ with $k_2^-/\sqrt{2}$ in the numerator without
affecting the UV poles. We have then 
%---------------
\begin{align}
%---------------
\frac{1}{2}
\int d{\rm PS}_{k_1} \int d{\rm PS}_{k_2}
\frac{k_2^- - k_1^+ (k_2^-)^2/k_1\cdot k_2
}{(k_1^0)^2 k_2^+ (k_1^0+k_2^0)^2}
2 \pi \delta(k_1^+ + k_2^+ - P^+ (1-z)).
%---------------
\end{align}
%---------------
We can rewrite the term involving $k_1 \cdot k_2$ in the denominator as 
%---------------
\begin{align}
%---------------
& \hspace{-5ex}
\int d{\rm PS}_{k_1} \int d{\rm PS}_{k_2}
\frac{k_1^+ (k_2^-)^2
}{(k_1^0)^2 k_2^+ (k_1^0+k_2^0)^2 k_1 \cdot k_2}
2 \pi \delta(k_1^+ + k_2^+ - P^+ (1-z))
\nonumber \\ &=
\int d{\rm PS}_{k_1} \int d{\rm PS}_{k_2}
\frac{k_2^- ( k_1 \cdot k_2 - k_1^- k_2^+ + \bm{k}_{1\perp} \cdot 
\bm{k}_{2\perp})
}{(k_1^0)^2 k_2^+ (k_1^0+k_2^0)^2 k_1 \cdot k_2}
2 \pi \delta(k_1^+ + k_2^+ - P^+ (1-z)).
%---------------
\end{align}
%---------------
The $k_1^- k_2^+$ term does not have a UV-divergent power count. 
In the case of the $ \bm{k}_{1\perp} \cdot \bm{k}_{2\perp}$ term, if we combine
the denominator factors $(k_1^0+k_2^0)^2$ and $k_1 \cdot k_2$ by using Feynman
parametrization and integrate over $\bm{k}_{2\perp}$, the shift in
$\bm{k}_{2\perp}$ that is necessary in eliminating the term linear in
$\bm{k}_{2\perp}$ in the denominator will produce up to a $\bm{k}_{1\perp}^2$ 
term in the numerator, while $\bm{k}_{2\perp}$ disappears in the numerator. 
Similarly to the $k_1^- k_2^+$ term, this term does not have a
UV-divergent power count. Hence, the only UV-divergent contribution comes from
the $k_1 \cdot k_2$ term. Then the sum of UV-divergent contribution to this
diagram can be written as 
%---------------
\begin{align}
%---------------
\frac{1}{2} \int d{\rm PS}_{k_1} \int d{\rm PS}_{k_2}
\frac{k_2^- - k_2^-}{(k_1^0)^2 k_2^+ (k_1^0+k_2^0)^2}
2 \pi \delta(k_1^+ + k_2^+ - P^+ (1-z)),
%---------------
\end{align}
%---------------
which vanishes. Hence, we conclude that the diagram 
${\cal A}_{5}^* {\cal A}_{7}$ is UV finite. 
Therefore, the only doubly real diagram that contains a UV divergence is 
${\cal A}_2^* {\cal A}_5$, which we considered in the previous section. 

The analysis of the NLO diagrams in this section shows that the threshold
double logarithms, which arise from double UV poles, can only arise from the
planar diagrams that we considered in the previous section. While the double UV
pole only appears in the virtual-real planar diagram, the cancellation of the
mixed double pole requires the corresponding doubly real diagram. 
All other diagrams can produce at most single UV poles, 

The analysis of the doubly real diagrams also serve as a double check of the
virtual-real diagrams: if a doubly real diagram contains a UV pole, the
corresponding virtual-real diagram can contain a double UV pole because the
virtual loop integral can contain additional UV divergences that arise from
the unbound $k^+$ integral. The nonappearance of UV divergences in doubly real
diagrams other than the planar diagram we considered in the previous section is
consistent with our analysis that the only double UV pole arises from the
planar virtual-real diagram.

%==============================================================================
\subsubsection{Results and discussion}
%==============================================================================

The $^3P^{[8]}$ soft function at NLO is given by 
%---------------
\begin{align}
\label{eq:s3p8_nloresult1}
%---------------
S_{^3P^{[8]}} (z)
&=
-(1-\epsilon)
\frac{(N_c^2-4) (N_c^2-1)}{N_c}
\frac{1}{\pi m^2 P^+} 
\nonumber \\ &\quad \times
\bigg[
\frac{(\pi \mu'^2/m^2)^\epsilon
\Gamma(1+\epsilon)}{(1-z)^{1+2 \epsilon}}
- \frac{\alpha_s C_A (\pi \mu'^2/m^2)^{2\epsilon}}{2 \pi} 
\frac{\epsilon_{\rm UV}^{-2} + O(\epsilon^{-1})}{(1-z)^{1+4 \epsilon}}
+O(\alpha_s^2) \bigg],
%---------------
\end{align}
%---------------
where we neglect contributions at order $\alpha_s$ that do not produce
threshold double logarithms. 
At order $\alpha_s$, the expansion in powers of $\epsilon$ produces many terms;
since we are only interested in identifying the threshold double logarithms, 
we can discard contributions that are unrelated to leading singularities at 
$z \to 1$. 
The factor $(1-\epsilon) (\pi \mu'^2/m^2)^{2\epsilon} \epsilon_{\rm UV}^{-2}$
gives\footnote{If we trace back the origin of the $\epsilon_{\rm UV}^{-2}$
contribution and keep all the $\epsilon$-dependent gamma functions, 
which amounts to the replacement 
$\epsilon_{\rm UV}^{-2} \to \Gamma(1+3 \epsilon) \Gamma(1-\epsilon)
\Gamma(\epsilon_{\rm UV}) \Gamma(2 \epsilon_{\rm UV})/[ \Gamma(1+\epsilon)
\Gamma(2+2 \epsilon)]$, the $\gamma_{\rm E}$ does not appear in the expansion
in powers of $\epsilon$, 
as is usual in the calculation in the $\overline{\rm MS}$ scheme.}
%---------------
\begin{align}
\label{eq:doublepoleexpansion}
%---------------
(1-\epsilon) (\pi \mu'^2/m^2)^{2\epsilon} \epsilon_{\rm UV}^{-2} 
&= \frac{1}{\epsilon_{\rm UV}^2} 
+ \frac{2 \log \frac{\mu^2}{4 m^2} +2 \gamma_{\rm E}-1}{\epsilon_{\rm UV}} 
+ O(\epsilon^0). 
%---------------
\end{align}
%---------------
while the factor $(1-z)^{-1-4 \epsilon}$ yields
%---------------
\begin{align}
\label{eq:1minuszexpansion}
%---------------
\frac{1}{(1-z)^{1+4 \epsilon}} = - \frac{\delta(1-z)}{4 \epsilon_{\rm IR}} 
+ \frac{1}{(1-z)_+} - 4 \epsilon \left[ \frac{\log(1-z)}{1-z} \right]_+
+ 8 \epsilon^2 \left[ \frac{\log^2 (1-z)}{1-z} \right]_+
+ O(\epsilon^3) .
%---------------
\end{align}
%---------------
We can see that the $\epsilon_{\rm UV}^{-2}$ term in
Eq.~(\ref{eq:doublepoleexpansion}) and the order-$\epsilon^2$ term in
Eq.~(\ref{eq:1minuszexpansion}) produces the double logarithmic term 
$\left[ \frac{\log^2 (1-z)}{1-z} \right]_+$. 
This threshold double logarithm is associated with the singular contribution in
the anomalous dimension of $S_{^3P^{[8]}} (z)$
that is proportional to $\left[ \frac{\log (1-z)}{1-z} \right]_+$, 
which comes from the 
$\epsilon_{\rm UV}^{-2}$ term in Eq.~(\ref{eq:doublepoleexpansion})
and the order-$\epsilon$ term in Eq.~(\ref{eq:1minuszexpansion}). 
Equivalently, the singular contribution in the anomalous dimension can also be
obtained from the $\epsilon_{\rm UV}^{-1} \log \mu^2$ term in
Eq.~(\ref{eq:doublepoleexpansion})
and the order-$\epsilon$ term in Eq.~(\ref{eq:1minuszexpansion}). 
The IR pole in Eq.~(\ref{eq:1minuszexpansion}) produces mixed poles like 
$1/(\epsilon_{\rm IR} \epsilon_{\rm UV}^2)$, which will cancel when we consider
the mixing between the $^3S_1^{[8]}$ and $^3P_J^{[8]}$ states, similarly to
the tree-level result. Since the most singular term in the $^3S_1^{[8]}$ soft
function at NLO is proportional to $\left[ \frac{\log (1-z)}{1-z} \right]_+$,
this does not affect the double logarithmic term in the $^3P^{[8]}$ soft
function. If we collect only the most singular terms at each order in
$\alpha_s$, $S_{^3P^{[8]}} (z)$ is given at NLO accuracy by 
%---------------
\begin{align}
\label{eq:s3p8_nloresult2}
%---------------
S_{^3P^{[8]}} (z) |_{\rm threshold}
&=
\frac{(N_c^2-4) (N_c^2-1)}{N_c}
\frac{1}{\pi m^2 P^+}
\bigg\{
- \frac{1}{(1-z)_+} 
\nonumber \\ & \quad 
+ \frac{\alpha_s C_A}{\pi}
\bigg[ 
4 \left( \frac{\log^2 (1-z)}{1-z} \right)_+
\nonumber \\ & \quad 
+ \frac{1}{\epsilon_{\rm UV}} \left( \frac{-2 \log (1-z)}{1-z} \right)_+
+ \frac{1}{2 \epsilon_{\rm UV}^2} \frac{1}{(1-z)_+} 
\bigg]
+O(\alpha_s^2) \bigg\},
%---------------
\end{align}
%---------------
whose Mellin transform is 
%---------------
\begin{align}
\label{eq:s3p8_nloresult3}
%---------------
\tilde{S}_{^3P^{[8]}} (N) |_{\rm threshold} 
&=
\frac{(N_c^2-4) (N_c^2-1)}{N_c}
\frac{1}{\pi m^2 P^+}
\bigg[
- \log N 
\nonumber \\ & \quad
+ \frac{\alpha_s C_A}{\pi}
\left(
\frac{4}{3} \log^3 N 
- \frac{\log^2 N}{\epsilon_{\rm UV}} 
+ \frac{\log N}{2\epsilon_{\rm UV}^2} 
\right)
+O(\alpha_s^2) \bigg]
\nonumber \\
& = 
\frac{(N_c^2-4) (N_c^2-1)}{N_c}
\frac{-\log N}{\pi m^2 P^+}
\nonumber \\ & \quad \times 
\bigg[
1
+ \frac{\alpha_s C_A}{\pi}
\left(
- \frac{4}{3} \log^2 N 
+ \frac{\log N}{\epsilon_{\rm UV}} 
- \frac{1}{2 \epsilon_{\rm UV}^2} 
\right)
+O(\alpha_s^2) \bigg], 
%---------------
\end{align}
%---------------
at large $N$. 
If we compare with the corresponding result for $\tilde{S}_{^3S_1^{[8]}} (N)$
in Eq.~(\ref{eq:S3S8NLOMellin}), we see that while the $^3P^{[8]}$ soft
function has the same anomalous dimension at large $N$ as the $^3S_1^{[8]}$
soft function, that is, the cusp anomalous dimension in the adjoint
representation. However, the coefficient of the double logarithmic term at
order $\epsilon^0$ is different from the $^3S_1^{[8]}$ case; 
this happens because the $^3P^{[8]}$ soft function already has a $\log N$
behavior at leading nonvanishing order in $\alpha_s$.

%==============================================================================
\subsection{\boldmath $^3P_J^{[1]}$}
%==============================================================================

The calculation of the soft functions for the $^3P_J^{[1]}$ channel is very 
similar to the $^3P^{[8]}$ case. The isotropic soft function
$S_{^3P^{[1]}}(z)$ involves almost the same diagrams as the $^3P^{[8]}$ soft
function, and differs only by color factors. Similarly to the
leading-order
case, explicit calculation shows that the anisotropic soft function 
$S_{^3P^{[1]}}^{TT}(z)$ at NLO also carries an additional factor of $\epsilon$,
so that anisotropic contribution cannot produce threshold double logarithms. 

%==============================================================================
\subsubsection{Isotropic contribution}
%==============================================================================

Similarly to the $^3P^{[8]}$ case, we first consider the planar diagrams. 
The virtual correction to the matrix element of the operator ${\cal W}_\beta^b
(^3P^{[1]})$ between the vacuum and the one-gluon state yields 
%---------------
\begin{align}
\label{eq:3p1virtual}
%---------------
& \hspace{-5ex}
\langle g^{a,\alpha} (l) | {\cal W}^{b}_\beta (^3P^{[1]}) | 0\rangle
|_{\textrm{virtual}}
\nonumber \\
&=
i g^2 C_A \delta^{ab}
p^\mu
( l_\mu \delta_\beta^{\alpha} - l_\beta \delta^{\alpha}_\mu )
\int_k
\frac{-i}{k^2+i \varepsilon}
p \cdot n
\frac{i}{-k \cdot n+i \varepsilon}
\nonumber \\ & \quad \times
\left[
\left( \frac{i}{l \cdot p+i \varepsilon}\right)^2
\frac{i}{(l+k) \cdot p +i \varepsilon}
+
 \frac{i}{l \cdot p+i \varepsilon}
\left(\frac{i}{(l+k) \cdot p +i \varepsilon} \right)^2
\right].
%---------------
\end{align}
%---------------
Apart from the overall color tensor, this is exactly the same as the
same diagram for the $^3P^{[8]}$ soft function. 
The UV-divergent contribution of the virtual planar diagram to the soft
function $S_{^3P^{[1]}}(z)$ is then given by 
%---------------
\begin{align}
\label{eq:S3p1_virtual}
%---------------
S_{^3P^{[1]}}(z) |_{\textrm{virtual, UV}}
&= - g^2 (d-2) (N_c^2-1)
\nonumber \\ & \quad \times
C_A \int d{\rm PS}_l
\frac{2 \pi \delta(l^+ - P^+(1-z))I(l)}{(l \cdot p)^2} +{\rm c.c.} ,
%---------------
\end{align}
%---------------
which is the same as Eq.~(\ref{eq:S3p8_virtual}) with the replacement
$(N_c^2-4)/N_c \to 1$ in the overall color factor. 
Now we consider the real planar diagram. 
the matrix element of the operator ${\cal W}^{b}_\beta (^3P^{[1]})$
between the vacuum and a two-gluon state from the gluon attachment on the
timelike Wilson line is given by
%---------------
\begin{align}
\label{eq:3p1real_a}
%---------------
& \hspace{-5ex} \langle g^{a_1,\mu_1}(k_1) g^{a_2,\mu_2}(k_2)
| {\cal W}^{b}_\beta (^3P^{[1]}) | 0
\rangle|_{\textrm{real (a)}}
\nonumber \\
&=
-ig
\left[
\frac{i}{p \cdot k_1+i \varepsilon}
\left(\frac{i}{p \cdot (k_1+k_2)+i \varepsilon} \right)^2
+
\left(
\frac{i}{p \cdot k_1+i \varepsilon} \right)^2
\frac{i}{p \cdot (k_1+k_2)+i \varepsilon}
\right]
\nonumber \\ & \quad \times
p^\mu
(k_1{}_\mu \delta^{\mu_1}_\beta - k_1{}_\beta \delta^{\mu_1}_\mu )
p^{\mu_2}
f^{a_2 a_1 b},
%---------------
\end{align}
%---------------
and the one from the gluon attachment on the lightlike Wilson line is
%---------------
\begin{align}
\label{eq:3p1real_b}
%---------------
& \hspace{-5ex} \langle g^{a_1,\mu_1}(k_1) g^{a_2,\mu_2}(k_2)
| {\cal W}^{b}_\beta (^3P^{[1]}) | 0
\rangle|_{\textrm{real (b)}}
\nonumber \\ &=
-i g
\frac{i}{n \cdot k_2 + i\varepsilon}
\left( \frac{i}{p \cdot k_1 +i \varepsilon} \right)^2
p^\mu
(k_1{}_\mu \delta^{\mu_1}_\beta - k_1{}_\beta \delta^{\mu_1}_\mu )
n^{\mu_2}
f^{a_1 a_2 b}.
%---------------
\end{align}
%---------------
Similarly to the virtual planar diagram, these results differ from the 
$^3P^{[8]}$ case only by overall color tensors. 
The UV-divergent part of the real planar diagram is given by 
%---------------
\begin{align}
\label{eq:S3p1_real_lightcone}
%---------------
S_{^3P^{[1]}} (z) |_{\textrm{real, UV}} &=
- \int d {\rm PS}_{k_1} \int d {\rm PS}_{k_2}
2 \pi \delta (k_1^++k_2^+-P^+ (1-z))
\nonumber \\ & \quad \times
g^2 (d-2) \frac{(N_c^2-1) (N_c^2-4)}{N_c} 
\frac{C_A p^+}
{ k_2^+ (k_1 \cdot p)^2 [ (k_1+k_2) \cdot p]}
+{\rm c.c.}, 
%---------------
\end{align}
%---------------
which is equal to Eq.~(\ref{eq:S3p8_real_lightcone}) with the replacement 
$(N_c^2-4)/N_c \to 1$. 

The UV power counting of the remaining diagrams is carried out in exactly the
same way as the $^3P^{[8]}$ case. The double virtual diagrams also vanish in
the $^3P^{[1]}$ case, because they are proportional to a vanishing color factor
$f^{abc} \delta^{bc} = 0$. 
For diagrams that do not vanish trivially due to color factors, the UV power
counting arguments for the $^3P^{[8]}$ soft function apply exactly the same
way. 
In the doubly real diagrams, 
the amplitude-level diagrams 3 and 4 do not appear in the
color-singlet case because the timelike Wilson lines in $S_{^3P^{[1]}} (z)$
terminate at the field-strength tensors. 
The explicit expressions for the amplitude-level results can be obtained from
the $S_{^3P^{[8]}} (z)$ case by the replacements 
$d^{bcd} \to \delta^{bd}$ and $\delta^{yc} \to 1$.
The result is that the contribution from the doubly real diagrams to
$S_{^3P^{[1]}} (z)$ do not involve UV divergences, except for the planar
diagram. 
Hence, the NLO threshold double logarithm in $S_{^3P^{[1]}} (z)$ is same
as $S_{^3P^{[8]}} (z)$, with the replacement $(N_c^2-4)/N_c \to 1$.

%==============================================================================
\subsubsection{Anisotropic contribution}
%==============================================================================

It remains to show that the anisotropic contribution 
$S_{^3P^{[8]}}^{TT} (z)$ does not contain threshold double logarithms. 
The strategy of the calculation is similar to the isotropic case. 
The UV-divergent contribution to the virtual planar diagram involves the same
one-loop integral $I(l)$, and is given by 
%---------------
\begin{align}
%---------------
S_{^3P^{[1]}}^{TT} (z)|_{\textrm{virtual, UV}} &=
-g^2 (N_c^2-1) C_A
\int d{\rm PS}_l \frac{2 \pi \delta(l^+ - P^+ (1-z))}{(l \cdot p)^{2}} I(l)
\nonumber \\ & \quad \times
\left(
\frac{p^2}{p_+^2}
\frac{l_+^2 p^2 -2 p^+ l^+ l \cdot p}{( l \cdot p)^2}
+ \frac{d-2}{d-1} \right)+{\rm c.c.}. 
%---------------
\end{align}
%---------------
Similarly to the isotropic case, the mixed double pole in $I(l)$ is canceled
by the soft IR divergence in the real planar diagram, so that the contribution
from the planar diagrams to $S_{^3P^{[8]}}^{TT} (z)$ is given by 
%---------------
\begin{align}
%---------------
S_{^3P^{[1]}}^{TT} (z)|_{\textrm{planar, UV}} &=
g^2 (N_c^2-1) C_A
\frac{\mu'^{2 \epsilon}}{(4 \pi)^{2-\epsilon}} 
\frac{2^{1-2 \epsilon} (p^2)^\epsilon}{\epsilon_{\rm UV}^2}
\int d{\rm PS}_l \frac{2 \pi \delta(l^+ - P^+ (1-z))}{(l \cdot p)^{2+2 \epsilon}} 
\nonumber \\ & \quad \times
\left(
\frac{p^2}{p_+^2}
\frac{l_+^2 p^2 -2 p^+ l^+ l \cdot p}{( l \cdot p)^2}
+ \frac{d-2}{d-1} \right).
%---------------
\end{align}
%---------------
The phase-space integral is almost the same as the leading-order case, except
for an additional factor of $(l \cdot p)^{-2 \epsilon}$ in the integrand. 
The result reads 
%---------------
\begin{align}
%---------------
S_{^3P^{[1]}}^{TT} (z)|_{\textrm{planar, UV}} &= g^2
\frac{3 (1-\epsilon) (1-4 \epsilon) \Gamma(3\epsilon_{\rm UV})}{2 (3-2 \epsilon)
\Gamma(4+2 \epsilon)} \frac{\mu'^{2 \epsilon}}{\pi^{3-2 \epsilon}
P^+ (p^2)^{1+2 \epsilon}} \frac{1}{(1-z)^{1+4 \epsilon}}.
%---------------
\end{align}
%---------------
Hence, the 
planar diagram contribution to 
$S_{^3P^{[8]}}^{TT} (z)$ involves at most a single logarithm. 

As was in the isotropic contribution, the doubly virtual diagrams vanish for
the anisotropic contribution, as $S_{^3P^{[1]}}^{TT} (z)$ involves the same
color tensor as $S_{^3P^{[1]}} (z)$, they are proportional to a vanishing color
factor. 
Similarly to the calculation of the planar diagram, 
the virtual-real diagram contributions to $S_{^3P^{[1]}}^{TT} (z)$
always involve an explicit factor of $\epsilon$, which renders the single-UV
poles in the virtual one-loop diagrams finite. 
The doubly real diagrams need calculation, which can be done by using the
amplitude-level results for $S_{^3P^{[8]}} (z)$ with the replacements 
$d^{bcd} \to \delta^{bd}$ and $\delta^{yc} \to 1$. 
By explicit calculation, we find that the potentially UV-divergent integrals
that appear in the doubly real diagrams for $S_{^3P^{[1]}}^{TT} (z)$ are the
same ones that appeared in $S_{^3P^{[8]}} (z)$, which are in fact UV finite. 
Hence, the only UV-divergent contributions to $S_{^3P^{[1]}}^{TT} (z)$
come from the planar diagrams, which only produce single logarithms.

%==============================================================================
\subsubsection{Results and discussion}
%==============================================================================

As we have shown by explicit calculation, the threshold double logarithm 
in the isotropic contribution $S_{^3P^{[1]}} (z)$ is equal to the
$S_{^3P^{[8]}} (z)$, with the replacement $(N_c^2-4)/N_c \to 1$ in the overall
color factor. Thus if we collect only the contributions relevant for threshold
double logarithms, we have 
%---------------
\begin{align}
\label{eq:s3p1_nloresult2}
%---------------
S_{^3P^{[1]}} (z) |_{\rm threshold}
&=
\frac{ N_c^2-1 }{\pi m^2 P^+}
\bigg\{
- \frac{1}{(1-z)_+}
%\nonumber \\ & \quad
+ \frac{\alpha_s C_A}{\pi}
\bigg[
4 \left( \frac{\log^2 (1-z)}{1-z} \right)_+
\nonumber \\ & \quad
+ \frac{1}{\epsilon_{\rm UV}} \left( \frac{-2 \log (1-z)}{1-z} \right)_+
+ \frac{1}{2 \epsilon_{\rm UV}^2} \frac{1}{(1-z)_+}
\bigg]
+O(\alpha_s^2) \bigg\},
%---------------
\end{align}
%---------------
whose Mellin transform is
%---------------
\begin{align}
\label{eq:s3p1_nloresult3}
%---------------
\tilde{S}_{^3P^{[1]}} (N) |_{\rm threshold}
& =
\frac{
N_c^2-1
}{\pi m^2 P^+}
(-\log N)
\nonumber \\ & \quad \times
\bigg[
1
+ \frac{\alpha_s C_A}{\pi}
\left(
- \frac{4}{3} \log^2 N
+ \frac{\log N}{\epsilon_{\rm UV}}
- \frac{1}{2 \epsilon_{\rm UV}^2}
\right)
+O(\alpha_s^2) \bigg].
%---------------
\end{align}
%---------------
The anisotropic contribution $S_{^3P^{[1]}}^{TT} (z)$ does not contain
threshold double logarithms, because the NLO correction only involves single UV
poles. This implies that the coefficient of the threshold double logarithmic
correction factor in the $^3P_J^{[1]}$ fragmentation functions are independent
of the total spin $J$ of the $Q \bar Q$ state, as was found in the explicit
calculation of the fragmentation function at NLO in Ref.~\cite{Zhang:2020atv}. 
We also expect
that, from the soft factorization of polarized $^3P_J^{[1]}$ fragmentation
functions, that the threshold double logarithmic correction factor 
is also independent of the $Q \bar Q$ helicity. 

The calculation of the isotropic soft function $S_{^3P^{[1]}} (z)$ can be
checked against the calculation in Ref.~\cite{Bodwin:2019bpf}. 
The calculation Ref.~\cite{Bodwin:2019bpf} at leading nonvanishing power in
the expansion in $q$,
corresponds to the first moment of the isotropic soft function 
$\tilde{S}_{^3P^{[1]}} (N=1) = \int_0^1 dz \, S_{^3P^{[1]}} (z)$. 
The sum of diagrams ${\cal C}$ and ${\cal D}$ in Ref.~\cite{Bodwin:2019bpf}
corresponds to the contribution from the planar diagrams in this work. 
In Ref.~\cite{Bodwin:2019bpf}, the UV divergences in the phase-space integral
have been regulated by using a cutoff, while the remaining UV divergences are
regulated dimensionally. This approach was used in Ref.~\cite{Bodwin:2019bpf}
because the aim was to isolate the IR divergences in the NRQCD LDMEs;
otherwise, the diagrams would have vanished due to the scaleless nature of loop
corrections to NRQCD LDMEs. 
This makes it easier for us to compare the results between
Ref.~\cite{Bodwin:2019bpf} and our calculation of $S_{^3P^{[1]}} (z)$. 
In Ref.~\cite{Bodwin:2019bpf}, it was found that only the diagrams ${\cal C}$
and ${\cal D}$ contained double UV poles, consistently with our results for
planar diagrams. The vanishing of diagrams ${\cal A}$ and ${\cal B}$, as well
as the nonappearance of UV divergences in diagrams ${\cal E}$ and ${\cal F}$
in Ref.~\cite{Bodwin:2019bpf} are also consistent with our analysis of
nonplanar diagrams.

%==============================================================================
\subsection{Summary of NLO results}
%==============================================================================

The results for the threshold behavior of the NLO calculations of the soft 
functions are shown in Eqs.~(\ref{eq:S3S8NLOMellin}), (\ref{eq:s3p8_nloresult3}), 
and (\ref{eq:s3p1_nloresult3}) for the $^3S_1^{[8]}$, $^3P^{[8]}$, and
$^3P_J^{[1]}$ channels, respectively. 
The large-$N$ behavior of the results can be written in the form 
%------------
\begin{subequations}
\begin{align}
%------------
\tilde{S}_{^3S_1^{[8]}} (N) |_{\rm threshold}
&=
\left[ 1 +
\frac{\alpha_s C_A}{\pi} \left( - \log^2 N 
+ \frac{\log N}{\epsilon_{\rm UV}} - \frac{1}{2 \epsilon_{\rm UV}^2} \right)
+ O(\alpha_s^2) \right]
\tilde{S}_{^3S_1^{[8]}}^{\rm LO} (N), 
\\
\tilde{S}_{^3P^{[8]}} (N)|_{\rm threshold}
&=
\left[ 1 +
\frac{\alpha_s C_A}{\pi} \left( - \frac{4}{3} \log^2 N
+ \frac{\log N}{\epsilon_{\rm UV}} - \frac{1}{2 \epsilon_{\rm UV}^2} \right)
+ O(\alpha_s^2) \right]
\tilde{S}_{^3P^{[8]}}^{\rm LO} (N), 
\\
\tilde{S}_{^3P^{[1]}} (N)|_{\rm threshold}
&= 
\left[ 1 +
\frac{\alpha_s C_A}{\pi} \left( - \frac{4}{3} \log^2 N
+ \frac{\log N}{\epsilon_{\rm UV}} - \frac{1}{2 \epsilon_{\rm UV}^2} \right)
+ O(\alpha_s^2) \right]
\tilde{S}_{^3P^{[1]}}^{\rm LO} (N). 
%------------
\end{align}
\end{subequations}
%------------
Since $C_A = N_c$, the NLO finite pieces exactly reproduce the threshold double
logarithms in the FFs given in Eq.~(\ref{eq:doublethresholdFF}).
The single pole term is exactly the large-$N$ behavior of the Mellin transform
of the $2 C_A z/(1-z)_+$ term in the gluon splitting function $P_{gg}(z)$. 
The double pole $1/\epsilon_{\rm UV}^2$, which does not depend on $N$ or the
final state, can be absorbed into the NLO Wilson coefficient of the soft
factorization formula. 
The universality of the single and double pole terms imply that the threshold
double logarithms come from a single universal origin, namely, the cusp
anomalous dimension that gives rise to the $2 C_A z/(1-z)_+$ term in $P_{gg}(z)$. 
However, unlike the pole terms the coefficients of the $\log^2 N$ term depend 
on the channel; because the threshold double logarithm is associated with an
infrared divergence, its coefficient depends on the final state.

%==============================================================================
\section{Resummation}
\label{sec:resum}
%==============================================================================

Since the threshold double logarithm always occurs from the planar diagram, 
the resummation is straightforward; an exchange of a gluon with momentum $l$ 
between the timelike and lightlike Wilson lines implies a loss of quarkonium
momentum by $l$, or equivalently, an increase in the momentum of the 
fragmenting gluon by $l$. Thus, the effect of adding this gluon exchange to the
tree-level soft function $S^{\rm LO}_{\cal N} (z)$ for channel $\cal N$ 
can be written in the form 
%---------------
\begin{align}
\label{eq:NLOKernel}
%---------------
\int_0^{K^+ (1-z)} \frac{dl^+}{l^+} {\cal K} (l^+) \left[ 
S^{\rm LO}_{\cal N} (z+l^+/K^+) - S^{\rm LO}_{\cal N} (z) \right], 
%---------------
\end{align}
%---------------
where ${\cal K} (l^+)$ is a kernel that can be determined from explicit
calculation of the gluon exchange diagram. The subtraction term 
$- S^{\rm LO}_{\cal N} (z)$ subtracts the $l^+ = 0$ singularity so that only
the finite piece remains in Eq.~(\ref{eq:NLOKernel}). 
By using change of variables, 
this integral can be rewritten as a convolution integral that is diagonalized 
in Mellin space. We define $z'$ through $z/z' \equiv z+l^+/K^+$, 
so that Eq.~(\ref{eq:NLOKernel}) can be rewritten as 
%---------------
\begin{align}
\label{eq:NLOKernel2}
%---------------
\int_0^z \frac{dz'}{z'} 
\left[ \frac{{\cal K}(P^+ (1-z')/z')}{1-z'} \right]_+ 
S^{\rm LO}_{\cal N} (z/z') 
= \left[ \frac{{\cal K}(P^+ (1-z)/z)}{1-z} \right]_+ 
\otimes S^{\rm LO}_{\cal N} (z). 
%---------------
\end{align}
%---------------
From our NLO calculation of the soft functions we can identify the Mellin
transform of the kernel $\left[ {\cal K}(P^+ (1-z)/z)/(1-z) \right]_+$ as 
%---------------
\begin{align}
%---------------
J_{^3S_1^{[8]}}^N = \frac{\alpha_s C_A}{\pi} 
\int_0^1 \frac{dz}{z} z^N
\left[ \frac{-2 \log (1-z)}{1-z} \right]_+, 
%---------------
\end{align}
%---------------
for the $^3S_1^{[8]}$ channel, and 
%---------------
\begin{align}
%---------------
J_{^3P^{[8]}}^N &= \frac{4 \alpha_s C_A}{3 \pi} 
\int_0^1 \frac{dz}{z} z^N
\left[ \frac{-2 \log (1-z)}{1-z} \right]_+, 
\\
J_{^3P^{[1]}}^N &= \frac{4 \alpha_s C_A}{3 \pi} 
\int_0^1 \frac{dz}{z} z^N
\left[ \frac{-2 \log (1-z)}{1-z} \right]_+, 
%---------------
\end{align}
%---------------
for the $^3P^{[8]}$ and $^3P^{[1]}$ channels. Note that 
$J_{^3P^{[8]}}^N = J_{^3P^{[1]}}^N = \dfrac{4}{3} J_{^3S_1^{[8]}}^N$,
and at large $N$, $J_{^3S_1^{[8]}}^N \sim
- \dfrac{\alpha_s C_A}{\pi} \log^2N$. 
These expressions follow from the identities
%---------------
\begin{align}
%---------------
\left[ \frac{-2 \log (1-z)}{1-z} \right]_+ \otimes \delta(1-z)
&= 
\left[ \frac{-2 \log (1-z)}{1-z} \right]_+, 
\\
\left[ \frac{-2 \log (1-z)}{1-z} \right]_+ \otimes \frac{1}{(1-z)_+}
&= 
\left[ \frac{3 \log^2 (1-z)}{1-z} \right]_+ - \frac{\pi^2}{3} \frac{1}{(1-z)_+}
\nonumber \\ & \quad 
+ 2 \zeta(3) \delta(1-z) + \frac{2 \log z \log(1-z)}{1-z}. 
%---------------
\end{align}
%---------------
It is straightforward to generalize this result to a ladder of an arbitrary 
number insertions of
planar gluons. Because the insertions of planar gluons onto the temporal
Wilson line are time ordered, in Mellin space the gluon insertions 
exponentiate in the form~\cite{Laenen:2000ij}
%---------------
\begin{align}
%---------------
\tilde{S}^{\rm resum}_{\cal N} (N)
&= \exp \left[ J^N_{\cal N} \right] \tilde{S}^{\rm LO}_{\cal N} (N).
%---------------
\end{align}
%---------------
Since the $z \to 1$ behavior of the fragmentation function is contained in the
soft function, the same expression holds for $D^{\rm LO} (z)$. 
%---------------
\begin{align}
\label{eq:DLOresum}
%---------------
\tilde{D}^{\rm resum}_{g\to Q \bar Q({\cal N})} (N)
&= \exp \left[ J^N_{\cal N} \right] \tilde{D}^{\rm LO}_{g\to Q \bar Q({\cal
N})} (N).
%---------------
\end{align}
%---------------
This expression contains all of the leading double threshold logarithms 
$\tilde{D}^{\rm LO}_{g\to Q \bar Q({\cal N})} (N) 
\times \alpha_s^n \log^{2n} N$ 
in the all-orders result for the fragmentation function, while at leading order
in $\alpha_s$ coincides with $\tilde{D}^{\rm LO}_{g\to Q \bar Q({\cal N})} (N)$. 
We can also combine the NLO correction term in the fixed-order calculation of
the fragmentation function to obtain the expression
%---------------
\begin{align}
\label{eq:DNLOresum}
%---------------
\tilde{D}^{\rm FO+resum}_{g\to Q \bar Q({\cal N})} (N)
&= \exp \left[ J^N_{\cal N} \right] 
\left( \tilde{D}^{\rm FO}_{g\to Q \bar Q({\cal N})} (N) - 
J^N_{\cal N} \tilde{D}^{\rm LO}_{g\to Q \bar Q({\cal N})} (N) \right), 
%---------------
\end{align}
%---------------
where $\tilde{D}^{\rm FO}_{\cal N} (N)$ is the fixed-order fragmentation
function computed to NLO in $\alpha_s$, while the last term in the parenthesis
subtracts the NLO double logarithm in the fixed-order result to avoid double
counting. In both Eqs.~(\ref{eq:DLOresum}) and (\ref{eq:DNLOresum}), 
the Mellin-space formulas for resummed fragmentation functions vanish as $N \to
\infty$, because the factor $\exp \left[ J^N_{\cal N} \right]$ vanishes faster
than any power of $N$. As a result, the resummed fragmentation functions in $z$
space are regular functions that vanish at $z = 1$. 

Note that there is some freedom in choosing the exact form of the Mellin-space
kernel $J^N_{\cal N}$. Since we are only interested in resumming the leading
double threshold logarithms, any form that has the same double logarithmic 
large $N$ behavior will reproduce the 
$\tilde{D}^{\rm LO}_{g\to Q \bar Q({\cal N})} (N) \times
\alpha_s^n \log^{2n} N$ terms in the resummed expression. 
The form that we choose is similar to the one that is commonly used in the
Mellin-space resummation of threshold logarithms in perturbative QCD
calculations; it has the advantage that the $z$-space expression for the kernel 
$J^N_{\cal N}$ vanishes quickly as $z \to 0$, so that only the behavior of the
fragmentation function near the threshold $z =1$ is affected by resummation. 
It also preserves the total fragmenting probability 
$\int_0^1 dz \, D(z)$, because $J^N_{\cal N}$ vanishes at $N=1$. 

Our resummation formula in Eq.~(\ref{eq:DNLOresum}) differs from the usual ones 
found in literature in that we do not formulate it as a solution of an 
evolution equation (see, for example, Refs.~\cite{Kidonakis:1997gm,
Kidonakis:1998bk, Kidonakis:1998nf, Bonciani:2003nt}). 
In the usual treatment of threshold logarithms, resummation is carried out by 
evolving the
renormalization scale of the fragmentation function from a $z$-dependent scale 
$\mu_{i} = \mu_0 (1-z)$ to a $z$-independent fixed scale $\mu_f = \mu_0$. 
This is not suitable for the fragmentation functions we consider in this work, 
because the $D_{g \to Q \bar Q({\cal N})} (z)$ that we compute are NRQCD
short-distance coefficients that are determined through perturbative matching 
at the hard scale, while physics
below the scale of the heavy quark mass are contained in the NRQCD
long-distance matrix elements. 
Computing the fragmentation function at a $z$-dependent renormalization scale
is thus inconsistent with the perturbative matching procedure. 
Instead, our resummation formula in
Eq.~(\ref{eq:DNLOresum}) allows resummation of threshold logarithms while
working at a fixed hard scale, consistently with the perturbative matching
procedure in NRQCD. 

The resummation formula in Eq.~(\ref{eq:DNLOresum}) first appeared in
Ref.~\cite{Chung:2024jfk}, which was used to explain the large-$p_T$
measurement of prompt $J/\psi$ production rates in Ref.~\cite{ATLAS:2023qnh}. 
One unexpected consequence of the resummation formula that was found in
Ref.~\cite{Chung:2024jfk} is that the vanishing of the resummed FFs at the
threshold $z=1$ brings in reduction of the sizes of the individual SDCs 
$d \sigma_{Q \bar Q ({\cal N})}$ for the color-octet channels. Because of this, 
the strong sensitivity on the color-octet matrix elements of the quarkonium
production rates computed in the NRQCD factorization formalism is fairly
reduced. This is evident in the calculation of $\chi_{c1} (3872)$ production
rates in Ref.~\cite{Lai:2025tpw}, based on the hypothesis that it is a
hidden-heavy tetraquark where the $c \bar c$ are mostly in the $^3S_1^{[8]}$
state~\cite{Grinstein:2024rcu, Braaten:2024tbm, Brambilla:2024imu}. 
In Ref.~\cite{Lai:2025tpw} it was found that it is necessary to use the
resummed SDCs in order to explain the prompt $\chi_{c1} (3872)$ production
rates at the LHC, as the fixed-order calculation leads to an overestimation of
the cross sections, especially when compared to the nonprompt production rate. 
In Refs.~\cite{Wang:2025drz, Wang:2026dul}, resummed FFs were used to compute the transverse
momentum distributions of $J/\psi$ and $\psi(2S)$ within jets. 
Use of the resummed FFs in Eq.~(\ref{eq:DNLOresum}) is crucial in calculations
of momentum distributions, as without resummation it is not possible to obtain
smooth distributions that are well behaved all the way up to the threshold 
$z=1$.

%==============================================================================
\section{Conclusions}
\label{sec:conclusions}
%==============================================================================

In this paper, we presented an all-orders analysis of threshold double
logarithms that appear in quarkonium fragmentation functions in the
nonrelativistic QCD factorization formalism. As is known from
existing calculations, fixed-order calculations of quarkonium fragmentation
functions involve singularities at the kinematical threshold due to emission of
soft gluons, with the strongest singularities coming from the threshold double
logarithms. While these singularities are integrable, this makes it impossible
to ensure the positivity of quarkonium production rates in fixed-order
perturbation theory, because in this case the fragmentation functions can never
be positive definite for any choice of nonperturbative NRQCD matrix elements.
This problem can only be resolved by resumming the threshold double logarithms
to all orders in perturbation theory. 

Resummation of threshold double logarithms can be carried out through soft
factorization, which helps identify and isolate the source of threshold
singularities. This procedure, which is based on standard methods of 
factorization in perturbative QCD, is discussed in detail in 
Sec.~\ref{sec:softfac}. 
Calculation of the soft functions at leading and next-to-leading
orders in the strong coupling are shown in Secs.~\ref{sec:softLO} and 
\ref{sec:softNLO}, respectively. 
Through these calculations we showed that the threshold singularities in the
fragmentation functions are indeed completely contained in the soft functions, 
and the threshold double logarithms can only arise from a specific type of
planar diagram. As shown in Sec.~\ref{sec:resum}, this allows us to resum the
threshold double logarithms to all orders in the strong coupling, in a similar
manner as the usual resummation of Sudakov double logarithms. 
While the result for the resummed fragmentation function was first presented in
an earlier publication in Ref.~\cite{Chung:2024jfk}, the detailed derivation 
and calculation of the soft function is shown here for the first time. 
The analysis presented in this work also makes it clear that the resummation 
formula for quarkonium fragmentation functions also applies to polarized
fragmentation functions, which was argued without proof in 
Ref.~\cite{Chung:2024jfk}. 

As was shown in Ref.~\cite{Chung:2024jfk}, only after threshold double
logarithms are resummed to all orders is it possible to construct quarkonium
fragmentation functions that are positive definite. This is integral in
establishing a proper description of prompt $J/\psi$ production rates at very
large transverse momentum. 
The calculation presented in this work provides the detailed derivation of the 
resummation formalism used in Ref.~\cite{Chung:2024jfk}. 
The use of resummed fragmentation functions was crucial in phenomenological
studies of large-$p_T$ production rates of $J/\psi$ in
Ref.~\cite{Chung:2024jfk}, prompt production rates of $\chi_{c1}(3872)$ in
Ref.~\cite{Lai:2025tpw}, and the transverse momentum distribution of $J/\psi$
and $\psi(2S)$ in jets in Refs.~\cite{Wang:2025drz, Wang:2026dul}.

While the problem of singularities in quarkonium fragmentation functions is
essentially resolved by resumming threshold logarithms at leading double
logarithmic level, the accuracy of resummation may be improved by computing the
soft functions at single pole level, and also at higher orders in the strong
coupling. Currently the calculation at single-pole accuracy has only been done
for the $^3S_1^{[8]}$ channel, due to complications involving rapidity
divergences that appear in the $^3P^{[8]}$ and $^3P^{[1]}_J$ channels. 
It may also be interesting to extend the resummation formalism to
next-to-leading power in the $1/N$ expansion; such an analysis may be useful 
in resolving the poor convergence of the velocity expansion in the
color-singlet channel, where the fragmentation functions are finite but nonzero
at threshold~\cite{Bodwin:2003wh, Bodwin:2012xc}.

\appendix
\section{MELLIN TRANSFORM}
\label{sec:mellin}

In this appendix we collect useful formulas on Mellin transforms that we use
throughout this paper. We define the Mellin transform of a function $f$ by 
%------------
\begin{align}
%------------
{\tilde f}(N) = \int_0^1 \frac{dz}{z} \, z^N f(z).
%------------
\end{align}
%------------
While in the general theory of Mellin transforms the upper limit of integration 
is $+\infty$, in this work we will only deal with functions that vanish outside
the kinematically allowed region $0\le z\le 1$, so that throughout this paper
the upper limit of the $z$ integration is always $1$. 
The Mellin transform diagonalizes the convolution 
%------------
\begin{align}
%------------
\int_0^1 \frac{dz}{z} \, z^N 
\left(f \otimes g\right) (z) 
= \tilde{f}(N) \tilde{g}(N), 
%------------
\end{align}
%------------
where $\otimes$ is defined by 
%------------
\begin{align}
%------------
\left(f \otimes g\right) (z) = \int^1_z \frac{dz'}{z'} f(z') g(z/z').
%------------
\end{align}
%------------
The original function $f(z)$ can be restored from ${\tilde f}(N)$ through the
inverse Mellin transform
%------------
\begin{align}
%------------
f(z) = \int_{c-i \infty}^{c+i \infty} \frac{dN}{2 \pi i} z^{-N}
{\tilde f}(N),
%------------
\end{align}
%------------
where $c$ is chosen so that all poles of ${\tilde f}(N)$ are on the left of the
integration contour.
It is evident that ${\tilde f}(N)$ must vanish faster than $1/N$ as
$N \to \infty$ for the contour integral to be well defined at $z=1$.
If ${\tilde f}(N)$ decreases like $1/N$, $f(z)$ at $z=1$ can still be found by
the limit $z \to 1^-$, and the inverse transform still gives a continuous
function of $z$. However, if ${\tilde f}(N)$ does not vanish as $N \to \infty$,
the inverse Mellin transform instead defines a distribution that is singular at
$z=1$. 
We list below the Mellin transforms of distributions that appear in this paper: 
%------------
\begin{subequations}
\begin{eqnarray}
%------------
\int_0^1 \frac{dz}{z} \, z^N \delta(1-z) &=& 1,
\\
\int_0^1 \frac{dz}{z} \, z^N \frac{1}{(1-z)_+} &=&
-H_{N-1}, 
\\
\int_0^1 \frac{dz}{z} \, z^N \left[ \frac{\log(1-z)}{1-z} \right]_+
&=& + \frac{1}{2} H_{N-1}^2 
- \frac{1}{2} \psi^{(1)} (N) + \frac{\pi^2}{12}, 
\\
\int_0^1 \frac{dz}{z} \, z^N \left[ \frac{\log^2(1-z)}{1-z} \right]_+
&=& - \frac{1}{3} H_{N-1}^3 - \frac{\pi^2}{6} H_{N-1} + H_{N-1} \psi^{(1)} (N)
\nonumber \\ && 
- \frac{1}{3} \psi^{(2)} (N) - \frac{2}{3} \zeta(3), 
%------------
\end{eqnarray}
\end{subequations}
%------------
where $H_N = \psi(N+1) + \gamma_{\rm E}$ is the harmonic number, $\psi(z) =
\Gamma'(z)/\Gamma(z)$ is the digamma function, $\gamma_{\rm E}$ is the
Euler-Mascheroni constant, and $\psi^{(n)} (z) = \frac{d^n}{dz^n} \psi(z)$ is
the polygamma function. From the large-$z$ asymptotic behavior of the digamma
function $\psi(z) \sim \log z$, we obtain the large-$N$ asymptotic behaviors 
of the plus distributions in Mellin space given by
%------------
\begin{align}
%------------
\int_0^1 \frac{dz}{z} \, z^N \left[ \frac{\log^{m-1}(1-z)}{1-z} \right]_+
\sim \frac{(-1)^{m}}{m} \log^m N, 
%------------
\end{align}
%------------
for $m = 1$, 2, 3 $\ldots$. 

%==============================================================================
\section{\boldmath HELICITY DECOMPOSITION FOR $^3P_J^{[1]}$ FFS}
\label{sec:helicitydecomp}
%==============================================================================

The polarized $^3P_J^{[1]}$ gluon FFs have first been computed in
Ref.~\cite{Cho:1994gb} in cutoff regularization. These results can be
translated to dimensional regularization based on the method developed in
Ref.~\cite{Braaten:1996rp}. We obtain
%---------------
\begin{align}
\label{eq:dg3P1pol}
%---------------
D^{\rm soft}_{g \to Q \bar Q (^3P_{J,h}^{[1]})}(z) &=
\frac{4 \alpha_s^2}{3 (d-1) m^5 N_c^2}
\bigg[
\left( \frac{Q_{J,h}}{2 J+1} - a_{J,h}\log \frac{\mu_\Lambda}{2 m} \right) 
\delta(1-z) 
\nonumber \\ & \hspace{22ex} 
+
\frac{a_{J,h}}{(1-z)_+}
+ \frac{P_J^h (z)}{2J+1}
\bigg]
+O(\alpha_s^3),
%---------------
\end{align}
%---------------
where 
%---------------
\begin{subequations}
\begin{align}
%---------------
a_{1,0} &= a_{1,1} + a_{1,-1} = \frac{1}{2}, 
\\
\frac{Q_{1,0}}{3} &= \frac{1}{8}, \quad 
\frac{Q_{1,1} + Q_{1,-1}}{3} = 0, \\
a_{2,0} &= \frac{1}{10}, \quad
a_{2,1}+a_{2,-1} = \frac{3}{10}, \quad
a_{2,2}+a_{2,-2} = \frac{3}{5}, 
\\
\frac{Q_{2,0}}{5} &= \frac{1}{40}, \quad
\frac{Q_{2,1}+Q_{2,-1}}{5}=0, \quad
\frac{Q_{2,2}+Q_{2,-2}}{5} = \frac{3}{20}, 
\\
\frac{P_1^0 (z)}{3} &= - \frac{z}{4} - \frac{z^2}{2},
\quad 
\frac{P_1^1 (z)+ P_1^{-1} (z)}{3} = - \frac{z^2}{2}, \\
\frac{P_2^0(z)}{5} &= - \frac{1}{20 z^4} 
\big[ z (2 z^5+5 z^4+36 z^3-468 z^2+864 z-432 )
\nonumber \\ & \hspace{15ex} 
+ 216 (z-2) (z-1)^2 \log (1-z)
\big], 
\\
\frac{P_2^1 (z)+ P_2^{-1} (z)}{5} &= 
- \frac{3}{10 z^4} 
\big[
z (z^5+4 z^4-56 z^3+152 z^2-192 z+96 )
\nonumber \\ & \hspace{15ex} 
+ 24 ( z^4-5 z^3+10 z^2-10 z+4 ) \log (1-z)
\big ],\\
\frac{P_2^2 (z)+ P_2^{-2} (z)}{5} &= 
-\frac{3}{5 z^4} 
\big[ z (z^5-7 z^4+25 z^3-37 z^2+24 z-12 )
\nonumber \\ & \hspace{15ex} 
+3 (z^5-6 z^4+14 z^3-16 z^2+10 z-4 ) \log
   (1-z) \big]. 
%---------------
\end{align}
\end{subequations}
%---------------
These expressions agree with the results shown in Ref.~\cite{Ma:2015yka}.

%==============================================================================
\section{LIGHT-CONE COORDINATES}
%==============================================================================

We use the following definition for light-cone coordinates 
%------------
\begin{align}
%------------
a_\pm = a^\pm = \frac{1}{\sqrt{2}} (a^0 \pm a^3), 
%------------
\end{align}
%------------
so that 
%------------
\begin{align}
%------------
a \cdot b &= a^+ b^- + a^- b^+ - \bm{a}_\perp \cdot \bm{b}_\perp. 
%------------
\end{align}
%------------
The $d$-dimensional integral is given by 
%------------
\begin{align}
%------------
\int d^dk &= \int_{-\infty}^{+\infty} dk^+ \int_{-\infty}^{+\infty} dk^- 
\int d^{d-2} k_\perp. 
%------------
\end{align}
%------------
The $k_\perp$ integral can be computed by using 
%------------
\begin{align}
\label{eq:kperpint}
%------------
\int d^{d-2} k_\perp =
\frac{2 \pi^{\frac{d-2}{2}}}{\Gamma( \frac{d-2}{2})}
\int_0^\infty |\bm{k}_\perp|^{d-3} d |\bm{k}_\perp|
= \frac{2 \pi^{1-\epsilon}}{\Gamma( 1-\epsilon)}
\int_0^\infty |\bm{k}_\perp|^{1-2 \epsilon} d |\bm{k}_\perp|
%------------
\end{align}
%------------
for $d=4-2 \epsilon$.

\begin{acknowledgments}
We thank Geoffrey Bodwin for fruitful discussions on resummation of threshold 
logarithms. 
The work of H.~S.~C. is supported by the 
Basic Science Research Program through the National Research Foundation of
Korea (NRF) funded by the Ministry of Education (Grant No. RS-2023-00248313)
and a Korea University grant. 
This work is also supported in part by the National
Research Foundation of Korea (NRF) under Grants No. RS-2025-24222969 (J.~L.)
and No. RS-2025-24533579 (U-R.~K.).

All authors contributed equally to this work.
\end{acknowledgments}

\bibliography{threshold.bib}

\end{document}